\PassOptionsToPackage{square,comma,numbers,sort&compress,super}{natbib}
\documentclass[12pt,twocolumn,twoside]{pnas-new}
\usepackage{balance} % balance last page of references
\usepackage{fancyvrb}

\usepackage{adjustbox}
\usepackage{setspace}

\usepackage{stackengine,scalerel}

\templatetype{pnasresearcharticle} % Choose template 
\usepackage{booktabs}
\usepackage{subfig}
\renewcommand{\marginpar}[1]{}

\usepackage{mathtools}
\usepackage{times}
\usepackage{latexsym}
\usepackage{multirow}
\usepackage{arydshln}

\usepackage{graphicx}
\usepackage{ulem}	% to define \uline
\usepackage{examples}
\usepackage{float}

\usepackage{pdfpages}
\usepackage{xspace}
\usepackage{amssymb,amsmath,epsfig}
\usepackage{mathpartir}
\usepackage{amsthm}
\usepackage{mathrsfs}
\usepackage{algorithmicx}
\usepackage[noend]{algpseudocode}

\usepackage{hhline}

\newcommand{\nts}{\ensuremath{\text{\it nt}}\xspace}
\newcommand{\pairs}{\ensuremath{\text{pairs}}\xspace}
\newcommand{\pairings}{\ensuremath{\text{pairings}}\xspace}
\newcommand{\unpaired}{\ensuremath{\text{unpaired}}\xspace}

\newcommand{\ppv}{\ensuremath{\text{PPV}}\xspace}
\newcommand{\sens}{\ensuremath{\text{Sensitivity}}\xspace}

\newcommand{\notes}[1]{}%{\it {\small {#1}}}}

\newcommand{\ith}[1]{\ensuremath{i^{{th}}}}

\newcount\permx
\newcount\permy
\def\permdot#1#2{
\permx=#1 \advance\permx by-1
\permy=#2 \advance\permy by-1
\psframe[fillcolor=black, fillstyle=solid]
(\permx,\permy)(#1, #2)
}

\newcommand{\candidates}{\ensuremath{\mathit{candidates}}\xspace}

\newcommand{\myboxmath}[1]{\ensuremath{\fcolorbox{white}{lightgray}{\!\ensuremath{#1}\!}}\xspace}

\newcommand{\vecx}{\ensuremath{\mathbf{x}}\xspace}
\newcommand{\vecy}{\ensuremath{\mathbf{y}}\xspace}

\newcommand{\subscript}[2]{_{{#1}, {#2}}}
\newcommand{\Qf}[2]{\ensuremath{Q\subscript{{#1}}{{#2}}}\xspace}
\newcommand{\Qhat}{\ensuremath{\widehat{Q}}\xspace}
\newcommand{\Qhatf}[2]{\ensuremath{\Qhat\subscript{{#1}}{{#2}}}\xspace}

\newcommand{\vecystar}{\ensuremath{\vecy^*}\xspace}

\newcommand{\smallnt}[1]{\ensuremath{_{\mbox{\tiny PP}}}\xspace}

\iffalse

\else

\fi

\newcommand{\smallurl}[1]{{\scriptsize \url{#1}}}

\newcommand{\pop}{\ensuremath{\text{\sc pop}}\xspace}

\newcommand{\nskip}{\ensuremath{\text{\sc skip}}\xspace}

\newcommand{\md}{\ensuremath{\text{\tt .}}}

\newcommand{\mla}{\ensuremath{\text{\tt (}}}
\newcommand{\mra}{\ensuremath{\text{\tt )}}}

\newcommand{\ml}{\mla}
\newcommand{\mr}{\mra}

\newcommand{\ecoli}{{\em E.~coli}\xspace}

\newcommand{\linearfold}{{LinearFold}\xspace}
\newcommand{\linearpartition}{{LinearPartition}\xspace}
\newcommand{\linearpartitionc}{{LinearPartition-C}\xspace}
\newcommand{\linearpartitionv}{{LinearPartition-V}\xspace}

\newcommand{\linearfoldv}{{LinearFold-V}\xspace}
\newcommand{\contrafold}{{CONTRAfold}\xspace}

\newcommand{\viennarnafold}{{Vienna RNAfold}\xspace}
\newcommand{\viennarnaplfold}{{Vienna RNAplfold}\xspace}
\newcommand{\rnafold}{{RNAfold}\xspace}

\newcommand{\panel}[1]{\large \sf {#1}}

\newcommand{\linear}{\ensuremath{\mathrm{linear}}\xspace}
\newcommand{\vienna}{\ensuremath{\mathrm{vienna}}\xspace}
\newcommand{\ensenergy}{\ensuremath{\Delta G^\circ_\text{ensemble}}\xspace}
\newcommand{\ensenergyvienna}{\ensuremath{\Delta G^{\circ\ \text{vienna}}_\text{ensemble}}\xspace}
\newcommand{\ensenergylinear}{\ensuremath{\Delta G^{\circ\ \text{linear}}_\text{ensemble}}\xspace}
\newcommand{\ddg}{\ensuremath{\Delta\Delta G^{\circ}_\text{ensemble}}\xspace}

\newcommand{\RMSD}{\ensuremath{\text{\sc rmsd}}\xspace}

\newcommand{\mtwo}[1]{\multirow{2}{*}{{#1}}}

\newcommand{\threshknot}{\ensuremath{\text{ThreshKnot}}\xspace}
\newcommand{\probknot}{\ensuremath{\text{ProbKnot}}\xspace}

\fancypagestyle{plain}{%
\fancyhf{} % clear all header and footer fields
\fancyfoot[C]{\sffamily\fontsize{9pt}{9pt}\selectfont\thepage} % except the center

}
\title{\linearpartition: Linear-Time Approximation of RNA Folding Partition Function and Base Pairing Probabilities}

\author[a]{He Zhang}
\author[b]{Liang Zhang}
\author[c,d,e]{David H.~Mathews}
\author[a,b,$\clubsuit$]{Liang Huang}

\affil[a]{Baidu Research USA, Sunnyvale, CA 94089, USA}
\affil[b]{School of Electrical Engineering \& Computer Science,
  Oregon State University, Corvallis, OR 97330, USA}
\affil[c]{Dept. of Biochemistry \& Biophysics}
\affil[d]{Center for RNA Biology}
\affil[e]{Dept. of Biostatistics \& Computational Biology, University of Rochester Medical Center, Rochester, NY 14642, USA}

\leadauthor{Zhang} 

\iftrue%%%%%%%%%%%%%5
\correspondingauthor{
\vspace{-1.4cm}
}
\fi

\begin{abstract}
\vspace{-0.2cm}
RNA secondary structure prediction 
is widely used to understand RNA function.
Recently, there has been a shift away from the classical minimum free energy (MFE) methods to partition function-based 
methods that account for folding ensembles and can therefore estimate structure and base pair probabilities. 
However, the classical partition function algorithm scales cubically with sequence length, 
and 
is therefore a slow calculation
for long sequences.
This slowness is even more severe than cubic-time MFE-based methods due to a larger constant factor in runtime.
Inspired by the success of our recently proposed \linearfold algorithm that 
predicts the approximate MFE structure in linear time,
we %address this issue by proposing
design a similar linear-time heuristic algorithm,
\linearpartition, to approximate the partition function 
and base pairing probabilities, which is shown to be
orders of magnitude faster than \viennarnafold and \contrafold
(e.g., 2.5 days vs.~1.3~minutes on a sequence with length 32,753 \nts).
More interestingly, %Although \linearpartition is approximate,
the resulting 
base pairing probabilities are even better correlated with the ground truth structures.
\linearpartition also leads
to a small accuracy improvement when used for downstream structure prediction
on families with the longest length sequences (16S and 23S rRNA),
as well as a substantial improvement on long-distance base pairs (500+ \nts apart).
\\[0.2cm]
See {\scriptsize \url{http://github.com/LinearFold/LinearPartition}} for code and {\scriptsize \url{http://linearfold.org/partition}} for server.
  \vspace{-1cm} 
\end{abstract}

\dates{}
\doi{\vspace{-0.5cm} Corresponding author: {liang.huang.sh@gmail.com}} % \qquad (includes supplementary information)}

\begin{document}

% Optional adjustment to line up main text (after abstract) of first page with line numbers, when using both lineno and twocolumn options.
% You should only change this length when you've finalised the article contents.
\verticaladjustment{-2pt}

\maketitle
\thispagestyle{firststyle}
\ifthenelse{\boolean{shortarticle}}{\ifthenelse{\boolean{singlecolumn}}{\abscontentformatted}{\abscontent}}{}

% If your first paragraph (i.e. with the \dropcap) contains a list environment
% (quote, quotation, theorem, definition, enumerate, itemize...), the line after
% the list may have some extra indentation. If this is the case, add \parshape=0
% to the end of the list environment.

%%%%%% INTRO %%%%%%
\vspace{-0.3cm}
\label{sec:intro}
\vspace{-0.1cm}
% !TEX root = main.tex

% state the problem
% RNA secondary structure prediction is a well-known problem, and it has been used for medical design. 
% Compared with MFE-based methods, partition function-based methods attract more and more attention due to their higher accuracy and ability to predict pseudoknots.
% Recently, 
% For example, almost every week a new ncRNA is found to be regulated in a particular disease, or a new class of noncoding
% transcripts is uncovered by a transcriptomic study, or a new
% article heralds a paradigm shift that lncRNAs will bring to
% our understanding of biology 
% Each year, noncoding RNA 
% Ribonucleic acid (RNA) 

\section{Introduction}

% \begin{figure}[t]
% \center
% \begin{tabular}{cc}
% \hspace{-3cm}\panel{A} & \hspace{-3cm}\panel{B} \\
% \includegraphics[scale=0.3]{figs/gold_RNAstructure}
% &
% \includegraphics[scale=0.3]{figs/mfe_RNAstructure}\\
% \hspace{-3cm}\panel{C} & \hspace{-3cm}\panel{D}\\[-0.4cm]
% \includegraphics[scale=0.085]{figs/tRNA_circular}
% &
% \includegraphics[scale=0.4]{figs/tRNA_heatmap_dark}

% \end{tabular}
% \caption{
% Comparison of MFE-based method and partition function-based method. 
%     {\bf A}: ground truth secondary structure of {\it E.~coli} tRNA$^\textit{Gly}$; 
%     {\bf B}: the corresponding MFE structure. 
%     Structural difference are denoted with blue in ground truth structure and red in MFE structure;
%     {\bf C}: the corresponding circular representation.
%     Ground truth base pairs are denoted with dash blue lines. 
%     Base pair probabilities are denoted with red solid lines and line shade is proportional to probability value.
%     {\bf D}: the corresponding heatmap representation.
%     MFE structure (lower triangle) misses some ground truth base pairs (blue cross), 
%     while base pairing probability matrix (upper triangle) covers these correct base pairs. 
% \label{tRNA}}
% \end{figure}

% For past decades, our understanding of ribonucleic acid (RNA) is changing. 
% New proofs reveal that 
% noncoding 
% RNAs %(ncRNAs) 
% Ribonucleic acids (RNAs)
RNAs
are involved in multiple processes, 
such as catalyzing reactions or guiding RNA modifications~\cite{Eddy:2001,Doudna+Cech:2002,Bachellerie+:2002}, 
% and regulating a particular disease~\cite{Kung+:2013},
and their functionalities are highly related to structures.
% but determining the structure using experimental methods is costly and time-comsuming. 
%%%%%%%%5
% from proposal
% Therefore, being able to %rapidly 
% determine the structure is %extremely 
% useful and desired.
% given the overwhelming pace of increase in genomic data (about 1021 base-pairs per year \cite{stephens+:2015}) %[97] 
% and given the small percentage of sequences that have experimentally determined structure. 
% While experimental assays still constitute the most reliable way to determine structures, they are prohibitively costly, slow, and difficult.
However, 
% both 
structure determination techniques, such as X-ray crystallography~\cite{Zhang+Adrian:2014}, 
Nuclear Magnetic Resonance (NMR)~\cite{Zhang+Keane:2019}, 
and 
% chemical probing methods 
cryo-electron microscopy~\cite{Lyumkis:2019}, 
% ~\cite{Ziehler+Engelke:2001},
though reliable and accurate,
are extremely slow and costly.
% considering the exponentially increasing genomic data (about $10^{21}$ base-pairs per year \cite{stephens+:2015}) 
% and undetermined structures.
% and therefore computational prediction provides an attractive alternative.
%%%%%%%%%%%%
% Due to such limitations, fast and accurate RNA structure prediction is required and desired,
% Due to such limitations, 
% for many RNA tasks 
Therefore,
fast and accurate computational prediction of RNA structure is useful and desired. %required. 
Considering full RNA %tertiary 
structure prediction is challenging~\cite{Miao+:2017},
% \cite{mccaskill:1990},
% even more difficult than protein folding \cite{mccaskill:1990},
% as an alternative
many studies focus on predicting secondary structure,
% the double helices folding structure formed by self-complementary nucleotides
the set of canonical base pairs in the structure 
(A-U, G-C, G-U base pairs)~\cite{Tinoco+Bustamante:1999},
% RNA secondary structure prediction
as it is well-defined, 
% in mathematics formation, 
and provides detailed information to help understand 
% RNA's mechanism of functionality,
the structure-function relationship, and
% functionality 
% as well as further 
%The secondary structure additionally
is a basis to predict full tertiary structure~\cite{Flores+Altman:2010,Seetin+Mathews:2011}.

\begin{figure}%[t]
%  \center
  \vspace{-0.5cm}
\begin{tabular}{ccc}
% \hspace{-3cm}\panel{A} & \\[-0.6cm]
% % \\[-0.6cm]
  % \multicolumn{2}{c}{\includegraphics[scale=0.7]{figs/heatmap_fig1A}}
  \\[-0.1cm]  
  \hspace{-.3cm}
  \raisebox{1.7cm}{\panel{A}}
  \hspace{-.3cm}
  %\hspace{-3.5cm}
  \raisebox{-1.cm}{\includegraphics[scale=0.3]{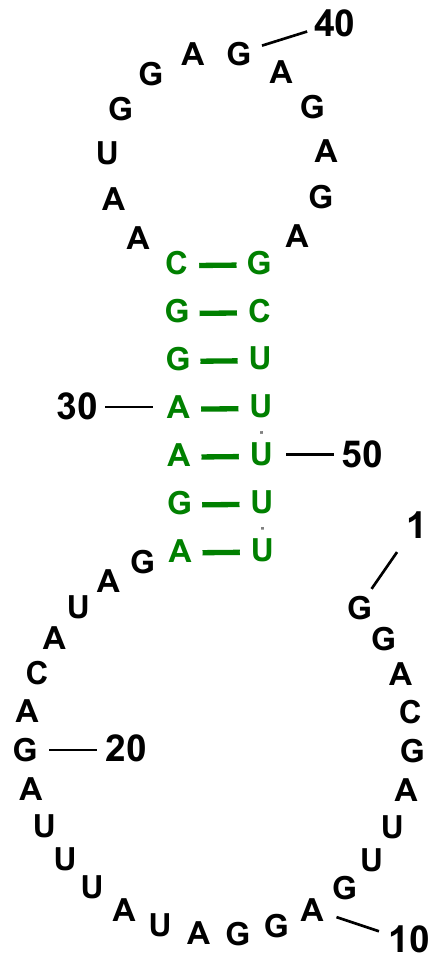}}
  &
  \hspace{.0cm}\raisebox{.8cm}{\panel{B}}
\raisebox{-2.2cm}{\hspace{-.7cm}\includegraphics[scale=0.3]{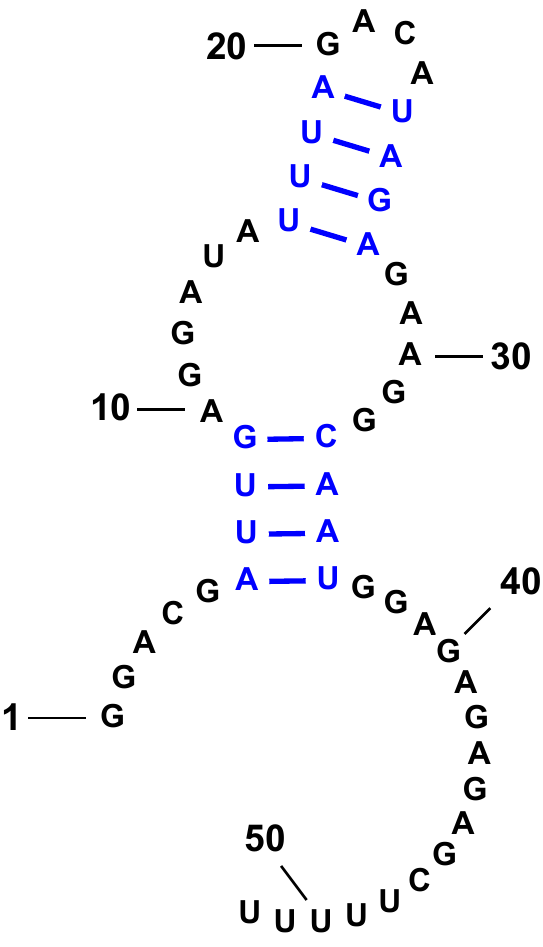}}
&
  %\hspace{-0.5cm}
\hspace{-0.3cm}\raisebox{2.cm}{\mtwo{{\raisebox{4.5cm}{\panel{C}} \hspace{-0.3cm} \includegraphics[scale=0.5]{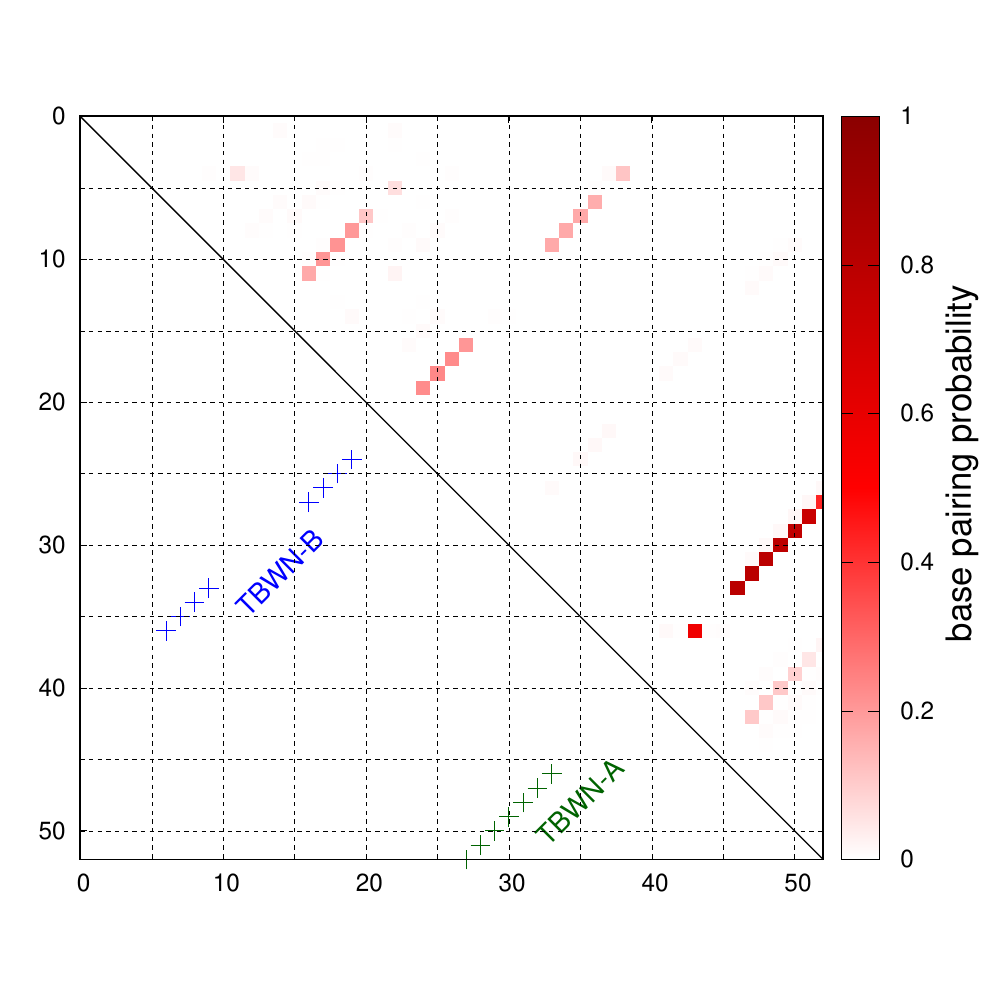}}}}
\\[0.2cm]
%\hspace{-1cm}
% \includegraphics[scale=0.4]{figs/tRNA_heatmap_dark}
\end{tabular}
\\[0.2cm]
\panel{D}
\\[-0.2cm]
%\centering
  \begin{tabular}{@{}c@{ }|@{}c@{ }|l@{\!}c@{ }|l@{\!}c@{}}
    & \ span & \multicolumn{2}{c|}{minimum free energy} & \multicolumn{2}{c}{partition-function} \\
    \hline
    bottom-up, {\em exact} &$\infty$ &    Zuker\cite{zuker+stiegler:1981} & $O(n^3)$ & McCaskill\cite{mccaskill:1990} & $O(n^3)$ \\    
    \hline
    \mtwo{local folding} & \mtwo{$L$}  &  \mtwo{Localfold\cite{lange+:2012}} & \mtwo{$O(nL^2)$} & {RNAplfold\cite{bernhart+:2006}} & \mtwo{$O(nL^2)$}\\
                  &           &   &       &   Rfold\cite{kiryu+:2008}        &\\
    \hline    
    left-to-right, {\em exact}  & \mtwo{$\infty$} & \mtwo{\linearfold\cite{huang+:2019}} & $O(n^3)$  & \mtwo{\linearpartition} & $O(n^3)$\\
     \quad + {\em beam pruning} &     && $O(n b\log b)$ && $O(n b^2)$
    %% left-to-right & \mtwo{\linearfold, $O(n b\log b)$} & \linearpartition, $O(n b^2)$\\[-0.1cm]
    %% + {\em beam pruning} & 
  \end{tabular}
  %% \caption{Comparison between classical, local, and left-to-right algorithms for MFE and partition function calculation. 
  %%   \linearfold and \linearpartition enjoy linear runtime thanks to left-to-right order which enables heuristic beam pruning,
  %%   and both become exact $O(n^3)$ algorithms without pruning. % size $b$ is $+\infty$.
  %%   ``Span'' denotes the maximum pair distance allowed ($\infty$ means no limit);
  %%   it is a small constant in local methods (e.g., default $L$=70 in RNAplfold).
  %%   \label{tab:overall}
  %% } 
\caption{
  %An example % of Tebowned RNA
%  illustrates 
% some RNAs 
  %that
  An RNA can fold into multiple structures at equilibrium.
  {\bf A--B}:~Two 
  %such 
  secondary structures of Tebowned RNA: TBWN-A and TBWN-B~\cite{Cordero+Das:2015}.
{\bf C}: upper triangle shows the estimated base pairing probability matrix for this RNA using \viennarnafold,
where darker red squares represent higher probility base pairs;
the lower triangle shows the two different structures; %TBWN-A and TBWN-B  %(in green)  (in blue),
%at equilibrium;
%(green crosses for TBWN-A and blue for TBWN-B base pairs);
    %    {\bf C}: TBWN-B secondary structure.
{\bf D:} Comparison between classical, local, and left-to-right algorithms for MFE and partition function calculation. 
    \linearfold and \linearpartition enjoy linear runtime because of a left-to-right order that  enables heuristic beam pruning,
    and both become exact $O(n^3)$ algorithms without pruning. % size $b$ is $+\infty$.
    ``Span'' denotes the maximum pair distance allowed ($\infty$ means no limit);
    it is a small constant in local methods (e.g., default $L$=70 \nts in RNAplfold).
\label{fig:overview}}
\vspace{-0.3cm}
\end{figure}

% Secondary structure prediction problem, 
% though still difficult, 
% is well-defined in mathematics formation, and can be suitable modeled with the decomposable substructures. 
% Utilizing this decomposable nature, 
RNA secondary structure prediction is NP-complete~\cite{Pedersen+:2000},
but nested (i.e., pseudoknot-free) secondary structures can be predicted with
cubic-time dynamic programming algorithms. 
% based on an important paradigm free energy minimization 
Commonly, the minimum free energy (MFE) 
structure is predicted~\cite{nussinov+jacobson:1980, zuker+stiegler:1981}.
% when a single structure is expected.
% and some prediction systems based on these algorithms, such as RNAstructure \cite{mathews+turner:2006}, \contrafold \cite{do+:2006} and \viennarnafold \cite{lorenz+:2011}, 
% have greatly improved the accuracy of prediction and are widely used.
% MEA -> partition function 
% For a given sequence, predicting the structure of minimum free energy (MFE) under certain free energy model by dynamic programming is a classical method for RNA secondary structure prediction. 
% cted.Free energy minimization is an important paradigm for predicting structure when a single structure is expe
% In the absence of many homologous sequences, the accuracy of MFE structure is 73\% in average \cite{mathews:2004}.
% However, this method assumes all thermodynamic parameters are correct and the energy model is perfect, which are both no the case in reality.
% Also, this method neglects the facts that multiple conformations exits at equilibrium \cite{mathews:2004}.
% This method 
% MFE method gives a practical solution to predict a single secondary structure, 
At equilibrium, the MFE structure is the most populated structure, 
but it 
% it neglects the fact that 
is a simplification because 
multiple conformations exist 
% at equilibrium 
as an equilibrium ensemble 
for one RNA sequence~\cite{mathews:2004}.
For example, many mRNAs {\textit {in vivo}} form a dynamic equilibrium 
and fold into a population of structures~\cite{Long+:2007, Lu+:2008, Tafer+:2008, lai+:2018}; 
% as well as abandons all pseudoknotted structures.
% Many RNA sequences, for example mRNAs, exist in a thermodynamic ensemble of structures 
% \cite{lai+:2018}.%[53].
% As an alternative, partition function-based method \cite{mccaskill:1990} provides an ensemble of all pseudoknot-free structures, and based on it base pairing probabilities and structural entropy \cite{Huynen+} can be calculated.  
% As an alternative, 
Figure~\ref{fig:overview}A--B shows the example of Tebowned RNA
which folds into more than one structure at equilibrium.
% ~\cite{Cordero+Das:2015}.
% TBWN-A, which has a long helices at 3'-end, 
% is the majority structure and accounts for $56 \pm 16\%$ of ensemble.
% While TBWN-B, which has two short helices at 5'-end, takes up $27 \pm 12\%$ of ensemble.
% Besides TBWN-A and TBWN-B illustrated in Figure~\ref{fig:overview}, Tebowned RNA can also fold into the state of TBWN-C 
% with a smaller ensemble proportion of $17 \pm 11\%$.
In this case, the prediction of one single structure, such as the MFE structure, 
is not expressive enough to capture multiple states of RNA sequences %Tebowned RNA %folding.
at equilibrium.

% Rather than predicting one single stucture, 
% partition function-based methods estimate the folding in a different point of view by
% taking into account ensemble of structures at equilibrium with Boltzmann Distribution.  
% Partition function, the sum of equilibrium constants for all possible secondary structures,
% is the normalization terms for calculating given secondary structure in the boltzmann ensemble.
Alternatively, we can compute the partition function, 
which is the sum of the equilibrium constants for all possible secondary structures,
and is the normalization term for calculating the probability of a secondary structure in the Boltzmann ensemble.
% Starting from the partition function,
% these methods are able to model mix of conformations,
% and further 
% we can % also
The partition function calculation can also be used to 
calculate base pairing probabilities of each nucleotide $i$ 
paired with each of possible nucleotides $j$~\cite{mccaskill:1990, mathews:2004}. 
% Base pairs with high probabilities 
% %in the matrix 
% indicate strong confidence of pairing in prediction,
% and are more likely to be paired in ground truth structures
% \cite{mathews:2004, Zuber+:2018}. 
% Figure~\ref{fig:overview}A 
In Figure~\ref{fig:overview}C,
the upper triangle presents the base pairing probability matrix of Tebowned RNA using \viennarnafold, 
showing that base pairs in TBWN-A have higher probabilities (in darker red) than
base pairs in TBWN-B (in lighter red).
This is consistent with the experimental result, i.e.,
TBWN-A is the majority structure that accounts for $56 \pm 16\%$ of the ensemble, 
while TBWN-B takes up $27 \pm 12\%$~\cite{Cordero+Das:2015}. % of the ensemble

\begin{figure}[t]
\vspace{-0.9cm}
\center
\includegraphics[width=.42\textwidth]{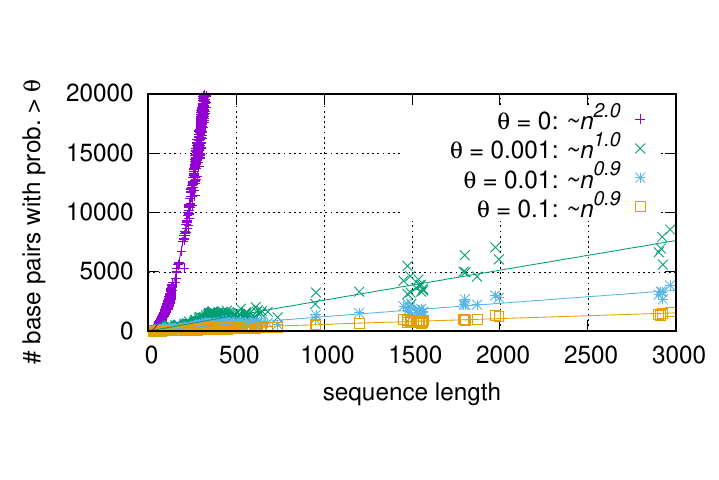}\\[-1cm]
\caption{
  Although the total number of possible base pairings scales $O(n^2)$ with the sequence length $n$
  (using the probability matrix from \viennarnafold as an example),
  with any reasonable threshold $\theta$, the number of surviving pairings (in colors for different $\theta$) grows linearly,
  suggesting our approximation, only computing $O(n)$ pairings, is reasonable.
  \label{fig:linearpairs}}
\vspace{-0.7cm}
\end{figure}

In addition to model multiple states at equilibrium, 
base pairing probabilities are used for downstream prediction methods, 
such as maximum expected accuracy (MEA)~\cite{Knudsen+Hein:2003, do+:2006}, 
to assemble a structure with improved accuracy compared with the MFE structure~\cite{lu+:2009}.
% As a by-product, the pair probabilities also enable maximum expected accuracy (MEA) structure prediction 
% \cite{do+:2006, lu+:2009}. % [29, 62].
Other downstream prediction methods, 
such as 
% HotKnot \cite{Ren+:2005}, % not based on partition function
ProbKnot~\cite{bellaousov+mathews:2010}, 
ThreshKnot~\cite{Zhang+:2019},
DotKnot~\cite{Sperschneider+Datta:2010} 
and IPknot~\cite{sato+:2011},
use base pairing probabilities to predict pseudoknotted structures with heuristics,
which is beyond the scope of standard cubic-time algorithms.
Additionally, the partition function 
% can also be applied to do stochastic sampling based on the ensemble distribution
is the basis of stochastic sampling, 
in which structures are sampled with their probability of occurring 
in the Boltzmann ensemble~\cite{Ding+Lawrence:2003, mathews:2006}.

% Moreover, although partition function-based method excluded pseudoknotted structures during dynamic programming process, 
% it is able to predict pseudoknotted base pairs and structure by using pair probability matrix, and pseudoknotted prediction systems such as HotKnot \cite{Ren+:2005}, ProbKnot \cite{bellaousov+mathews:2010}, DotKnot \cite{Sperschneider+Datta:2010} and IPknot \cite{Sato+:2011} all take pair probability matrix as inputs. 
% Figure~\ref{tRNA} compares MFE-based and partition function-based methods.
% Furthermore, single structure prediction based on partition function calculation, such as maximum expected accuracy (MEA) ThreshKnot \cite{do+:2006, threshknot}, achieves higher accuracy in average. 

% speed
Therefore, there has been a shift from the classical MFE-based methods to partition function-based ones.
These latter methods, 
as well as the prediction engines based on them, such as partition function-mode of RNAstructure~\cite{mathews+turner:2006}, 
\viennarnafold~\cite{lorenz+:2011}, 
and \contrafold~\cite{do+:2006},
are all based on the seminal algorithm that McCaskill pioneered~\cite{mccaskill:1990}.
It employs a dynamic program to capture all possible (exponentially many) nested structures,
but its $O(n^3)$ runtime still scales poorly for longer sequences. 
This slowness %of partition function-based methods
is even more severe than the $O(n^3)$-time MFE-based ones
%with the same $O(n^3)$ time complexity
due to a much larger constant factor.
For instance, for 
% {\it E.~coli} 
{\it H.~pylori} 23S rRNA (sequence length 2,968~{\it nt}),
\viennarnafold's %(version 2.4.11)
computation of the partition function and base pairing probabilities
is 9$\times$ slower than MFE (71 vs.~8 secs),
%% takes 8 seconds to find the MFE structure, 
%% but %36+37=
%% 71 seconds to compute the partition function
%% %and another 37 seconds for
%% and base pairing probabilities, %, respectively,
%% which is a 9x slowdown;
%it is even worse for
%to make things worse,
and \contrafold % is even slower,
%% which takes about 6 seconds to find the MFE structure, % prediction,
%% but 50+70=120 seconds to compute the partition function and base pairing probabilities, respectively,
%gives a
is even 20$\times$ slower (120 vs.~6~secs).
The slowness %(both $O(n^3)$-time and the constant factor)
prevents their applications to longer sequences.
% 132.248859 contrafold

% , which is more than 9 $\times$ slower. 
% because function-based method takes two-round cubic loops for inside-outside calculation.

To address this $O(n^3)$-time bottleneck, %alleviate the cubic-factor slowness, 
we present \linearpartition, 
which is inspired by our %efficient linearization idea of \linearfold \cite{huang+:2019}
recently proposed \linearfold algorithm~\cite{huang+:2019} 
that approximates the MFE structure in linear time.
Using the same idea, \linearpartition can approximate
the partition function and base pairing probability matrix in linear time.
% Recently, \linearfold \cite{huang+:2019} 
% presents the first linear-time and linear-space MFE-based RNA folding prediction system.
% For the same 
% % {\it E.~coli} 
% {\it H.~pylori} 23S rRNA, \linearfold spends about 2 seconds, leading to a 4$\times$ runtime decrease.
% % Overall, \linearfold achieves significant efficiency and scalability improvement and higher accuracy. 
% % the first linear-time MFE-based (approximate) algorithm for RNA folding, 
% % achieves significant efficiency and scalability improvement and higher accuracy 
% % than classical $O(n^3)$ MFE-based method, especially on long sequences. 
% Inspired by the efficient linearization idea of \linearfold, we present \linearpartition, 
% which can approximate the partition function and base pairing probability matrix in linear time, 
% addressing speed bottleneck in existing systems.
% Similar as \linearfold, 
Like \linearfold,
\linearpartition % incrementally 
scans the RNA sequence from 5'-to-3'
using a left-to-right dynamic program that runs in $O(n^3)$ time,
but unlike the classical bottom-up McCaskill algorithm~\cite{mccaskill:1990} with the same speed,
our left-to-right scanning makes it possible to apply the beam pruning heuristic~\cite{Huang+Sagae:2010} 
%to narrow down the search space, 
to achieve linear runtime in practice; see Fig~\ref{fig:overview}D.
 % with substructure of lower energy.
% and only retain states with top $b$ free energy of ensemble (inside score), 
% where $b$ is the beam size.
% Though introducing beam prune to filter some structures,
Although the search is approximate,
the well-designed heuristic ensures 
%that only states with worse ensemble free energy
the surviving structures capture the bulk of the free energy of the ensemble.
It is important to note that, unlike local folding methods in Fig.~\ref{fig:overview}D,
our algorithm does {\em not} impose any limit on the base-pairing distance;
in other words, it is a {\em global} partition function algorithm.
%and the resulting partition function is close to the exact version.

%  (inside score) 
% are pruned,
% and partition function is still similar as optimal algorithm.
% and the majority is catched.

More interestingly, as Figure~\ref{fig:linearpairs} shows,
even with the $O(n^3)$-time McCaskill algorithm, %(like the one implemented in \viennarnafold),
the resulting number of base pairings with reasonable probabilities (e.g., >0.001)
grows only linearly with the sequence length.
This suggests that our algorithm, which only computes $O(n)$ pairings by design,
is a reasonable approximation.

% \linearpartition, inherits efficiency and accuracy of \linearfold. 

% \linearpartition is 11$\times$ faster than \viennarnafold for the longest sequence ({\it H.~pylori} 23S rRNA, 2,968 nucleotides) in the ArchiveII dataset,
% and 256$\times$ faster for the longest sampled sequence (15,780 nucleotides) from RNAcentral that \viennarnafold can run.
\linearpartition is 2,771$\times$ faster than \contrafold for the longest sequence (32,753~{\it nt})
% 32,767 nucleotides) 
that \contrafold can run 
in the dataset (2.5 days vs.~1.3 min.).
% Meanwhile, \linearpartition leads to a small improvement on PPV and Sensitivity compared with \viennarnafold.
% % in both MEA and ThreshKnot prediction 
% % using the probability matrix computed in linear time.
% Surprisingly, \linearpartition achieves higher accuracy improvement on longer families (16S and 23S rRNA).
Interestingly, \linearpartition is orders of magnitude faster {\em without} sacrificing accuracy.
In fact, the resulting base pairing probabilities are even better correlated with ground truth structures,
and 
when applied to downstream structure prediction tasks,
%such as MEA and ThreshKnot (a thresholded version of ProbKnot),
they lead to a small accuracy improvement on longer families (small and large subunit rRNA),
as well as a substantial accuracy improvement on long-distance base pairs (500+ \nts apart).
%% Our work is the first major speedup on this important problem in 30 years, and
%% enables calculations on full length sequences such as mRNAs.
%% has numerous applications
%% such as  sampling.

% \begin{itemize}
% \item Present an alternative left-to-right dynamic programming fashion for partition function calculation.
% \item The first algorithm to achieve linear runtime and space for partition function and base pairing probability calculation.
% % \item . 
% \end{itemize}

% more striking than LinearFold
Although both \linearpartition and \linearfold use linear-time beam search,
the success of the former is arguably more surprising,
since rather than finding one single optimal structure, 
\linearpartition needs to sum up exponentially many structures
that capture the bulk part of the ensemble free energy.
% For users, 
\linearpartition also results in more accurate downstream structure predictions than \linearfold.
%and is able to serve tasks such as pseudoknot prediction and accessibility calculation.

% \label{sec:rnaintro}
% \input{rnaintro}

% \label{sec:linpar}
% \input{linpar}

\label{sec:algorithm}
% !TEX root = main.tex
\vspace{-0.4cm}
\section{The \linearpartition Algorithm}
\vspace{-0.1cm}

\newcommand{\pluseq}{\mathrel{+}=}

We denote $\vecx\!=\!x_1 ... x_n$ as the input RNA sequence of length $n$, and $\mathcal{Y(\vecx)}$ as the set of all possible secondary structures of $\vecx$.  
% $\vecy$ is a secondary structure 
% of $\vecx$ 
% in $\mathcal{Y(\mathbf x)}$. 
The partition function is: % $Q(\vecx)$ 
% of $\vecx$ 
%is then: %defined as:
\begin{equation}
	Q(\vecx)=\sum_{\vecy \in \mathcal{Y(\mathbf x)}} e^{-\frac{\Delta G^{\circ}({\vecy})}{RT}} \notag
\end{equation}
where $\Delta G^{\circ}({\vecy})$ is the  conformational Gibbs free energy change of structure $\vecy$, 
$R$ is the universal gas constant 
and $T$ is the thermodynamic temperature.
$\Delta G^{\circ}({\vecy})$ is calculated using loop-based Turner free-energy model~\cite{mathews+:1999, Mathews+:2004}, 
but for presentation reasons, % simplicity in presenting the algorithm
we use a revised Nussinov-Jacobson energy model, 
i.e., a free energy change of $\delta(\vecx, j)$ for unpaired base at position $j$ 
and a free energy change of $\xi(\vecx, i, j)$ for base pair of $(i,j)$.
For example, we can assign $\delta(\vecx, j)\!=\!1$ kcal/mol and $\xi(\vecx, i, j)\!=\!-3$ kcal/mol for CG pairs and $-2$ kcal/mol for AU and GU pairs. 
Thus, $\Delta G^{\circ}({\vecy})$
can be decomposed as:
\begin{equation}
	\Delta G^{\circ}({\vecy}) = \sum_{j \in \unpaired(\vecy)} \delta(\vecx, j) \ + \sum_{(i,j) \in \pairs(\vecy)} \xi(\vecx, i, j) \notag
\end{equation}
where ${\textrm {unpaired}}(\vecy)$ is the set of unpaired bases in $\vecy$, 
and ${\textrm {paired}}(\vecy)$ is the set of base pairs in $\vecy$.
% and $y_j$ denotes the position $j$ of $\vecy$.
%With the simplified model,
The partition function now decomposes as: % $Q(\vecx)$ is:
\begin{equation}
	Q(\vecx)=\sum_{\vecy \in \mathcal{Y(\vecx)}} (\prod_{j \in \unpaired(\vecy)} e^{-\frac{\delta(\vecx, j)}{RT}} \prod_{(i,j) \in \pairs(\vecy)} e^{-\frac{\xi(\vecx, i, j)}{RT}}) \notag
\end{equation}

%% We provide the pseudocode of our simplified linear-time partition function algorithm (based on the revised Nussinov-Jacobson energy model) in Figure~\ref{fig:algorithm},
%% illustrating how our algorithm linearizes partition function calculation. 

% \linearpartition scans from 5'-end to 3'-end (left-to-right), 
% calculating $Q_{0,j}$, which is the partition function from 5'-end to current step $j$.
%In order to define our algorithm,
We first define {\bf span} $[i,j]$ to be the subsequence $x_i ... x_j$
(thus $[1,n]$ denotes the whole sequence \vecx, and $[j, j\!-\!1]$ denotes the empty span between $x_{j-1}$ and $x_j$ for any $j$ in $1..n$).
We then define a {\bf state} to be a span associated with its partition function:\\[-0.4cm] % $\Qf{i,qj]$: 
\[
  [i,j]: \Qf{i}{j}
\]
%% where $i$ and $j$ are start and end points of the span ($i=0..n, j=1..n$ where $n$ is the sequence length), %is the index of an openning bracket, 
%% % $j$ is the index of current step, 
%% and $\Qf{i,j]$ is the partition function of span $[i,j]$. %state $\langle i,j \rangle$. 
%% % We require each state $\langle i,j \rangle$ only has at most one openning bracket at $i$.
%Each state $\langle i,j \rangle :
where % $\Qf{i}{j}$ 
\begin{center}
  \vspace{-0.6cm}
  $\displaystyle \Qf{i}{j} = \sum_{\vecy \in \mathcal{Y}(x_i ... x_j)} e ^{-\frac{\Delta G^{\circ}(\vecy)}{RT}}$
  \vspace{-0.1cm}
\end{center}
encompasses all possible substructures for span $[i, j]$, % $[i,j]$, i.e
which can be visualized as
%\begin{center}
%  \vspace{-0.5cm}
%    \hspace{1.6cm}
\raisebox{-0.2cm}{
  \hspace{-0.5cm}
    \begin{tabular}{lr@{\quad}} % adjust alignment with j
    \multicolumn{2}{c}{
      ${\myboxmath{\ \ \Qf{i}{j} \ \ }}$
    }
    \\
    $i$ & $j$
\end{tabular}}
.
%    \end{center}

%which can be visualized as
%\begin{center}
 % \end{center}

%% We require these substructures to have an open bracket at nucleotide $i$.
%% For example, ``\bml\md\md'' and ``\bml\bml\bmr'' 
%% % ``\md\md\md'' 
%% are valid states, 
%% while ``\bml\bml\md'' and ``\md\bml\md'' are invalid.
%% As special cases, states with $i=0$ can have none open brackets to 
%% allow unpaired substructures in 5'-end,
%% i.e., ``\md\md\md'' and ``\md\bml\bmr'' are valid for states $\langle 0,j \rangle : Q(0,j)$.

  \algrenewcommand\algorithmicindent{0.5em}%
  \algnewcommand\algorithmicforeach{\textbf{for each}}
\algdef{S}[FOR]{ForEach}[1]{\algorithmicforeach\ #1\ \algorithmicdo}
\begin{figure}[t]%[b]
% \begin{algorithm}[H]
% \algsetup{linenosize=\tiny}
  % \scriptsize
  % \newcommand{\pluseq}{\mathrel{+}=}
\center
\footnotesize
% \hspace{-0.23cm}\includegraphics[scale=.16]{figs/index} \\[-3.cm]
%\hspace{-0.23cm}\includegraphics[scale=.83]{figs/algorithm} \\[0.2cm]
\begin{algorithmic}[1]
  \newcommand{\INDSTATE}[1][1]{\State\hspace{#1\algorithmicindent}}
  \setstretch{1.05} % lhuang: usepackage setspace
\Function{LinearPartition}{$\vecx, b$} \Comment{$b$ is the beam size}
% \bindent
    \State $n \gets$ length of $\mathbf x$
    \State $Q \gets$ hash() \Comment{hash table: from span $[i,j]$ to $\Qf{i}{j}$}
    \State $\Qf{j}{j-1} \gets 1$ for all $j$ in $1...n$ \Comment{base cases} \label{line:base}
    \For{$j=1 ... n$}
    \ForEach {$i$ such that $[i,\,j-1]$ in $Q$} \Comment{$O(b)$ iterations}
        \smallskip 
            \State $\Qf{i}{j} \pluseq \Qf{i}{j-1} \cdot e^{-\frac{\delta(\vecx,j)}{RT}} $ \Comment{\nskip} \label{line:skip}
            \If{$x_{i-1}x_j$ in \{AU, UA, CG, GC, GU, UG\}}  \label{line:pair}
                % \State $Q_{i,\,j+1} \gets  C(i,\,j) \cdot e^{-\frac{\xi(\vecx,i,\,j)}{RT}} $
                \ForEach{$k$ such that $[k,\,i-2]$ in $Q$} \Comment{$O(b)$ iters} \smallskip 
                    \State $\Qf{k}{j} \pluseq {\Qf{k}{i-2} \cdot \Qf{i}{j-1} \cdot e^{-\frac{\xi(\vecx,i-1,j)}{RT}}} $ \Comment{\pop} \label{line:pop}
                    % \State $C(0,j+1) \pluseq {C(0,k) \cdot C(k,j+1) \cdot e^{-\frac{\xi(\vecx,i,\,j)}{RT}}} $
                \EndFor
                % \State $C(0,j+1) \pluseq C(0,i) \cdot Q_{i,j+1}$ \Comment{COMBINE}
            \EndIf
        \EndFor
        \State $\textsc {BeamPrune}(Q,j, b)$ \Comment{choose top $b$ out of $Q(\cdot,j)$} \label{line:beamprune}%{see Fig.~\ref{fig:beam_prune_alg}}
        \EndFor
        \vspace{-0.1cm}
    \State \Return $Q$ \Comment{partition function $Q(\vecx)=\Qf{1}{n}$}
% \eindent
\EndFunction
\end{algorithmic}
% \end{algorithm}
\caption{
Partition function calculation pseudocode of a simplified version of the \linearpartition %linear-time partition function calculation
algorithm (the inside phase).
See Fig.~\ref{fig:beam_prune_alg} for the pseudocode of beam pruning (line~\ref{line:beamprune}).
The base-pairing probabilities are computed with the combination of the outside phase
(Fig.~\ref{fig:outside}).
% as well as a beam prune algorithm. 
% Here we model hash tables following Python dictionaries, where $(i, j) \in C$ checks whether the key $(i, j)$ is in the hash $C$; 
% this is needed to ensure linear runtime. 
% Quick select algotithm is used in beam prune, 
% and we skip the details for quick select here since it is well known.
% Real \linearpartition system is much more involved, but the pseudocode illustrates the left-to-right partition function calculation idea using a Nussinov-like fashion.
The actual algorithm using the Turner model is available on \href{https://github.com/LinearFold/LinearPartition}{GitHub}.
%See Fig.~\ref{fig:beam_prune_alg} for {\sc BEAMPRUNE} function.
\label{fig:algorithm}}
\vspace{-.3cm}
% \end{figure*}
\end{figure}

For simplicity of presentation,
in the pseudocode in Fig.~\ref{fig:algorithm}, $Q$ is notated as a hash table,
mapping from  $[i,j]$ to %its partition function
$\Qf{i}{j}$;
see Supplementary Information Section~\ref{sec:si:algdetails} for details
of its efficient implementation. % to make sure the overall runtime is $O(nb^2)$.
As the base case, we set $\Qf{j}{j-1}$ to be 1 for all $j$,
meaning all empty spans have partition function of 1 (line~\ref{line:base}).
%% %to store and look up states.   
%% $\langle 0,1 \rangle:1.0$ represents the dummy head state, 
%% whose partition function $Q(0,1)$ is initialized with value $1$.
%% This represents a structure with no pairs, 
%% i.e., the random coil, 
%% which is the reference state (free energy change of 0) and thus has equilibrium constant of 1.
%% {\color{blue}?????? (added by Mathews, but state <0,1>:1 is a singleton not a coil)}
Our algorithm then scans the sequence from left-to-right (i.e., from 5'-to-3'),
and at each nucleotide $x_j$ ($j=1...n$), % (the length of the sequence), 
% state $\langle 0,j+1 \rangle$ can always be extended with ``\md'' from state $\langle 0,j \rangle$.
% with its 
% Then we define three actions, PUSH, SKIP and POP
% to search states $\langle \cdot,j+1 \rangle$:
we perform two actions: %, \nskip and \pop:
%three actions, PUSH, SKIP and POP, are performed:
% similar as in \linearfold but different 
\begin{itemize}
%% \item PUSH: create a new state $\langle j,j \rangle:1$ representing an open bracket at $j$,
%% whose partition function is 1. 

%% 	% \begin{equation*}
%% 	% 	\frac{\langle i,j \rangle : \Qf{i,j]}{\langle j,j+1 \rangle : 1.0}
%% 	% \end{equation*}

\item \nskip (line~\ref{line:skip}): We extend each %state $[i,\,j-1]: \Qf{i,\,j-1]$ to a new state $[i,j]: \Qf{i,j]$
span $[i,\,j\!-\!1]$ in $Q$ to $[i,j]$ %: \Qf{i,j]$
by adding an unpaired base $y_j\!=$``\md'' (in the dot-bracket notation) to the right of each substructure in $\Qf{i}{j-1}$,
updating $\Qf{i}{j}$: % as follows:
\[
\Qf{i}{j} \pluseq \Qf{i}{j-1} \cdot e^{-\frac{\delta(\vecx, j)}{RT}}
\]
which can be visualized as
%\begin{center}
%  \vspace{-0.7cm}
%  \qquad \qquad
  \begin{tabular}{lr@{\,\quad}} % adjust alignment with j
    \multicolumn{2}{c}{
      $\overbrace{\myboxmath{\ \ \Qf{i}{j-1} \ \ } \ \Huge\md\!\!\!\!}^{\Qf{i}{j}}$
    }
    \\
    $i$ & $j$
  \end{tabular}.
%\end{center}
%% \begin{center}
%%   {

%%     \begin{tabular}{@{}l@{\;}c@{}}
%%       %      \hline
%%       \multicolumn{2}{c}{$\Qf{i,j]$}\\[0.1cm]
%%       \hline
%%   \myboxmath{\quad \Qf{i,j-1]\quad} & {\large\md} \\
%% %  \hline
%%    \multicolumn{1}{l}{$i$} & $j$
%%     \end{tabular}
%% }
%% \end{center}

	% \begin{equation*}
	% 	\frac{\langle i,j \rangle : \Qf{i,j]}{\langle i,j+1 \rangle : \Qf{i,j] \cdot e^{-\frac{\delta(\vecx, j)}{RT}}}
	% \end{equation*}
\vspace{-0.2cm}
\item \pop (lines~\ref{line:pair}--\ref{line:pop}): If $x_{i-1}$ and $x_j$ are pairable,
we combine span $[i,j-1]$ in $Q$ %[i,j]$ 
with each combinable ``left'' span $[k,i-2]$ in $Q$ %: \Qf{k,i-1]$,
% and create a new state $\langle k,j \rangle:Q(k,j)$,
% where $Q(k,j+1)=Q(k,i) \cdot \Qf{i,j] \cdot e^{-\frac{\xi(\vecx, i, j)}{RT}}$.
and update the resulting span $[k,j]$'s partition function % $[k,j+1]: Q(k,j+1)$
%as follows: 
\[
\Qf{k}{j} \pluseq \Qf{k}{i-2} \cdot \Qf{i}{j-1} \cdot e^{-\frac{\xi(\vecx, i-1, j)}{RT}}.
\]
This means that every substructure in $\Qf{i}{j-1}$ can be combined
with every substructure in $\Qf{k}{i-2}$ and a base pair $(i-1, j)$
to form one possible substructure in $\Qf{k}{j}$:
\begin{center}
  \vspace{-0.4cm}
  \begin{tabular}{l@{}r@{}l@{}r@{\,\quad}} %adjust alignment with j
    \multicolumn{4}{c}      
                {$\overbrace{\myboxmath{\; \Qf{k}{i-2}\; } {\Large\ml} \;\; \myboxmath{\; \Qf{i}{j-1}\; } \; {\Large\mr}}^{\Qf{k}{j}}$} \\
  $k$ \hspace{1.15cm} & $i\!-\!1$ & \hspace{0.02cm} {$i$} &  $j$
\end{tabular}
\end{center}
%% \begin{center}
%% \begin{tabular}{l@{}c@{\;}l@{\;}c}
%%   {\myboxmath{\; \Qf{k,i-1]\; }} & \ml & \myboxmath{\quad \Qf{i,j]\quad} & \mr \\
%%   {$x_k$} & $x_{i-1}$ & {$x_i$} & $x_j$
%% \end{tabular}
%% \end{center}

	% \begin{equation*}
	% 	\frac{\langle k,i \rangle : Q(k,i) \ \ \ \ \ \langle i,j \rangle : \Qf{i,j]}{\langle k,j+1 \rangle : Q(k,i) \cdot \Qf{i,j] \cdot e^{-\frac{\xi(\vecx, i, j)}{RT}}}
	% \end{equation*}

% Note that the operator of updating $Q(k,j+1)$ is "+=" (see Figure~\ref{algorithm} line 11).

% \item COMBINE: for each close state $\langle i,j \rangle$, it can be combined with its prefix state $\langle 0,i \rangle$ and form state $\langle 0,j+1 \rangle$:
% 	\begin{equation*}
% 		\frac{\langle 0,i \rangle : Q_{0,i} \ \ \langle i,j \rangle : Q_{i,j}}{\langle 0,j+1 \rangle : Q_{0,i} \cdot Q_{i,j} \cdot e^{-\frac{\xi(\vecx, i, j)}{RT}}}
% 	\end{equation*}
\end{itemize}

Above we presented a simplified version of our left-to-right \linearpartition algorithm. % that resembles \linearfold\/ {\em without} beam pruning.
%Like \linearfold,
We have 
three nested loops, one for $j$, one for $i$, and one for $k$,
and each loop takes at most $n$ iterations; % (i.e., $k,i$ and $j$ are all bounded by $n$).
therefore, the time complexity {\em without} beam pruning is $O(n^3)$,
which is identical to the classical McCaskill Algorithm (see Fig.~\ref{fig:overview}D).
In fact, there is an alternative, bottom-up, interpretation of our left-to-right algorithm
that resembles the Nussinov-style recursion of the classical McCaskill Algorithm:
\vspace{-0.1cm}
\[
\Qf{k}{j} \!=\! \Qf{k}{j-1} \cdot e^{-\frac{\delta(\vecx, j)}{RT}} + \!\! \sum_{k<i\leq j} \!\! {\Qf{k}{i-2} \cdot \Qf{i}{j-1}   \cdot e^{-\frac{\xi(\vecx, i-1, j)}{RT}}}
\]
%% \begin{equation*}
%% 	\begin{split}
%% 		\Qf{k}{j} &= \Qf{k}{j-1} \cdot e^{-\frac{\delta(\vecx, j)}{RT}} \\
%% 		       &+ \sum_{k<i\leq j} {\Qf{k}{i-2} \cdot \Qf{i}{j-1}   \cdot e^{-\frac{\xi(\vecx, i-1, j)}{RT}}}
%% 	\end{split}
%% \end{equation*}

%% % beam prune
%% The pseudocode in Figure~\ref{fig:algorithm} shows that our \linearpartition algorithm has three nested loops, 
%% one for $j$, one for $i$, and one for $k$,
%% and each loop has at most $n$ iterations. % (i.e., $k,i$ and $j$ are all bounded by $n$).
%% Therefore, the  time complexity without beam pruning is $O(n^3)$, which is identical to the classical McCaskill Algorithm,
However, unlike the classical bottom-up McCaskill algorithm,
our left-to-right dynamic programming, inspired by \linearfold, % instead of bottom-up fashion;
%% this is similar to the $O(n^3)$ left-to-right dynamic programming algorithm in \linearfold.
%% This left-to-right dynamic programming
makes it possible to further apply the beam pruning heuristic
%a heuristic method,
to achieve linear runtime in practice.
% and achive linear runtime.
The main idea is, at each step $j$, among all possible spans $[i, j]$ that ends at $j$  (with $i=1...j$), 
we only keep the top $b$ most promising candidates (ranked by their partition functions \Qf{i}{j}).
%i.e., the top $b$ among all \Qf{i}{j}'s,
%those %$(i,j)$
%with higher partition functions, %value $Q_{i,j}$, 
%and remove the other ones
where $b$ is the beam size.
% We adopt quick select algorithm to ensure the process of selecting top $b$  candidates costs linear runtime.
With such beam pruning, 
we reduce the number of states from $O(n^2)$ to $O(nb)$,
and the runtime from $O(n^3)$ to $O(nb^2)$.
For details of the efficient implementation and runtime analysis, please refer to
Supplementary Information Section~\ref{sec:si:algdetails}.
Note $b$ is a user-adjustable constant ($b=$100 by default).

After the partition-function calculation, also known as the ``inside'' phase of the classical inside-outside algorithm~\cite{baker:1979},
we design a similar linear-time ``outside'' phase (see Supplementary Section~\ref{sec:outside})
to compute the base pairing probabilities:
\vspace{-0.1cm}
\[
p_{i,j} = \sum_{(i,j)\in \pairs(\vecy)} p(\vecy),
\]
where $p_{i,j}$ is the probability of nucleotide $i$ pairing with $j$,
which sums the probabilities of all structures that contain $(i,j)$ pair,
and $p(\vecy)=e^{-\frac{\Delta G^\circ(\vecy)}{RT}} / Q(\vecx)$
is the probability of structure \vecy % of the structure $\vecy$ in the ensemble.
in the ensemble.  %among all possible structures
%(or the probability of $i$ being unpaired when $j=N+1$).

\label{sec:results}
% !TEX root = main.tex
\vspace{-0.1cm}
\section{Results}
\vspace{-0.1cm}

\subsection{Efficiency and Scalability}
% !TEX root = main.tex

\begin{figure}[t]
\center
\begin{tabular}{cc}
\hspace{-4.5cm}{\panel{A}} & \hspace{-4.8cm}{\panel{B}} \\[-0.5cm]
\hspace{-.2cm}\includegraphics[scale=.9]{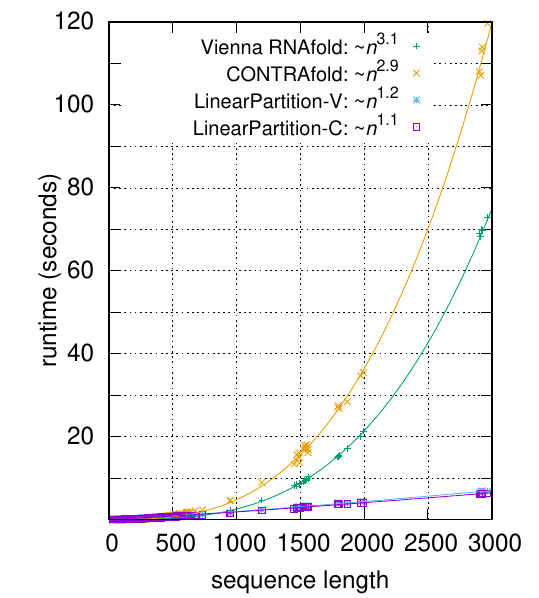}
&
\hspace{-.6cm}\includegraphics[scale=.9]{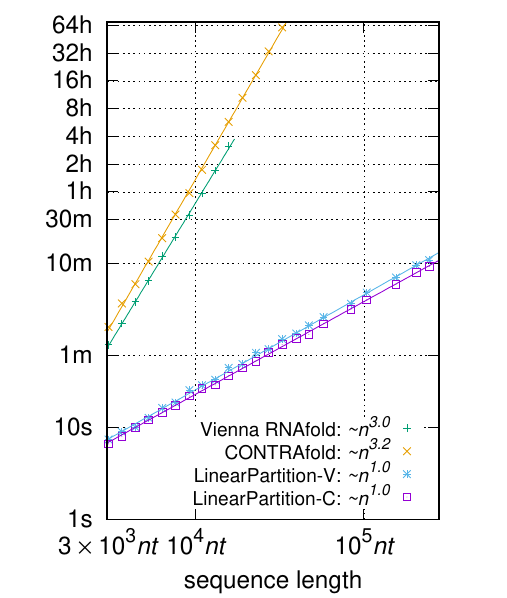} \\[-1.8cm]
\hspace{-4.5cm}\raisebox{4.cm}{\panel{C}} & \hspace{-5.5cm}{\includegraphics[scale=1.35]{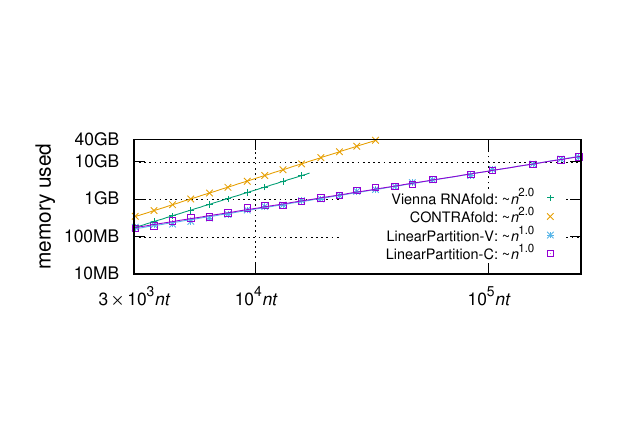}} \\[-2.1cm]
% \multicolumn{2}{c}{\includegraphics[scale=1.35]{figs/mem}} \\[-1.1cm]
\end{tabular}
\caption{Total runtime and memory usage of computing both the partition function and base pairing probabilities. % running speed and space comparisons.
% Runtime comparisons on ArchiveII and RNAcentral datasets, and memory usage comparison on RNAcentral dataset. 
{\bf A}: Runtime comparisons on the ArchiveII dataset; the curve-fittings were log-log in gnuplot with $n > 10^3$.
{\bf B}: Runtime comparisons on the RNAcentral dataset (log scale). %; the x-axis and y-axis are in log scale. 
%% Note that we show the total runtime of computing both the partition function and base pairing probabilities,
%% where the former takes about half of the time.
The partition function computation takes about half of the total time shown here.
% \viennarnafold overflows on the sequence of 19,071~\nts. 
% time limit is 24 hours.
{\bf C}: Memory usage comparisons on the RNAcentral dataset (log scale).
\label{fig:runtime}}
\vspace{-0.5cm}
\end{figure}

We present two versions of \linearpartition:
{\em \linearpartitionv}
using thermodynamic parameters~\cite{mathews+:1999,Mathews+:2004,xia+:1998}
%as implemented in
following \viennarnafold~\cite{lorenz+:2011},
and
{\em \linearpartitionc} using
the learning-based parameters from \contrafold~\cite{do+:2006}.
%% \linearpartitionv uses the experiment-based thermodynamic parameters~\cite{mathews+:1999,Mathews+:2004,xia+:1998}
%% as implemented in \viennarnafold~\cite{lorenz+:2011}.
%% % (Version 2.4.11) 
%% % (\url{https://www.tbi.univie.ac.at/RNA/download/sourcecode/2_ 4_x/ViennaRNA-2.4.11.tar.gz}).
%% \linearpartitionc uses the machine learning-based parameter values from \contrafold~\cite{do+:2006}.
% (Version 2.0.2; 
% (\url{http://contra.stanford.edu/}).
% \viennarnafold is a widely-used RNA structure prediction package,
% while \contrafold is a successful machine learning-based RNA structure prediction system.
% Both provide partition function and base pairing probabilities calculation based on 
% the classical cubic runtime algorithm.
% Our comparisons mainly focus on the systems with the same model, 
% i.e., \linearpartitionv vs. \viennarnafold and \linearpartitionc vs. \contrafold.
% In this way the differences are based on algorithms themselves rather than models.
% bugs in contrafold
% We found a non-trival bug in \contrafold by comparing our results to CONTRAfold, 
% which leads to overcounting in multiloops in the partition function calculation.
% We corrected the bug, and all experiments are based on this bug-fixed version of \contrafold.
We use %run all evaluations %experiments 
% (compiled by GCC 4.9.0) 
a Linux machine 
with 2.90GHz Intel i9-7920X CPU and 64G memory for benchmarks.
We use sequences from two datasets, ArchiveII~\cite{mathews+:1999, sloma+mathews:2016} and RNAcentral~\cite{rnacentral:2017}. See~\ref{sec:datasets} for details of the datasets.

\begin{figure*}[!htb]
\centering
  \vspace{-1.7cm}
  \begin{tabular}{cccc}
    % \raisebox{4.5cm}
    \raisebox{3.8cm}{\panel{A}} &
    {\hspace{-0.cm}\includegraphics[width=0.24\textwidth]{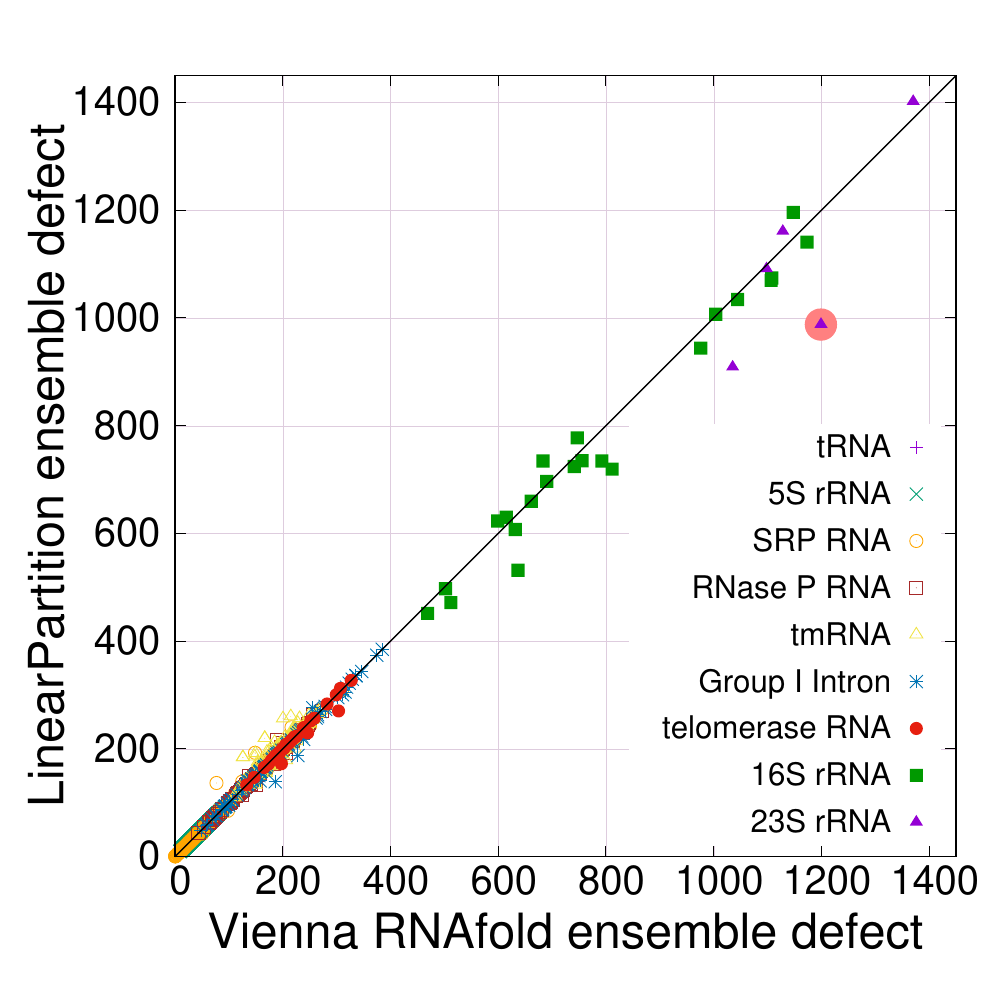}} &
    \raisebox{3.8cm}{\hspace{1.4cm}\panel{B}} &
    \raisebox{-0.3cm}{\hspace{-1.2cm}\includegraphics[width=0.5\textwidth]{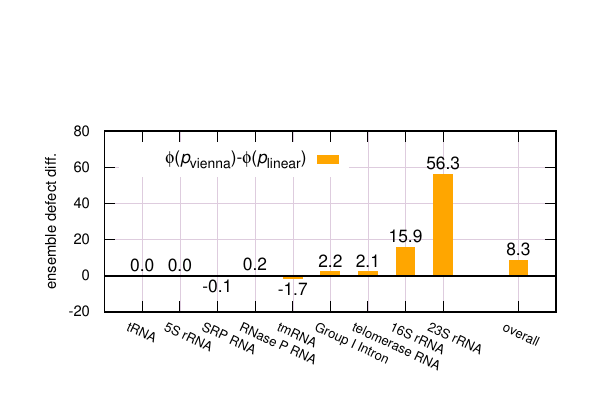}}
  \end{tabular}
  \\[-0.4cm]
%   \vspace{-.5cm}
  \begin{tabular}{ccccc}
&\hspace{-4.cm} \panel{C} & \hspace{-4.6cm}\panel{D} & \hspace{-4.6cm}\panel{E} & \hspace{-4.6cm}\panel{F}\\[-0.3cm]
\raisebox{.9cm}{\rotatebox{90}{\ecoli 23S rRNA}}&
\hspace{-0.2cm}\includegraphics[width=0.22\textwidth]{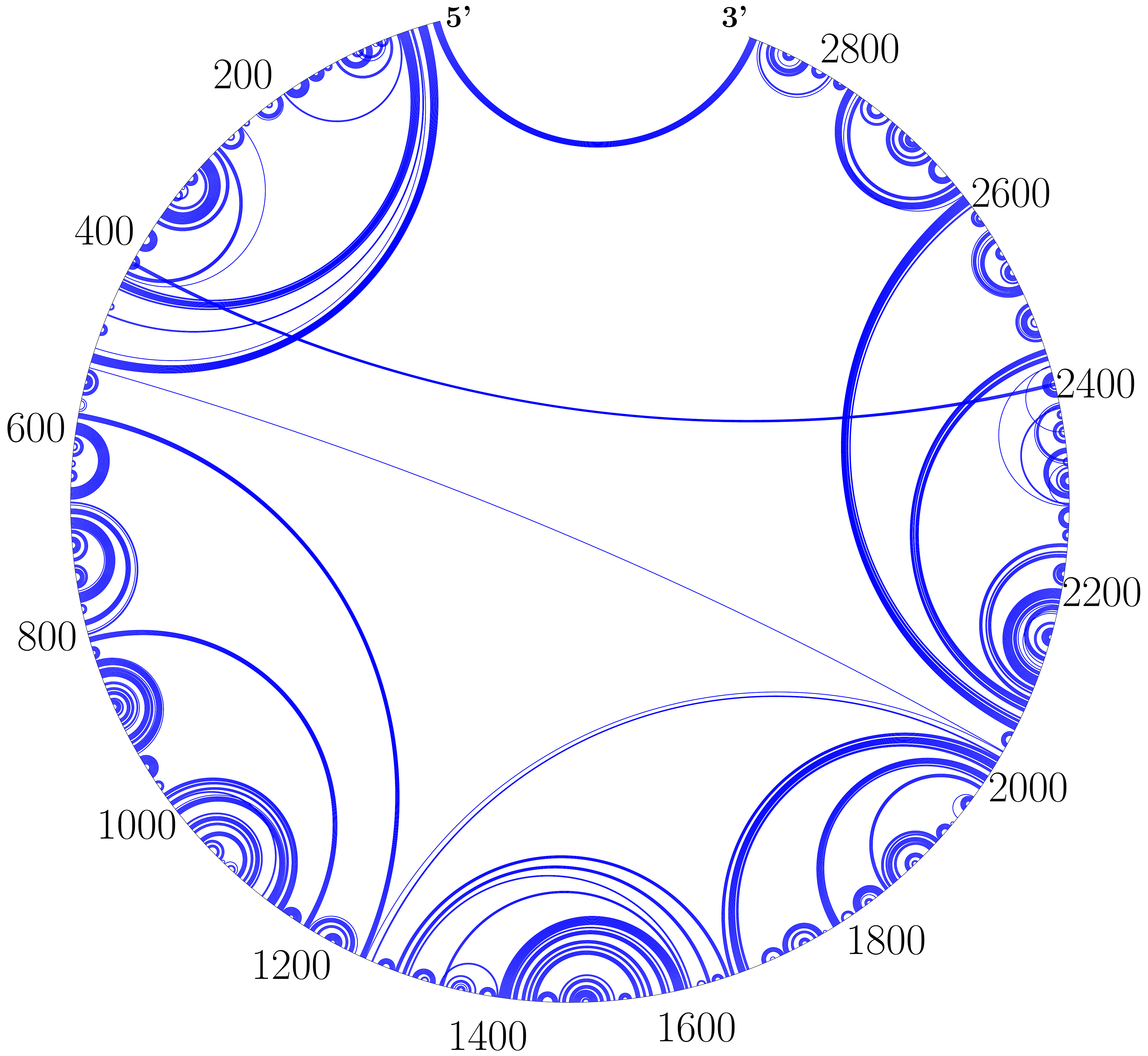} &
\hspace{-0.35cm}\includegraphics[width=0.22\textwidth]{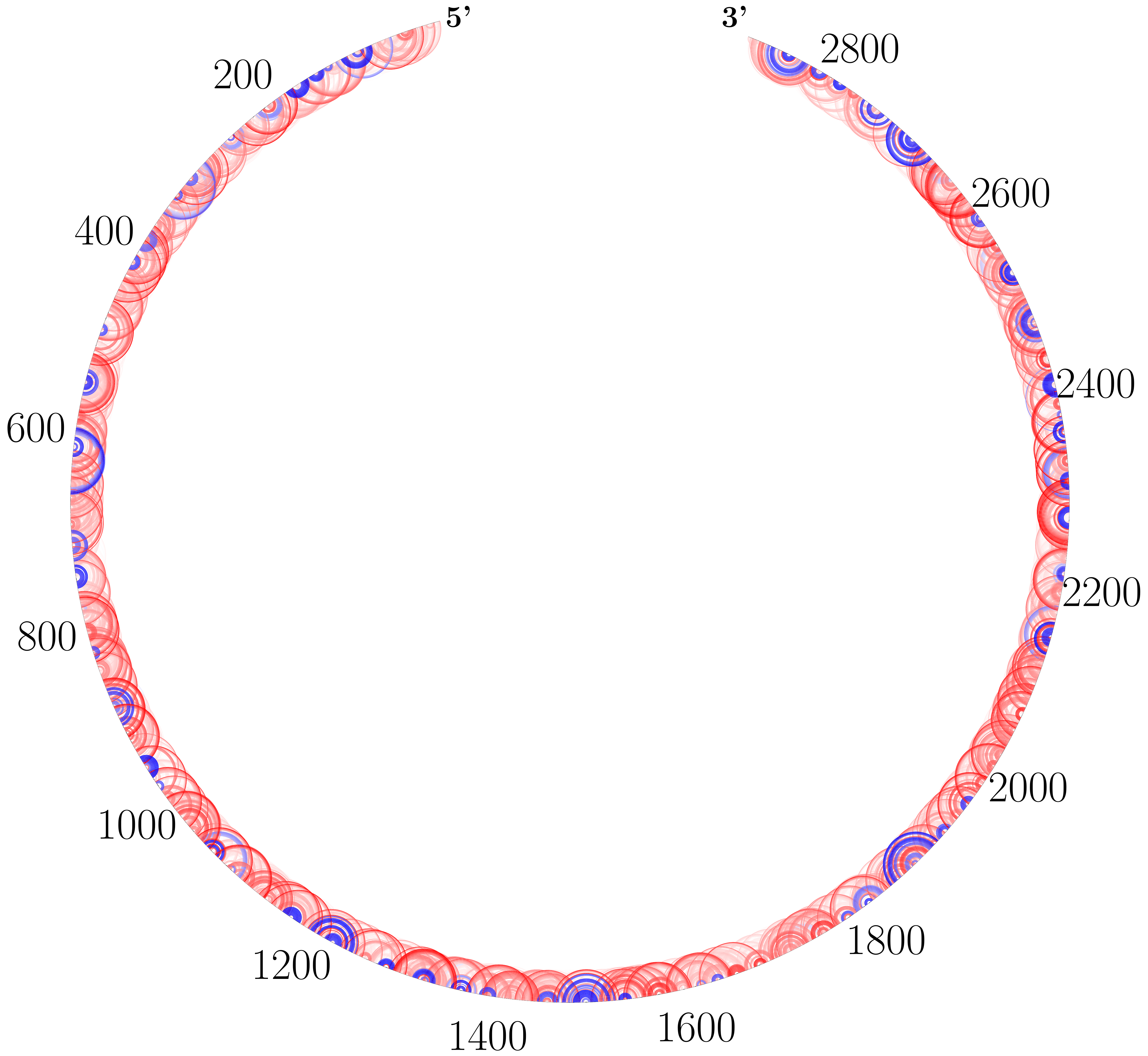} &
\hspace{-0.35cm}\includegraphics[width=0.22\textwidth]{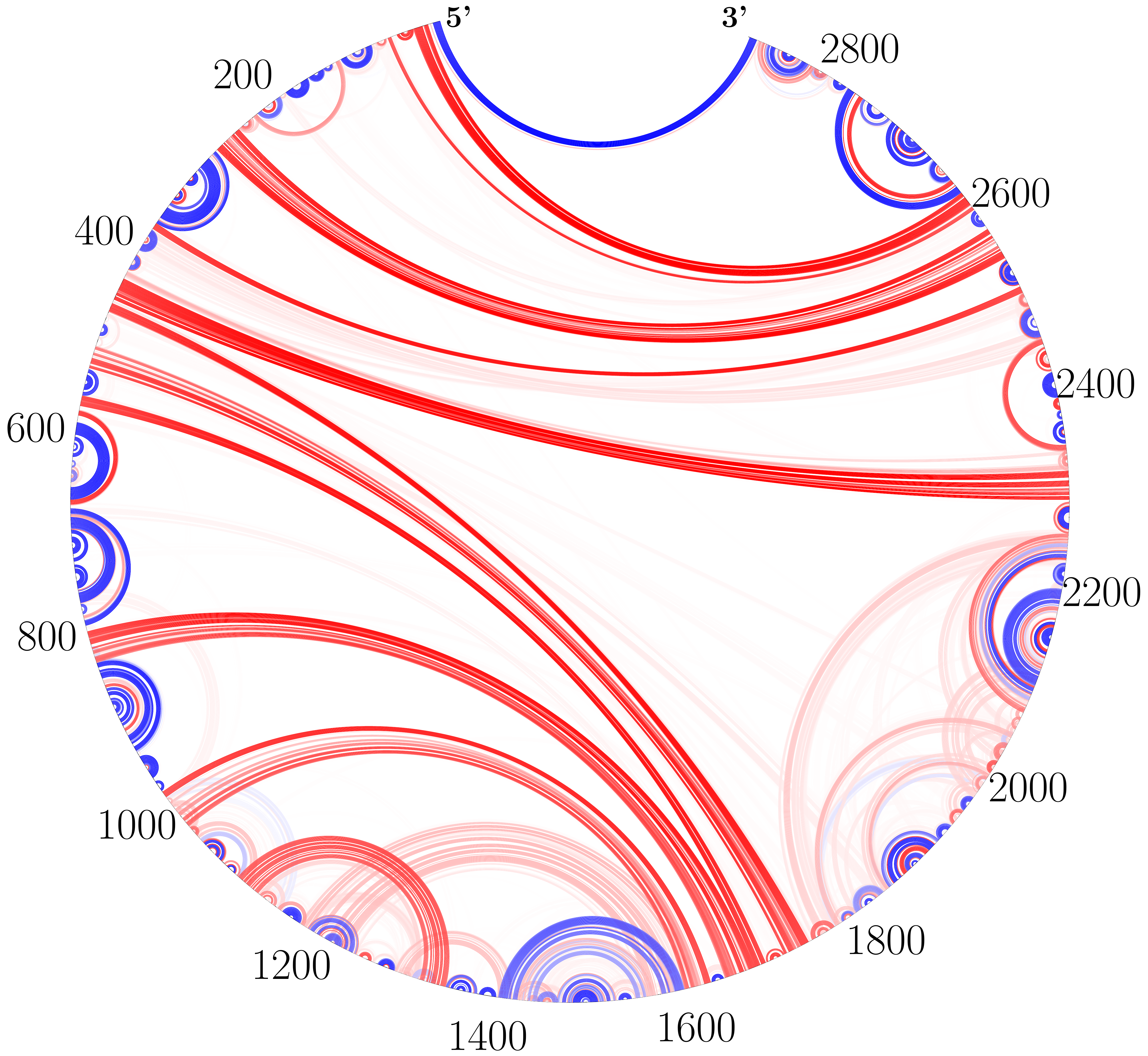} &
\hspace{-0.35cm}\includegraphics[width=0.22\textwidth]{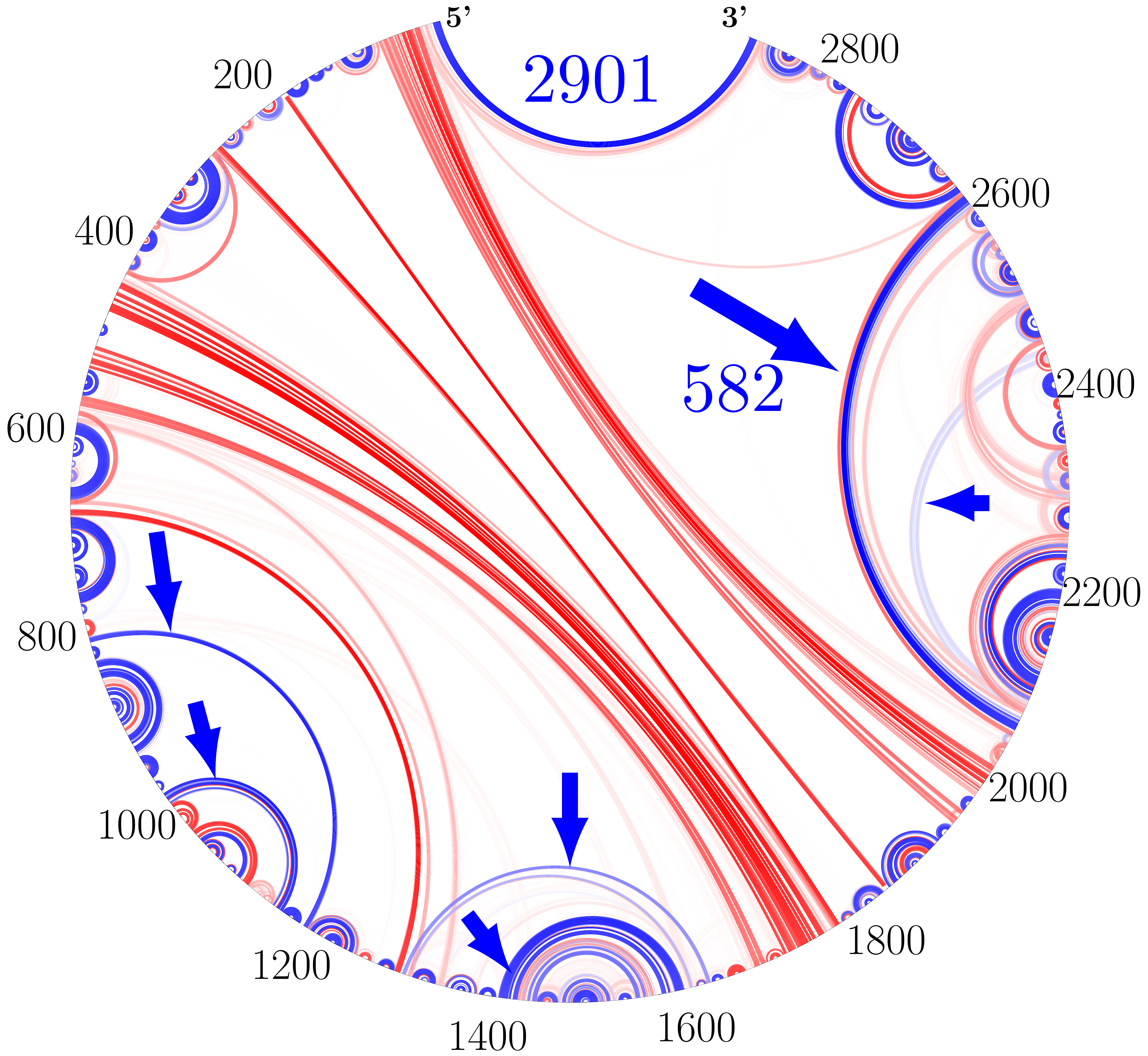} \\
&\hspace{-4.cm} \panel{G} & \hspace{-4.6cm}\panel{H} & \hspace{-4.6cm}\panel{I} & \hspace{-4.6cm}\panel{J}\\[-0.4cm]
\raisebox{.4cm}{\rotatebox{90}{{\it C.~ellipsoidea} Group I Intron}}&
\hspace{-0.2cm}\includegraphics[width=0.22\textwidth]{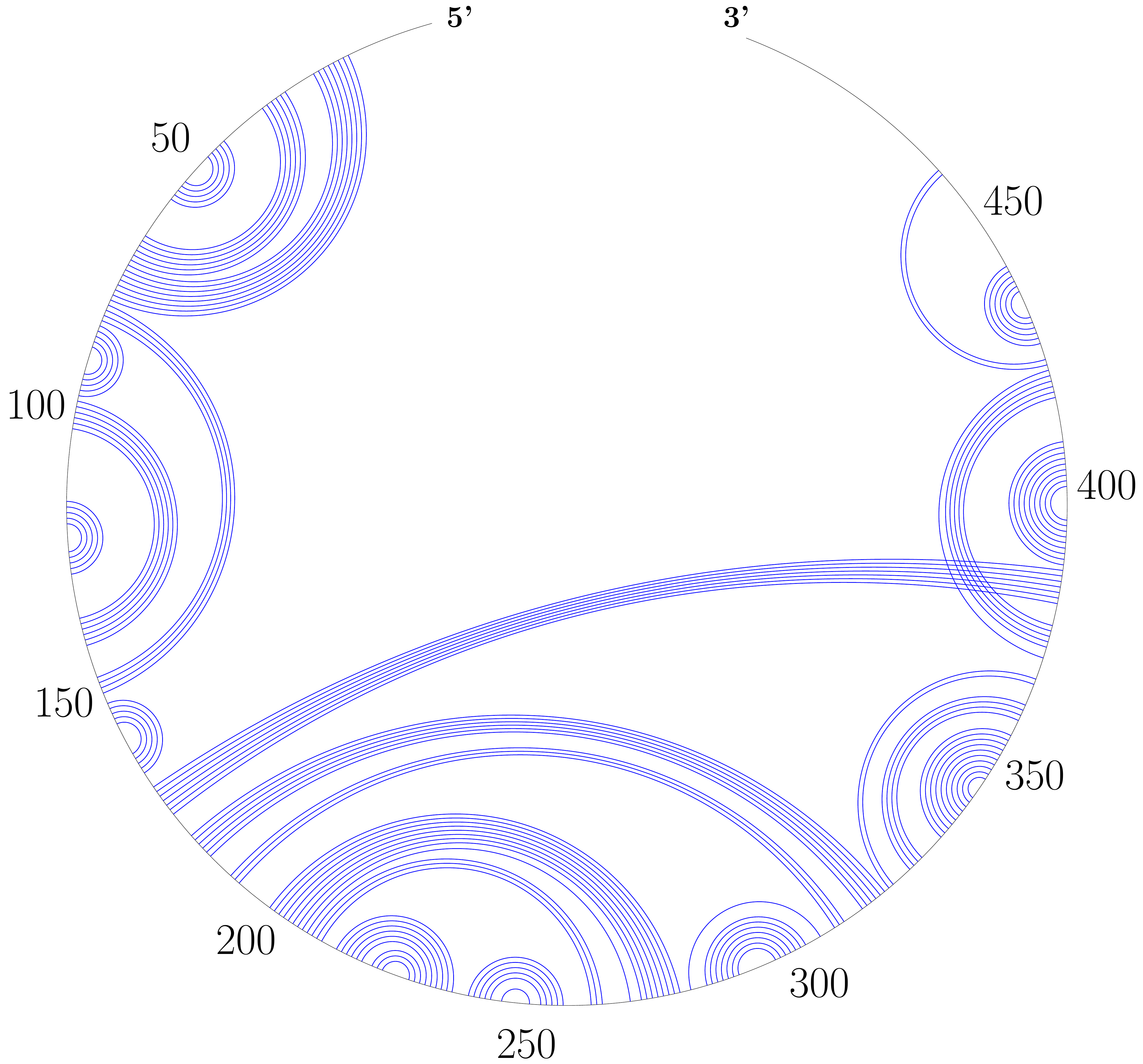} &
\hspace{-0.35cm}\includegraphics[width=0.22\textwidth]{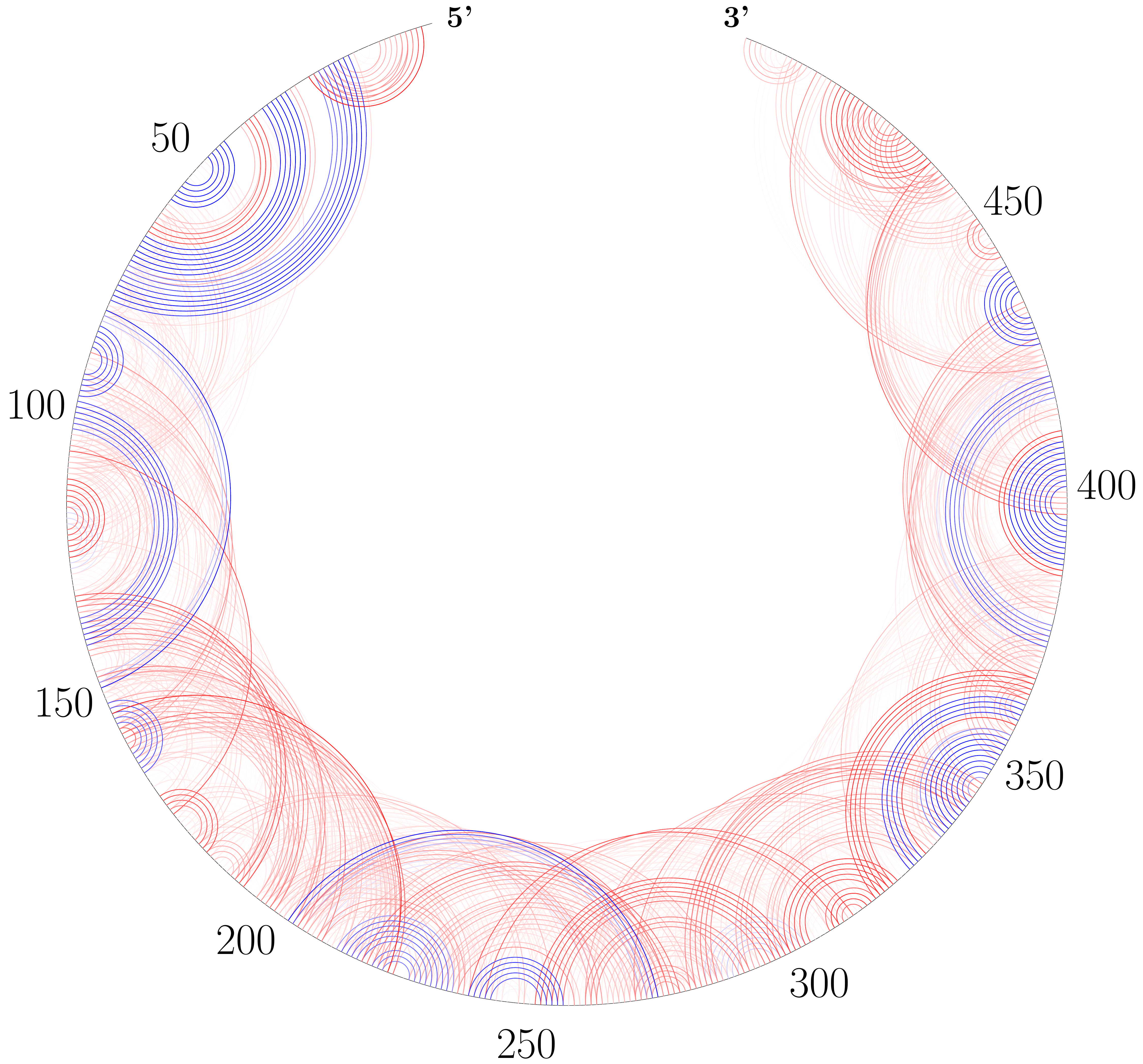} &
\hspace{-0.35cm}\includegraphics[width=0.22\textwidth]{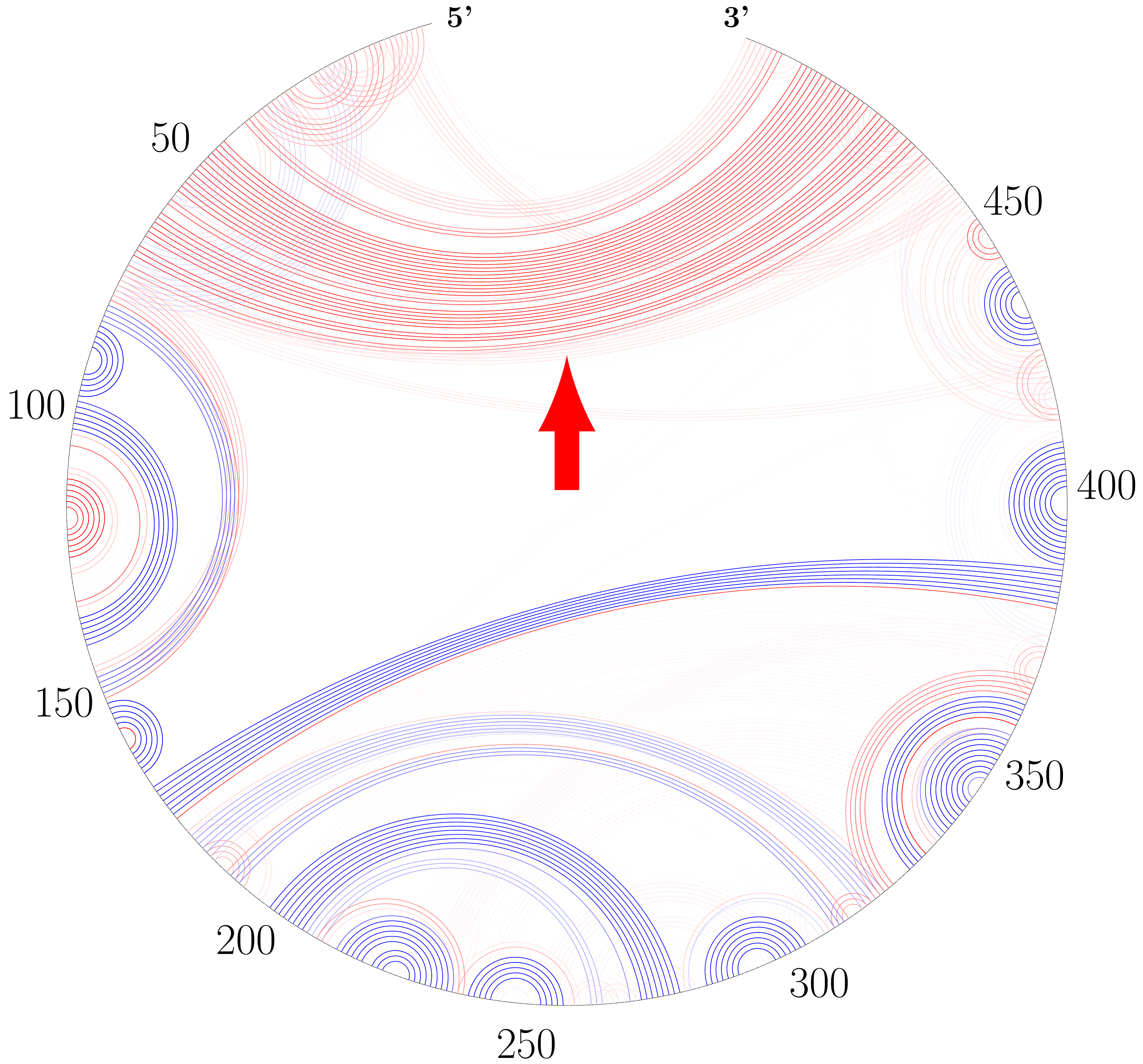} &
\hspace{-0.35cm}\includegraphics[width=0.22\textwidth]{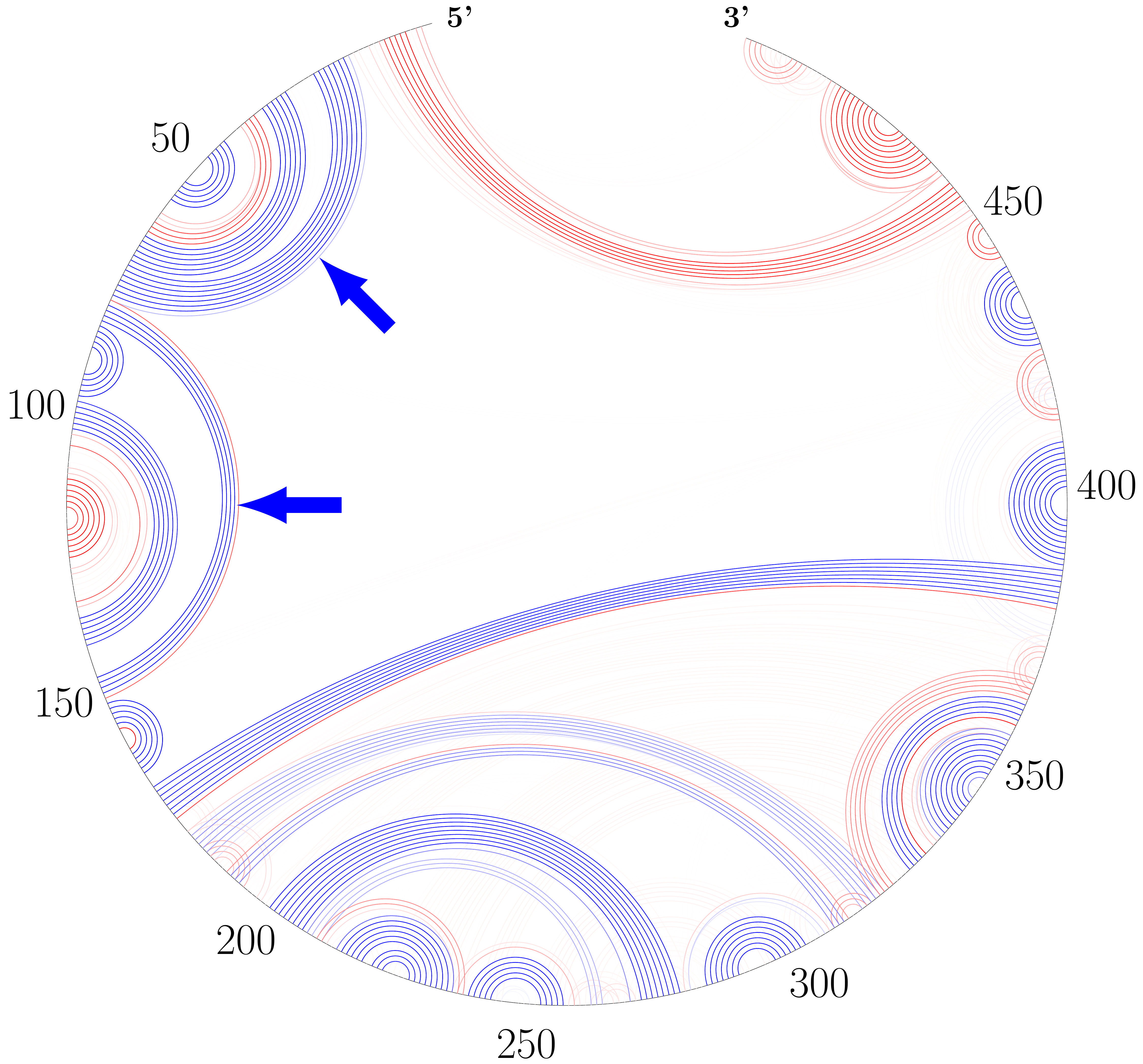}
\\[-0.2cm]
\end{tabular}
\caption{{\bf A}: Ensemble defect (expected number of incorrectly predicted nucleotides; lower is better) comparison between \viennarnafold and \linearpartition on the ArchiveII dataset.
  {\bf B}: Ensemble defect difference for each family.
  \linearpartition has lower ensemble defects for longer families:
   on average 56.3 less incorrectly predicted nucleotides on 23S rRNA and 8.3 less over all families.
  {\bf C--F}: An example of \ecoli 23S rRNA (shaded point in {\bf A}). 
  {\bf C}: Circular plot of the ground truth.
{\bf D--F}: Base pair probabilities from \viennarnaplfold (with default window size $70$), \rnafold and \linearpartition, respectively; 
Blue denotes pairs in the known structure and Red denotes predicted pairs not in the known structure.  
  The darkness of the line indicates pairing probability.
%  with the darkest lines close to a probability of 1. 
  {\bf G--J}: Circular plots of {\it C.~ellipsoidea} Group I Intron. 
%base pairs in the ground truth are in blue;
% blue and red indicate correct and incorrect base pairs,
% respectively, 
% and color darkness represents probabilities.
See Fig.~\ref{fig:example} 
% and~\ref{fig:circular_grp1} 
for another view of this example. % {\it C.~ellipsoidea} Group I Intron.
\label{fig:ensemble}
\vspace{-0.2cm}
}
\end{figure*}

Fig.~\ref{fig:runtime} compares the efficiency and scalability between the two baselines, 
\viennarnafold and \contrafold,
and our two versions, \linearpartitionv and \linearpartitionc.
% fair comparison, hack contrafold and vienna
To make the comparison fair, 
we disable the downstream tasks
(MEA prediction in \contrafold, and centroid prediction and visualization in \rnafold)
% which will be run together with partition function and base pairing probability calculation by default, 
which are by default enabled.
 % with partition function and base pairing probabilities calculation,
% for instance, 
%% These tasks are MEA structure generation in \contrafold,
%% as well as centroid structure generation and prediction visualization in \viennarnafold. 
Fig.~\ref{fig:runtime}A shows that both \linearpartitionv and \linearpartitionc
scale almost linearly with sequence length $n$.
The runtime deviation from exact linearity is due to the relatively short sequence lengths in the ArchiveII dataset, 
which contains a set of sequences with well-determined structures~\cite{sloma+mathews:2016}. 
%are relatively short
% (average length is 222.2~\nts).
Fig.~\ref{fig:runtime}A also confirms that the baselines scale cubically and the $O(n^3)$ runtimes are substantially slower than \linearpartition on long sequences. 
For the {\it H.~pylori} 23S rRNA sequence (2,968~\nts, the longest in ArchiveII), 
both versions of \linearpartition  take only 6 seconds,
while \rnafold and \contrafold take 73 and 120 seconds, resp.
%and  almost 120 seconds.

% Fig.~\ref{fig:runtime}B visualizes the runtimes on RNAcentral sampled sequences.
% We set the runtime limit for a sequence to be at most 24 hours.
We also notice that both \rnafold and \contrafold have limitations on even longer sequences.
% although \viennarnafold designs a scalar estimated from minimum free energy of the given sequence to avoid overflow, 
\rnafold scales the magnitude of the partition function 
using a constant estimated from the minimum free energy of the given sequence to avoid overflow,
% but is still easy to get overflow 
but overflows still occur 
on long sequences.
For example, it overflows %on partition function calculation 
on the 19,071~\nts sequence in the sampled RNAcentral dataset.
% so we only show the runtime before this sequence for \viennarnafold.
% Unlike \viennarnafold, \linearpartition adopts a log space partition function calculation
% as \contrafold and solve overflow issue fundamentally.
% \contrafold adopts logarithmic scale of partition function to solve overflow issue,
% but it cannot run on sequence longer than 32,753~\nts due to memory limitation. 
\contrafold stores the logarithm of the partition function to solve the overflow issue,
but cannot run on sequences longer than 32,767~\nts due to using {\tt unsigned short}
% in its implementation
to index sequence positions.
% Beyond these limitations, 
\linearpartition, like \contrafold, performs computations in the log-space,
but can run on all sequences in the RNAcentral dataset.
Fig.~\ref{fig:runtime}B compares the runtime of four systems on a sampled subset of RNAcentral dataset,
and shows that % only \linearpartition can finish all the examples,
on longer sequences the runtime of \linearpartition is exactly linear.
% but \viennarnafold only runs on the first 17 data points because of overflow issue. 
%Comparing the runtimes of
For the 15,780~\nts sequence, 
 % \viennarnafold can run,
the longest example shown for \rnafold, % in Fig.~\ref{fig:runtime}B,
%\viennarnafold takes more than 3 hours and
\linearpartitionv is 256$\times$ faster (more than 3 hours vs.~44.1 seconds). 
Note that \rnafold may not overflow on some longer sequences, 
where \linearpartitionv should enjoy an even more salient speedup.
% ???\contrafold stops at 17th or need to continue???
For the longest sequence that \contrafold can run (32,753~\nts) in the dataset,
\linearpartition is 2,771$\times$ faster (2.5 days vs.~1.3 min.).
%% % is 22,911~\nts, 
%% it takes 60.7 hours,
%% compared to \linearpartitionc's 52.4 seconds
%% (2,771$\times$ speedup).
% 32,767
% We find \contrafold has a sequence length limitation of 32,767~\nts due to using "unsigned short" in its implementation.
% We also test on an artificial sequence with length of 32,767~\nts, 
% the upper limit for \contrafold, 
% and the runtime is ?????
Even for the longest sequence in RNAcentral
(Homo Sapiens Transcript NONHSAT168677.1 with length 244,296~\nts~\cite{Zhao+:2016}),
both \linearpartition versions finish in $\sim$10 minutes.

Fig.~\ref{fig:runtime}C
%% compares the memory usage 
%% %of 4 systems 
%% on RNAcentral-sampled sequences. 
%% It 
shows that  \rnafold and \contrafold use $O(n^2)$ space
while \linearpartition uses $O(n)$.
%% With increasing length, the two baselines require much more memory space than \linearpartition.

%%%%%%%%%%%%%%%%

%%%%%%%%%%%%%%%%%

%\smallskip
Now that we have established the speed of  \linearpartition,
%in the next two subsections,
we move on to the quality of its output.
%i.e., the resulting Boltzmann distribution and base pairing probabilities.
%% We first study the correlation with ground truth structures in Sec.~\ref{sec:corr},
%% and then use base pairing probabilities for downstream structure predictions in Sec.~\ref{sec:acc}.

\vspace{-.1cm}
\subsection{Correlation with Ground Truth Structures}
\label{sec:corr}

% ensemble defect

We use {\em ensemble defect}~\cite{Zadeh+:2010} (Fig.~\ref{fig:ensemble}A--B)
to represent the quality of the Boltzmann distribution.
It is the expected number of incorrectly predicted nucleotides over the whole ensemble at equilibrium,
and formally,
for a sequence \vecx and its ground-truth structure \vecystar, the ensemble defect is% $\Phi(\vecx,\vecystar)$ %denotes the average number of incorrectly paired nucleotides at equilibrium, and is formalized as:
\begin{equation}
%\begin{split}
\vspace{-0.2cm}
\Phi(\vecx,\vecystar) = \sum_{\vecy \in \mathcal{Y}(\vecx)} p(\vecy) \cdot d(\vecy, \vecystar) % \\ 
%                 &= |\vecx| - 2 \sum_{(i,j) \in \mathrm{pairs(\vecystar)}} p_{i,j} - \sum_{j \in \mathrm{unpaired(\vecystar)}} q_{j}
%\end{split}
\label{eq:phi}
\end{equation}
where % $\phi$ is the folding model, 
%  $\vecystar$ is the groud truth secondary structure, 
%  $\mathcal{Y(\vecx)}$ is the ensemble of the sequence $\vecx$, 
%  and $|\vecx|$ is the sequence length. % ($N+1$ is for conveniently describing unpaired bases);
$p(\vecy)$ %=e^{-\frac{\Delta G^\circ(\vecy)}{RT}} / Q(\vecx)$
is the probability of structure \vecy % of the structure $\vecy$ in the ensemble.
in the ensemble %among all possible structures
$\mathcal{Y(\vecx)}$,
and $d(\vecy, \vecystar)$ is the distance between $\vecy$ and $\vecystar$, %,  defined as:
defined as the number of incorrectly predicted nucleotides in \vecy:
\vspace{-0.3cm}
 \[% \begin{equation*}
  \begin{split}
    d(\vecy,\vecystar)=  |\vecx|  & - |\pairs(\vecy) \cap \pairs(\vecystar)|\\
                        & - |\unpaired(\vecy) \cap\unpaired(\vecystar)|
  \end{split}
%  \label{eq:d}
 \]% \end{equation*}
 % \vspace{-0.2cm}
%% where  $\pairs(\vecystar)$ is the set of base pairs in $\vecystar$,
%% and $\unpaired(\vecystar)$ is the set of unpaired bases in $\vecystar$.
 The na\"ive calculation of Eq.~\ref{eq:phi} requires enumerating all possible % (exponentially many)
 structures in the ensemble, % which is intractable,
  but by plugging $d(\vecy, \vecystar)$ %Eq.~\ref{eq:d}
  into Eq.~\ref{eq:phi} % and with some derivations,
  we have~\cite{Zadeh+:2010}
\begin{equation}
\vspace{-0.2cm}
\begin{split}
\Phi(\vecx,\vecystar) = |\vecx| - 2\!\! \sum_{(i,j) \in \pairs(\vecystar)} p_{i,j} - \sum_{j \in \unpaired(\vecystar)} q_{j} \notag
\end{split}
\label{eq:phi2}
\vspace{-0.2cm}
\end{equation}
where $p_{i,j}$ is the probability of $i$ pairing with $j$ %, i.e., %(or the probability of $i$ being unpaired when $j=N+1$).
%% \(%[
%% p_{i,j} = \sum_{(i,j)\in \pairs(\vecy)} p(\vecy),
%% \)
and $q_{j}$ is the probability of $j$ being unpaired, i.e., $q_j = 1- \sum p_{i,j}$.
  % $S_{i,j}(s)$ is the a structure matrix with entries $S_{i,j}(s) \in \{0, 1\}$, i.e., if structure $s$ contains pair $(i,j)$, then $S_{i,j}(s) = 1$, otherwise $S_{i,j}(s) = 0$.
This means we can now use base pairing probabilities to compute the ensemble defect.

%  $d(\vecy, \vecystar)$ is the distance between $\vecy$ and $\vecystar$,  defined as:

\begin{figure}[!t]
  % \hspace{1.5cm}
\center
% \begin{tabular}{c|c|c|c}
\definecolor{darkgreen}{rgb}{0, 0.5, 0}
\newcommand{\bluetri}{{\color{blue} $\blacktriangle$}}
\newcommand{\greentri}{{\color{darkgreen} $\blacktriangle$}}
\newcommand{\redtri}{{\color{red} $\blacktriangle$}}
\newcommand{\bluecir}{{\color{blue} $\circ$}}
\newcommand{\greencir}{{\color{darkgreen} $\circ$}}
\newcommand{\redcir}{{\color{red} $\circ$}}
\begin{tabular}{cc}
\panel{A} & \hspace{-1cm}\panel{B} \\[-.7cm]
% \raisebox{8.7cm}{\hspace{-0.2cm}} %\includegraphics[width=0.4\textwidth]{figs/ensemble_defect} }
% &
% \hspace{-.2cm}\panel{A} & \hspace{-0.6cm}\panel{B} \\[-1cm] %& \hspace{-2.5cm}\panel{D}
	\multicolumn{2}{c}{
	  \hspace{-.3cm}
%          \vspace{-.2cm}
	  \raisebox{7.5cm}
	  {\includegraphics[width=4.4cm]{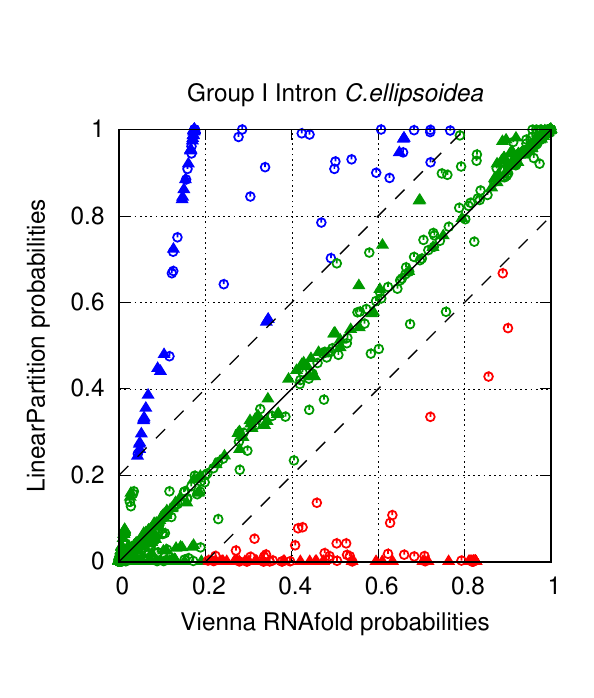}}
	  \hspace{-4.5cm}
	   %\raisebox{0cm}
	  {\includegraphics[width=0.5\textwidth]{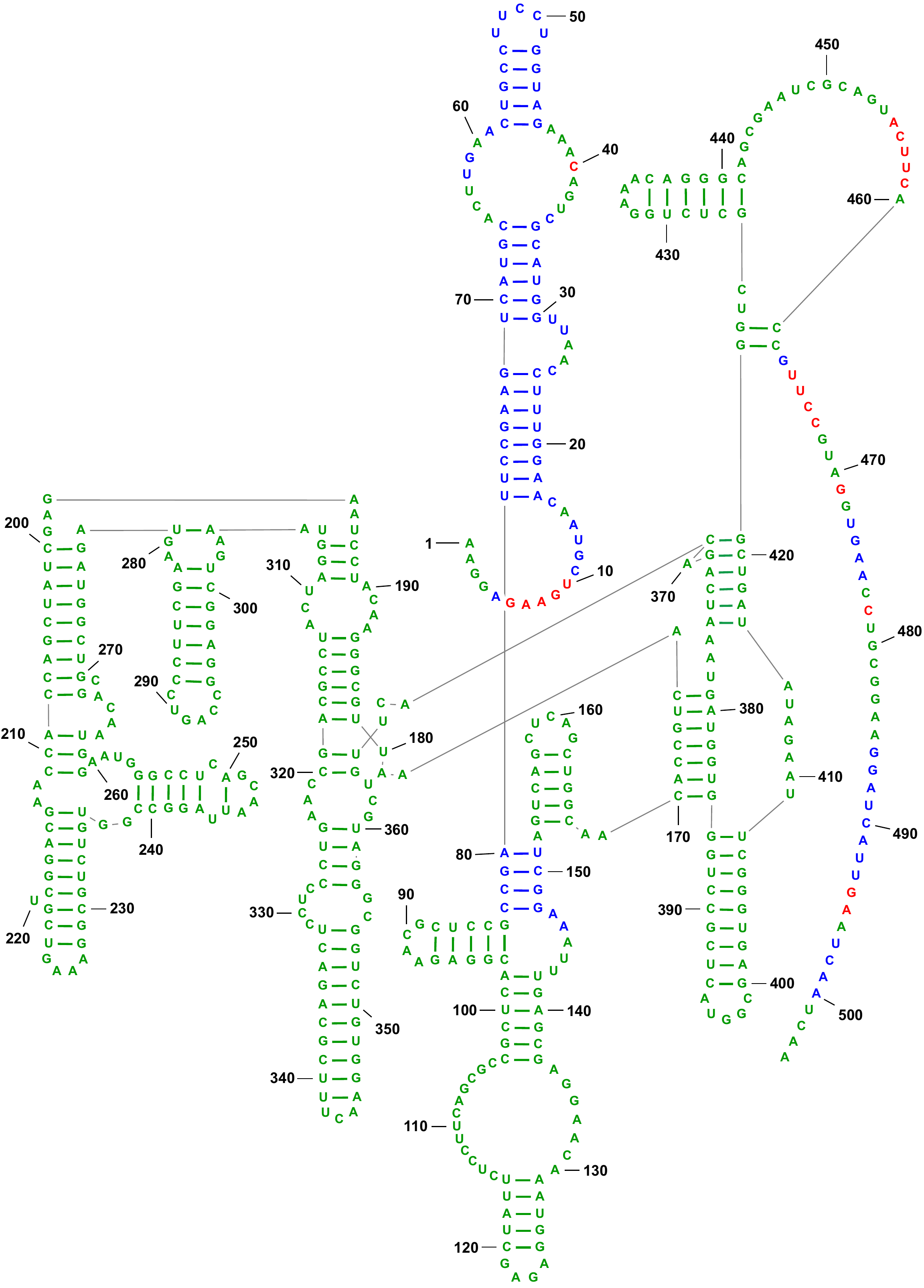}}
  	} \\[-1.6cm]
  	\hspace{-.0cm}\panel{C} &  \\[-1.cm]
  	\end{tabular}
% \hspace{-7cm}\panel{B} &\\[-.8cm]

% &&\raisebox{0cm}{\hspace{0.2cm}\includegraphics[width=0.25\textwidth]{figs/23s_gold}} \\
% &&\hspace{-4.5cm}\panel{E}\\[-.3cm]
% &&\raisebox{0cm}{\hspace{0.2cm}\includegraphics[width=0.25\textwidth]{figs/23s_vienna_plfold_example}} \\
% &&\hspace{-4.5cm}\panel{F}\\[-.3cm]
% &&\raisebox{0cm}{\hspace{0.2cm}\includegraphics[width=0.25\textwidth]{figs/23s_vienna_example}} \\
% &&\hspace{-4.5cm}\panel{G}\\[-.3cm]
% &&\raisebox{0cm}{\hspace{0.2cm}\includegraphics[width=0.25\textwidth]{figs/23s_example}} \\[-2.3cm]
% % } & \\[-.4cm]
% \hspace{-.2cm}\raisebox{1cm}{\panel{C}} & \\[-2.2cm]
% \hspace{-7cm}\panel{E} &\\[-.5cm]
% \multicolumn{2}{l}{% \hspace{-0.3cm}

\resizebox{0.5\textwidth}{!}
{
  \begin{tabular}{c}
%	\toprule
         % &     \multicolumn{2}{c}{total} & \multicolumn{2}{c}{correct} \\
			 % \midrule
\hspace{6.5cm}{\color{blue} $p_\linear \!-\! p_\vienna \!>\! 0.2$} \\[.3cm]
\hspace{6.4cm}{\color{darkgreen} $|p_\linear \!-\! p_\vienna| \!\leq \!0.2$} \\[.3cm]
\hspace{6.3cm}{\color{red} $p_\linear \!-\! p_\vienna \!<\! -0.2$}  \\[-2.4cm]
%	                 \bottomrule
  \end{tabular}
 }

% \resizebox{0.45\textwidth}{!}
% \panel{C}}
\vspace{.5cm}
{
  \hspace{-4.5cm}\begin{tabular}{r@{\; }r@{\quad}r@{\; }r}
%	\toprule
% \panel{C}} & &&\\
          \multicolumn{2}{c}{total} & \multicolumn{2}{c}{correct} \\
			 \midrule
55 \bluetri & 40 \bluecir  & 23 \bluetri & 37 \bluecir \\
  126,645 \greentri & 420 \greencir & 111 \greentri & 180 \greencir\\
 56 \redtri & 44 \redcir & 0 \redtri & 19 \redcir \\[1.9cm]
%	                 \bottomrule
  \end{tabular}
  }

% }
% &\\[-2cm]
% &\raisebox{5cm}{\hspace{-0.2cm}\includegraphics[width=0.25\textwidth]{figs/23s_gold}} \\
% \end{tabular}\\

% \begin{tabular}{cccc}
% \hspace{-4.4cm} \panel{F} & \hspace{-4.6cm}\panel{G} & \hspace{-4.6cm}\panel{H} & \hspace{-4.6cm}\panel{I}\\[-0.2cm]
% \hspace{-0.2cm}\includegraphics[width=0.25\textwidth]{figs/23s_gold} &
% \hspace{-0.35cm}\includegraphics[width=0.25\textwidth]{figs/23s_vienna_plfold_example.pdf} &
% \hspace{-0.35cm}\includegraphics[width=0.25\textwidth]{figs/23s_vienna_example} &
% \hspace{-0.35cm}\includegraphics[width=0.25\textwidth]{figs/23s_example.pdf}
% \end{tabular}
\vspace{-2cm}
\caption{
{\bf A--C}: An example of {\it C.~ellipsoidea} Group I Intron. 
{\bf A}: Solid triangles ({\small {\color{blue} $\blacktriangle$} {\color{darkgreen}$\blacktriangle$}  {\color{red}$\blacktriangle$}}) stand for base pairing probabilities and
unfilled circles ({\small {\color{blue} $\circ$} {\color{darkgreen}$\circ$}  {\color{red}$\circ$}})  stand for single-stranded probabilities.
{\color{blue} blue: $p_\linear \!-\! p_\vienna \!>\! 0.2$};
{\color{darkgreen} green: $|p_\linear \!-\! p_\vienna| \!\leq \!0.2$};
{\color{red} red: $p_\linear \!-\! p_\vienna \!<\! -0.2$};
%some high probability pairs and unpaired bases in \linearpartition have low probabilities in \viennarnafold (in blue), and some low probability ones in \linearpartition have high probabilities in \viennarnafold (in red); 
	{\bf B}: Ground truth structure colored with the above scheme; % from \viennarnafold and \linearpartition; 
	%pink binds around position 370 are pseudoknotted pairs.
	{\bf C}: Statistics of this example. 
	"total" columns are the total numbers of triangles and circles with different colors in {\bf A},
	while "correct" columns are the corresponding numbers %of such triangles and circles
        in the ground-truth structure  in {\bf B},
        which is better correlated with \linearpartition's probabilities than \viennarnafold's ({\color{blue} 23 blue pairs} and {\color{red} 0 red pairs}). %ground-truth structure.
	% Note that each triangle represents a pair of nucleotides.
%% {\bf D--G}: An example of \ecoli 23S rRNA. 
%% 	{\bf D}: circular plot of the ground truth.
%% {\bf E--G}: circular plots using the base pair probabilities from \viennarnaplfold (with default window size $70$), \rnafold and \linearpartition, respectively; 
%% base pairs in the ground truth are in blue;
%% the color shade of the lines are propotional to their probabilities.
	\label{fig:example}\\[-.7cm]
}
\end{figure}

\begin{figure*}[hbt!]
\center
\begin{tabular}{@{}c@{}c@{}c@{}c@{}}
\hspace{-4.6cm}\panel{A} & \hspace{-5.2cm}\panel{B} & \hspace{-5.2cm}\panel{C} & \hspace{-2.3cm}\panel{D} 
\\[-0.6cm]
\raisebox{0.35cm}{\includegraphics[scale=.8]{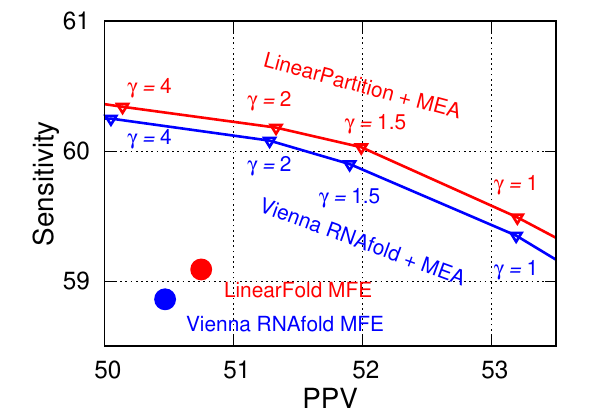}} 
&
%\hspace{-0.5cm}
\hspace{-0.3cm}\includegraphics[scale=.93]{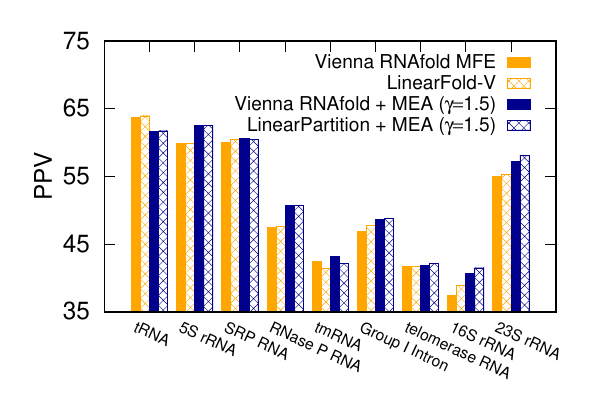} 
&
%\hspace{-0.5cm}
\hspace{-0.2cm}\includegraphics[scale=.93]{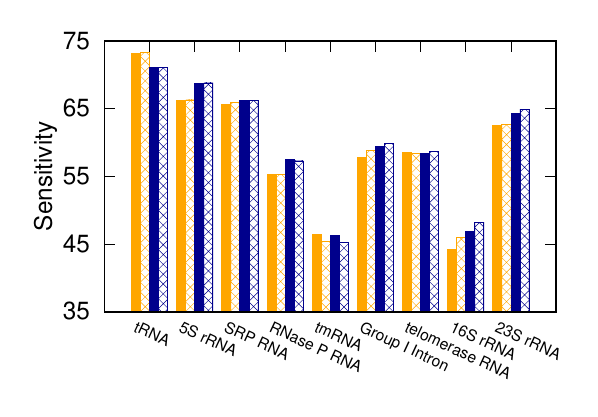}  
&
\raisebox{.5cm}{\hspace{0.2cm}\includegraphics[scale=.56]{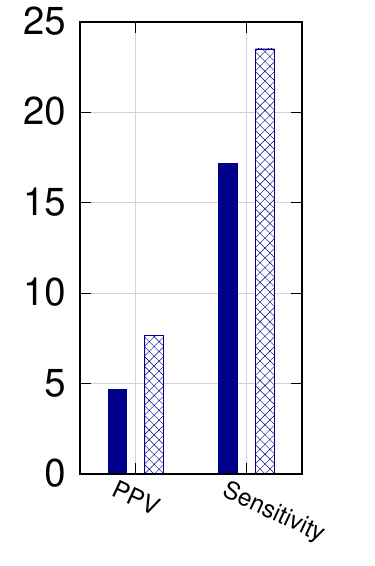}}
\\[-0.5cm]
\hspace{-4.6cm}\panel{E} & \hspace{-5.2cm}\panel{F} & \hspace{-5.2cm}\panel{G} & \hspace{-2.3cm}\panel{H} 
\\[-0.5cm]
\raisebox{0.35cm}{\includegraphics[scale=.8]{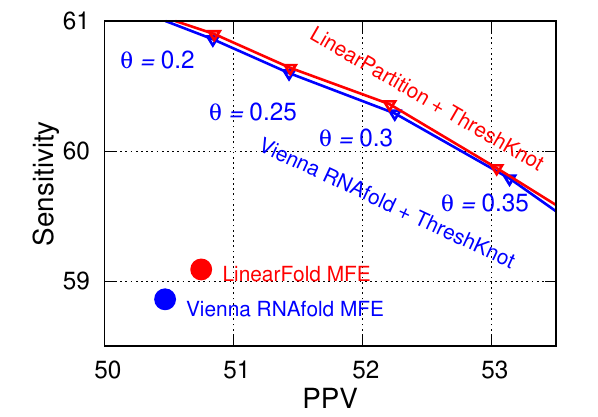}}
&
%\hspace{-0.5cm}
       \hspace{-0.3cm}{\includegraphics[scale=.93]{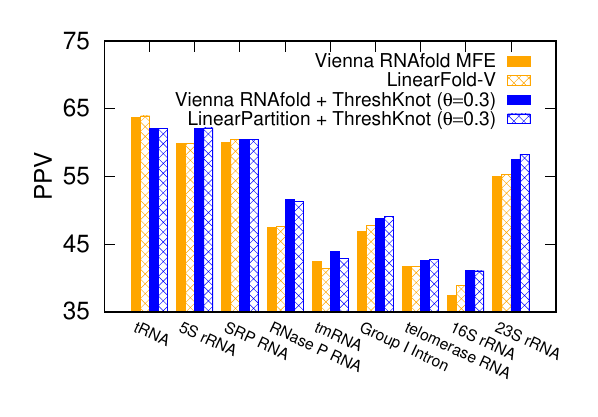}}
&
%       \hspace{-0.5cm}
      \hspace{-0.2cm}{\includegraphics[scale=.93]{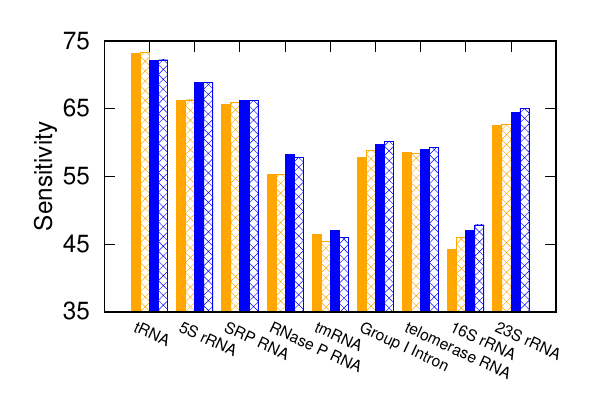}}
              &
\raisebox{.5cm}{\hspace{0.3cm}\includegraphics[scale=.56]{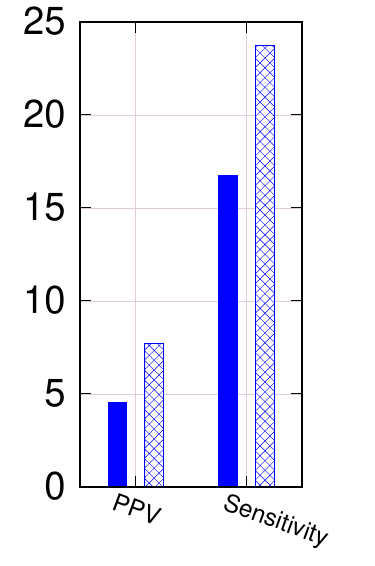} }
\vspace{-0.6cm}
\end{tabular} % A: specify MFE in label
\caption{Accuracy of downstream  predictions (MEA and \threshknot) using base pairing probabilities from \viennarnafold and \linearpartition
        on the ArchiveII dataset.
	{\bf A}: Overall PPV-Sensitivity tradeoff of MFE (single point) and MEA with varying $\gamma$
	%MFE is a single point, but
        (which can be tuned for higher sensitivity or PPV by adjusting $\gamma$).
	{\bf B} \& {\bf C}: PPV and Sensitivity comparisons of MEA structures for each family.
  {\bf D}: Accuracy comparison of long-distance base pairs (>500 \nts apart) in the MEA structures.
        {\bf E--H}: Same as {\bf A--D}, but using \threshknot predictions instead of MEA.
	%% {\bf D}: Overall PPV-Sensitivity tradeoff of \threshknot with varying threshold $\theta$.
	%% {\bf E} and {\bf F}: PPV and Sensitivity comparisons of ThreshKnot structures for each family.
        %	For easy comparison, {\bf A} and {\bf D} and also {\bf BCEF} are on the same scales.
        We conclude that MEA predictions based on \linearpartitionv are consistently better in both PPV and Sensitivity than
        those based on \viennarnafold for all $\gamma$'s,
        while \threshknot predictions based on those two are almost identical for all $\theta$'s.
        \linearpartitionv is  substantially better on long-distance base pairs in both MEA and \threshknot predictions.
	\label{mea}
        \vspace{-0.6cm}
}
\end{figure*}

Fig.~\ref{fig:ensemble}A--B employs ensemble defect to measure 
the average number %(Fig.~\ref{fig:ensemble_defect}A) 
%and ratio %(Fig.~\ref{fig:ensemble_defect}B) 
of incorrectly predicted nucleotides over the whole ensemble (lower is better).
% ~\cite{Dirks+:2004, Zadeh+:2010}.
\viennarnafold and \linearpartition have similar ensemble defects for short sequences, % are mostly similar between , % are the same or similar,
but \linearpartition has lower ensemble defects for longer sequences, esp.~16S and 23S rRNAs;
in other words, \linearpartition's ensemble has less expected number of incorrectly predicted nucleotides
(or higher number of correctly predicted nucleotides). %, i.e., better correlation with ground-truth structure).
In particular, on 16S and 23S rRNAs, \linearpartition has on average 15.9 and 56.3 more correctly predicted nucleotides than \rnafold,
and on average 8.3 more correctly predicted nucleotides over all families (Fig.~\ref{fig:ensemble}B).
Figs.~\ref{fig:ensemble_defect} show the relative ensemble defects (normalized by sequence lengths),
where the same observations hold, and \linearpartition has on avearge 0.4\% more correctly predicted nucleotides over all families.
In both cases, the differences on tmRNA (worse) and Group I Intron (better) are statistically significant ($p<0.01$).

This finding also implies that  \linearpartition's  base pairing probabilities 
are on average higher than \rnafold's for ground-truth base pairs,
and on average lower for incorrect base pairs.
We use two concrete examples to illustrate this.
First, we plot the ground truth structure of \ecoli 23S rRNA (2,904~\nts) in
Fig.~\ref{fig:ensemble}C,
and then plot the predicted base pairing probabilities
from the local folding tool \viennarnaplfold (with default window size 70), \rnafold, and \linearpartition
in Fig.~\ref{fig:ensemble}D--F, respectively.
We can see that local folding can only produce local pairing probabilities,
while \rnafold misses
most of the long-distance pairs from the ground truth (except the 5'-3' helix),
and includes many incorrect long-distance pairings (shown in red).
By contrast, \linearpartition successfully predicts many
long-distance pairings that \rnafold misses, the longest being 582~\nts apart (shown with arrows).
Indeed, the ensemble defect of this example confirms that \linearpartition's ensemble distribution
has on average 211.4 more correctly predicted nucleotides (over 2,904~\nts, or 7.3\%) than \rnafold's.

As the second example, we use %an RNA sequence,
{\it C.~ellipsoidea} Group I Intron (504~\nts).
First, in Fig.~\ref{fig:ensemble}G--J, we plot the circular plots in the same style as the previous example,
where \linearpartition is substantially better in predicting 4 helices in the ground-truth structure:
[17,24]--[72,79], [30,45]--[66,71], [44,48]--[54,58], and [80,83]--[148,151] (annotated with blue arrows).
%as an example to illustrate this.
% show \linearpartition gives more accurate predicted secondary structure than \viennarnafold.
%compare the base pairing probabilities generated by \viennarnafold and \linearpartition.
Next, in Fig.~\ref{fig:example}A, 
we plot the base pairs (in triangle) and unpaired bases (in circle) 
with \rnafold probability on x-axis and \linearpartition probability on y-axis.
We color the circles and triangles in blue where \linearpartition gives 0.2 higher probability 
than \rnafold %gives low probabilities 
(top left region),
the opposite ones (bottom right region) in red,
and the remainder (diagonal region, with probability changes less than 0.2)
in green.
Then in Fig.~\ref{fig:example}B,
we visualize the ground truth structure~\cite{Cannone+:2002} and color the bases as in Fig.~\ref{fig:example}A.
We observe that the majority of bases are in green, indicating that \rnafold and \linearpartition
agree  with for a majority of the structure features. 
But the blue helices (near 5'-end and [80,83]--[148,151], see also Fig.~\ref{fig:ensemble}J)
indicate that \linearpartition favors these correct substructures by giving them higher probabilities than \rnafold.
We also notice that all red features (where \rnafold does better than \linearpartition) are unpaired bases.
This example shows that although \linearpartition and \rnafold give different probabilities, % compared with \rnafold, 
it is likely that \linearpartition prediction structure is closer to the ground truth structure
(which will be confirmed by downstream structure predictions in Section~\ref{sec:acc}).
The ensemble defect of this example also confirms
that \linearpartition has on average 47.1 more correctly predicted nucleotides (out of 504~\nts, or 9.3\%)
than \rnafold.

Fig.~\ref{fig:example}C gives the statistics of this example.
We can see the green triangles in Fig.~\ref{fig:example}A, 
which denote similar %base pairing
probabilities between \rnafold and \linearpartition, 
are the vast majority. % and the total number is 126,645.
The total number of blue triangles,
for which \linearpartition gives higher base pairing probabilities,
is 55,
and among them 23 %base pairs
(41.8\%) are in the ground truth structure.
On the contrary,
56 triangles are in red,
but none of these \rnafold prefered base pairs are correct. %in the ground truth structure.
% Notice that all the red triangles are on $y=0$ line, 
% for which \linearpartition gives 0 probabilities,
% are in the ground truth structure.
For unpaired bases, 
\linearpartition also gives higher probabilities to more ground truth unpaired bases:
there are 40 blue circles,
among which 37 (92.5\%) are unpaired in the ground truth structure,
while only 19 out of the 44 red circles (43.2\%) are in the ground truth structure.
%% See also Fig.~\ref{fig:ensemble}G--J for another view of this example in the style of Fig.~\ref{fig:ensemble}C--F
%% using circular plots.

\vspace{-0.3cm}
\subsection{Accuracy of Downstream Predictions}%  using Base Pairing Probabilities}
\label{sec:acc}

% \begin{figure}[t!]
% \center
% \begin{tabular}{cc}
% \raisebox{2cm}{\panel{A}} & \includegraphics[scale=.7]{figs/new_MEA_diff_gamma_beam_inf_threshold_001} \\[-0.6cm]
% \raisebox{2cm}{\panel{B}} &
% \raisebox{-0.25cm}{\includegraphics[scale=.58]{figs/for_He_Zhang_MEA_gamma_B100}}\\[-0.6cm]
% \raisebox{3cm}{\panel{C}} & \hspace{-0.5cm}\includegraphics[scale=.78]{figs/ProbKnot_PPV}\\[-0.4cm]
% \raisebox{3cm}{\panel{D}} &\hspace{-0.5cm}\includegraphics[scale=.78]{figs/ProbKnot_Sens}

% \end{tabular} % A: specify MFE in label
% \caption{Accuracy comparison for two systems.
% 	{\bf A}: Overall MFE and MEA structure PPV-sensitivity tradeoff of two systems with varying $\gamma$. 
% 	{\bf B}: Overall ThreshKnot structure PPV-sensitivity tradeoff of two systems with varying threshold $p$.
% 	{\bf C-D}: PPV and sensitivity comparison for each family.
% 	\label{mea}
% }
% \end{figure}

An important application of the partition function
is to improve structure prediction accuracy (over MFE) using base pairing probabilities.
%These downstream prediction methods include MEA\cite{do+:2006}, \centroidfold~\cite{sato+:2009}, \dotknot~\cite{Sperschneider+Datta:2010}, \probknot~\cite{bellaousov+mathews:2010}, and \ipknot~\cite{sato+:2011},
%have been shown to outperform MFE in accuracy
Here we use two such ``downstream prediction'' methods, MEA\cite{do+:2006} and \threshknot~\cite{Zhang+:2019} which is a thresholded version of \probknot~\cite{bellaousov+mathews:2010},
and compare their results using base pairing probabilities from $O(n^3)$-time baselines and our $O(n)$-time \linearpartition.
%We next consider the accuracy of downstream structure prediction using \linearpartition-produced 
%% First we take base pairing probability matrices from \linearpartition and \viennarnafold (or \contrafold), 
%% feed them to the standard MEA algorithm separately, 
%% and compare the accuracies of prediction structures.
We use Positive Predictive Value (PPV, the fraction of predicted pairs in the known structure, a.k.a.~precision) 
and sensitivity (the fraction of known pairs predicted, a.k.a.~recall) 
as accuracy measurements for each family,
and get overall accuracy be averaging over families. 
% We use slipping method{}
When scoring accuracy, we allow base pairs to differ by one nucleotide in position~\cite{mathews+:1999}.
% ~\cite{sloma+mathews:2016}
We compare \rnafold and \linearpartitionv on the ArchiveII dataset in the main text, and provide the \contrafold vs.~\linearpartitionc comparisons
in the Supporting Information Figs.~\ref{fig:mea_lpc}--\ref{fig:threshknot_lpc}.

% Fig.~\ref{mea} shows that \linearpartition even leads to a small improvement in the downstream MEA predictoin using the probability matrix computed in linear time. 
% Fig.~\ref{mea} gives the accuracy comparison between 
Fig.~\ref{mea}A shows  
MEA predictions (\rnafold + MEA and \linearpartition + MEA) are more accurate than MFE ones (\rnafold MFE and \linearfoldv),
but more importantly,
\linearpartition + MEA  consistently outperforms \rnafold + MEA in both PPV and sensitivity
with the same $\gamma$, a hyperparameter that balances PPV and sensitivity in MEA algorithm.
%\linearpartition + MEA enjoys a small improvement in both PPV and sensitivity.

Figs.~\ref{mea}B--C detail the per-family PPV and sensitivity, respectively,
for MFE and MEA ($\gamma=1.5$) results from Fig.~\ref{mea}A.
% two MFE-based systems and two MEA-based systems for each family
%% for MEA and MFE structures using \rnafold and our \linearfold and \linearpartition.
%% % We can see
%% The MEA
%% % -based systems lead to accuracy improvements over MFE-based systems 
%% structure predictions are more accurate than MFE predictions
%% for most families.
\linearpartition + MEA has similar PPV and sensitivity as \rnafold + MEA on short families (tRNA, 5S rRNA and SRP),
but interestingly, is more accurate on longer families, especially the two longest ones, 16S rRNA (+0.86 on PPV and +1.29 on sensitivity) 
and 23S rRNA (+0.88 on PPV and +0.62 on sensitivity). 
%% We also performed a two-tailed permutation test to test the statistical significance, 
%% and observed that on tmRNA the MEA structures of \linearpartition is significantly worse ($p<0.01$) than \viennarnafold in both PPV and Sensitivity.

% ThreshKnot
%% ProbKnot is another partition function-based structure prediction method that adds a straightforward post-processing step 
%% % after partition function calculation
%% of base pairing probabilities to predict structures
%% and is simpler and faster than MEA~\cite{bellaousov+mathews:2010}.
% Beyond nested structures, ProbKnot can predict pseudoknots. 
\probknot is another downstream prediction method
that is simpler and faster than MEA;
it assembles 
%In ProbKnot, structures are composed of
base pairs with reciprocal highest pairing probabilities. 
Recently, we demonstrated \threshknot~\cite{Zhang+:2019}, 
a simple thresholded version of ProbKnot that only includes pairs that exceed the threshold, 
leads to more accurate %overall 
predictions that outperform MEA by filtering out unlikely pairs, i.e., those whose probabilities fall under a given threshold $\theta$.
%% It has been shown ThreshKnot can achieve better PPV and Sensitivity than the more involved MEA algorithm, 
%% % and can predict pseudoknots which is beyond MEA scope,
%% so we also compare ThreshKnot structure accuracy between \viennarnafold and \linearpartition.

Shown in Fig.~\ref{mea}E, \linearpartition + \threshknot is almost identical in overall accuracy %(slightly better in sensitivity)
to \rnafold + \threshknot at all $\theta$'s,
% in the downstream ThreshKnot prediction using the probability matrix computed in linear time.
%Figs.~\ref{mea}E and F show that \linearpartition + ThreshKnot
and is slightly better than the latter on long families
(+0.24 on PPV and +0.38 on sensitivity for Group I Intron, +0.12 and +0.37 for telomerase RNA, and +0.74 and +0.62 for 23S rRNA)
(Figs.~\ref{mea}F--G).
We also performed a two-tailed permutation test to test the statistical significance, 
and observed that on tmRNA, both MEA and \threshknot structures of \linearpartition are significantly worse ($p\!<\!0.01$) than their \rnafold-based counterparts in both PPV and Sensitivity.
%As was observed for MEA comparison, \linearpartition + ThreshKnot is significantly worse ($p<0.01$) than \viennarnafold on tmRNA.
%in PPV and Sensitivity.

\begin{figure}%[!b]
  \vspace{-.1cm}
\center
%% \iffalse
%% \begin{tabular}{cc}
%% \raisebox{4.6cm}{\panel{A}} & \hspace{-1cm}\includegraphics[width=0.3\textwidth]{figs/ensemble_xy} \\
%% % \raisebox{3.6cm}{\panel{B}}
%% \hspace{-1cm}\includegraphics[width=0.2\textwidth]{figs/ensemble_len}
%%  & \hspace{-1cm}\includegraphics[width=0.2\textwidth]{figs/ensemble_len_norm} \\[-0.3cm]
%% \end{tabular}
%% \fi
%% \iffalse
%% \begin{tabular}{ccc}
%% {\panel{A}} & {\panel{B}} & {\panel{C}} \\
%% \hspace{-1cm}\includegraphics[width=0.3\textwidth]{figs/ensemble_xy} &
%% \hspace{-.8cm}\includegraphics[width=0.3\textwidth]{figs/ensemble_len} &
%% \hspace{-.8cm}\includegraphics[width=0.3\textwidth]{figs/ensemble_len_norm} \\[-0.3cm]
%% \end{tabular}
%% \fi
\begin{tabular}{cc}
% {\panel{A}} & {\panel{B}}\\
\hspace{-0.2cm}\raisebox{2.9cm}{\panel{A}\hspace{-2.2cm}}\raisebox{2.5cm}{\multirow{2}{*}{\includegraphics[width=0.47\textwidth]{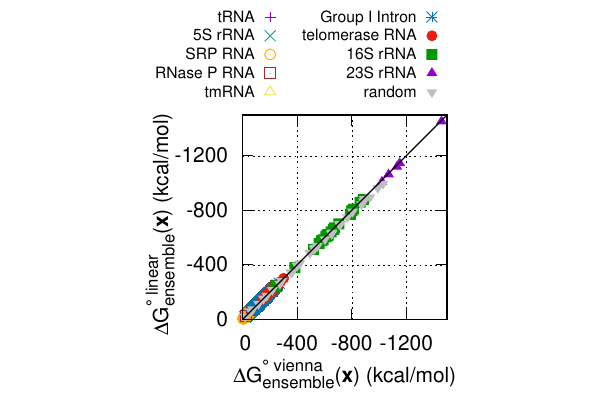}}} & 
\hspace{-2.4cm}\raisebox{2.9cm}{\panel{B}}\hspace{-.35cm}\includegraphics[scale=0.74]{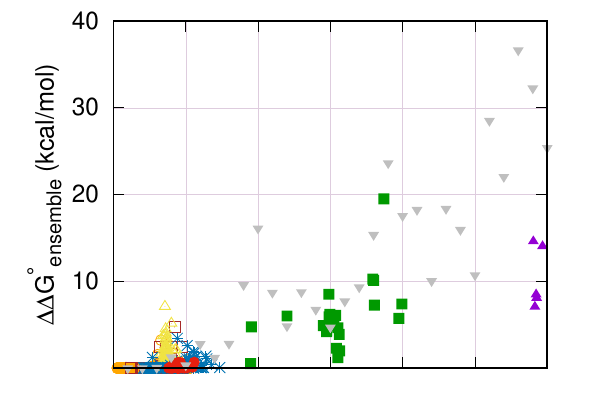} \\[-.5cm]
% &{\panel{A}}\\
&\hspace{-2.4cm}\raisebox{2.6cm}{\panel{\color{white}C}}\hspace{-.3cm}\includegraphics[scale=0.79]{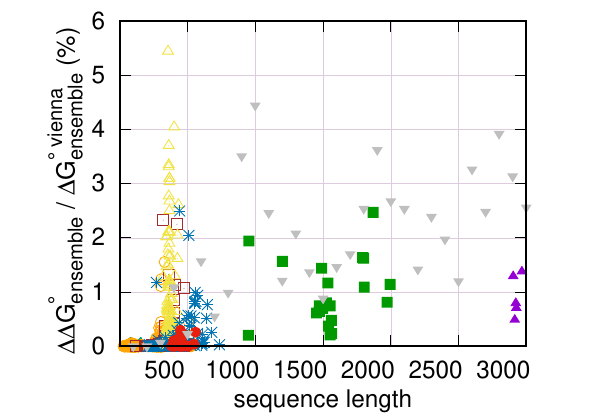} \\[-0.4cm]
\end{tabular}
\caption{
  Approximation quality of partition function  on ArchiveII dataset and random sequences. %. shorter than 3,000~\nts.
  {\bf A}: The x and y axes are
  ensemble folding free energy changes $\ensenergy(\vecx)$ of \viennarnafold %($-RT\log Q_\vienna(\vecx)$)
  and \linearpartition, % ($-RT\log Q_\linear(\vecx)$),
  respectively.
  {\bf B}: 
  Difference of ensemble folding free energy change (top), ${\ddg(\vecx)}$, between \rnafold and \linearpartition.
  and the relative differences (bottom),  ${\ddg(\vecx)}/{\ensenergyvienna(\vecx)}$, in percentages.
  %% of 
  %% ensemble folding free energy changes,
%  $\frac{ \ensenergyvienna(\vecx) - \ensenergylinear(\vecx)}{\ensenergylinear(\vecx)}$.
  \label{fig:partition}
  \vspace{-0.2cm}
}
\end{figure}

Fig.~\ref{mea}D \& H show that \linearpartition-based predictions are subtantially better than \rnafold's (in both PPV and sensitivity) for long-distance base pairs (those with 500+ \nts apart),
which are well known to be challenging for the current models.
Fig.~\ref{fig:distance} details the accuracies on base pairs with different distance groups.

Figs.~\ref{fig:mea_lpc}--\ref{fig:threshknot_lpc} show similar comparisons between 
\contrafold and \linearpartitionc using MEA and ThreshKnot prediction, %, separately. 
% Fig.~\ref{fig:mea_lpc} compares MEA structures ($\gamma>1.5$) accuracy based on these two systems and Fig.~\ref{fig:threshknot_lpc} compares ThreshKnot structures ($p>0.2$). 
with similar results to Fig.~\ref{mea}, i.e., downstream structure prediction using \linearpartitionc is as accurate as using \contrafold, and  (sometimes significantly) more accurate on longer families.

\iftrue
%<<<<<<< HEAD
\begin{figure*}[!h]
%\vspace{-0.2cm}
%% =======
%% \begin{figure*}[!ht]
%% \vspace{-0.5cm}
%% >>>>>>> e07b35b96f33f564a07e21f062362ad3acd7bbad
\center
\begin{tabular}{cc|c}
\hspace{-5.5cm}\panel{A} & \hspace{-6.cm}\panel{B} & \hspace{-5.2cm}\panel{E} \\[-0.5cm]
\hspace{-0.5cm}\includegraphics[width=0.33\textwidth]{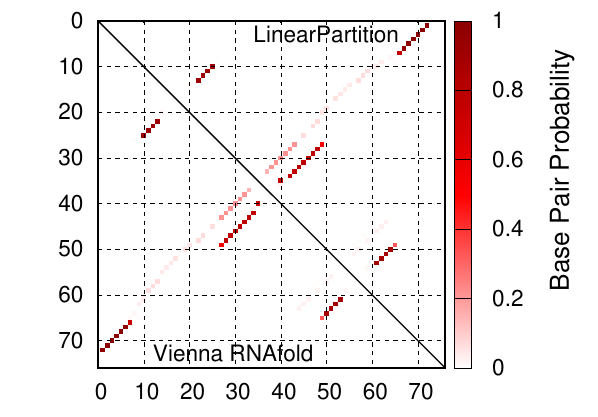}
&
% \hspace{-0.8cm}{\includegraphics[width=0.33\textwidth]{figs/rmsd_family_plus_random_b2}}
\hspace{-1.1cm}{\includegraphics[width=0.33\textwidth]{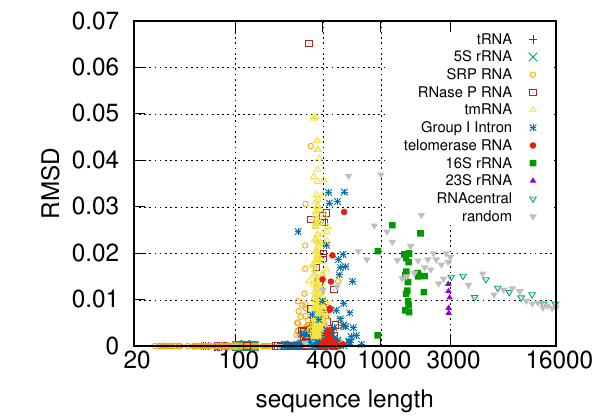}}
% \\[-0.2cm]
&
% \panel{C} & \panel{D}\\[-0.5cm]
\hspace{0.2cm}\includegraphics[width=0.33\textwidth]{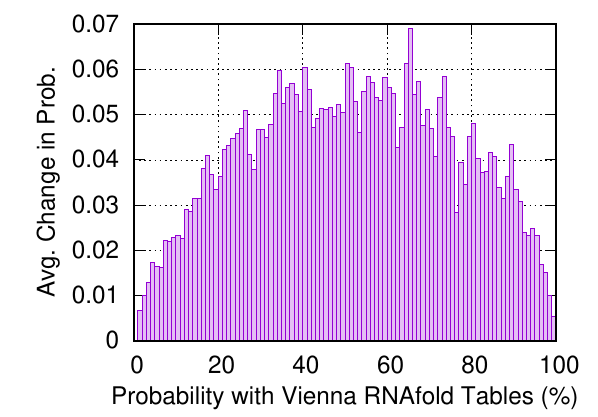} \\[-0.1cm]

\hspace{-5.5cm}\panel{C} & \hspace{-6.cm}\panel{D} & \hspace{-5.2cm}\panel{F} \\[-0.5cm]
\hspace{-.8cm}\raisebox{-0.2cm}{\includegraphics[width=0.35\textwidth]{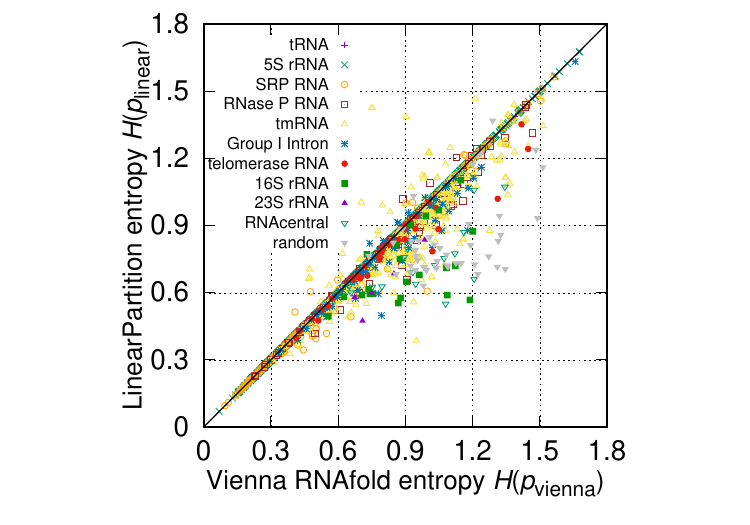}}
&
\hspace{-0.7cm}\includegraphics[width=0.33\textwidth]{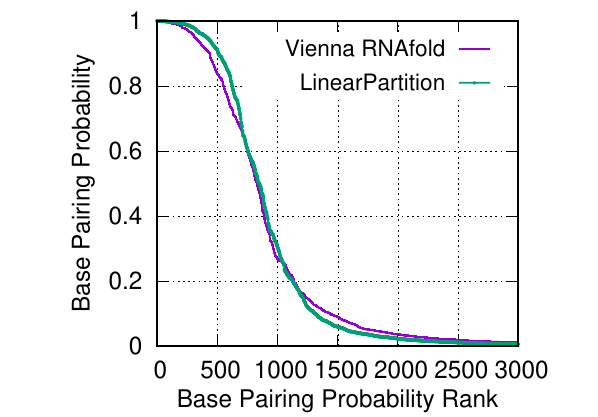} 
&
\hspace{-0.0cm}\includegraphics[width=0.34\textwidth]{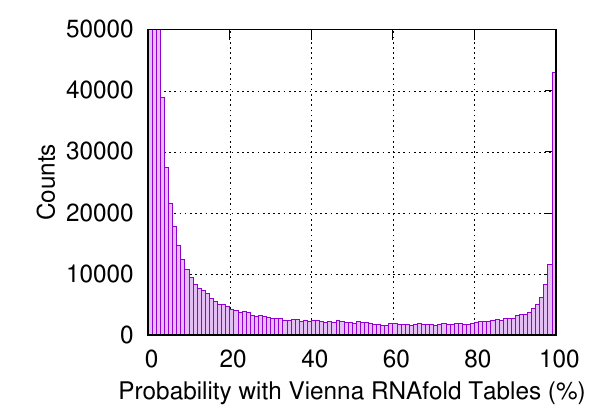}  \\[-0.2cm]
\end{tabular}
\caption{
Comparison of base pairing probabilities from \viennarnafold and \linearpartition.
	{\bf A}: \linearpartition (upper triangle) and Vienna RNAfold (lower triangle) result in identical base pairing probability matrix for \ecoli tRNA$^\textit{Gly}$.
	{\bf B}: The root-mean-square deviation, $\RMSD(p_\vienna, p_\linear)$, is relatively small between \linearpartition and Vienna RNAfold; all tRNA and 5S rRNA sequences \RMSD is close to 0 (e.g., RMSD$<10^{-5}$).
	{\bf C}: Average positional structural entropy $H(p)$ comparison; \linearpartition has noticeably lower entropy.
	{\bf D}: \linearpartition starts higher and finishes lower than \viennarnafold in a sorted probability curve for \ecoli 23S rRNA,
        suggesting lower entropy.
	{\bf E}: Mean absolute value of change in base pairing probabilities between \viennarnafold and \linearpartition; these changes are averaged within every probability bin.
	{\bf F}: Pair probability distribution of \viennarnafold. 
	Note that the y-axis is limited to 50,000 counts, and the counts of first three bins (with probability smaller than 3\%) are far beyond 50,000.
	\label{fig:rmsd}
}
\vspace{-0.3cm}
\end{figure*}
\fi

\vspace{-0.2cm}
\subsection{Approximation Quality (Default Beam Size)}

\linearpartition uses beam pruning to ensure $O(n)$ runtime, % and linear space, % linearity, 
thus is approximate compared with standard $O(n^3)$-time algorithms.
% Fig. 4A–B show that our \linearpartition algorithm can indeed approximate the partition function reasonably well.
We now investigate its approximation quality at the default beam size~100.

First, in Fig.~\ref{fig:partition}, we measure the approximation quality of the partition function calculation,
%and specifically,
in particular,  the ensemble folding free energy change (also known as ``free energy of the ensemble'')
which reflects the size of the partition function, 
\[
\ensenergy (\vecx) = -RT \log Q(\vecx).
\]
Fig.~\ref{fig:partition}A shows that the \linearpartition 
% free energy of ensemble 
estimate for the ensemble folding free energy change
is close to the \rnafold estimate 
on the ArchiveII dataset and randomly generated RNA sequences.
The similarity shows that little magnitude of the partition function is lost by the beam pruning. 
For short families, free energy of ensembles between \linearpartition and \rnafold are almost the same.
For 16S and 23S rRNA sequences and long random sequences (longer than 900 nucleotides), 
\linearpartition gives a lower magnitude ensemble free energy change, 
but the difference,
\[
\ddg(\vecx) = \ensenergyvienna(\vecx) - \ensenergylinear(\vecx) \, \geq 0
\]
is smaller than 20 kcal/mol for 16S rRNA, 
15 kcal/mol for 23S rRNA,
and 37 kcal/mol for random sequences (Fig.~\ref{fig:partition}B). 
The maximum difference for random sequence is larger than natural sequences (by 17.2 kcal/mol).
This likely reflects the fact that random sequences tend to fold less selectively to probable structures~\cite{Fu+:2015},
and the beam is therefore pruning structures in random that would contribute to the overall folding stability.
% Fig.~\ref{ensemble} confirms \linearpartition approximation quality of partition function is good.
Fig~\ref{fig:partition}C shows the ``relative'' differences in ensemble free energy changes,
 ${\ddg(\vecx)}/{\ensenergyvienna(\vecx)}$,
are also very small: only up to 2.5\% and 1.5\% for 16S and 23S rRNAs, and up to 4.5\% for random sequences.

Next, in Fig.~\ref{fig:rmsd}, we measure the approximation quality of base pairing probabilities using root-mean-square deviation (\RMSD)
between two probability matrices $p$ %(from cubic algorithms, for example, \viennarnafold) 
and $p'$ %(from our algorithm, for example, \linearpartition)
%(i.e., $p_\vienna$ and $p_\linear$)
over the set of all possible Watson-Crick and wobble pairs on a sequence $\vecx$.
We  define
\begin{equation}
  \vspace{-0.4cm}
\begin{split}
  {\pairings}(\vecx) \!=\!\{  &1\leq i <j\leq |\vecx| \bigm\vert j-i>3 \\[-0.1cm]
    &  \vecx_i\vecx_j \in \{\text{\small CG, GC, AU, UA, GU, UG}\}\!\} \notag
\end{split}
\end{equation}
\vspace{-0.2cm}
and:\vspace{-0.05cm} 
\begin{equation}
\RMSD(p,p')\!=\!\!\!\sqrt{\frac{1}{|{\pairings}(\vecx)|} \sum_{(i,j) \in {{\pairings}(\vecx)}}{\!\!\!\!\!(p_{i,j}\!-\!p'_{i,j})}^2}\notag
\end{equation}

\begin{figure*}[!h]
  \vspace{-0.7cm}
\center
\begin{tabular}{cccc}
% \hspace{-5.cm} \panel{A} & \hspace{-5.2cm}\panel{B}  \\[-0.2cm]
\raisebox{3.7cm}{\panel{A}} & 
\hspace{-0.5cm}\includegraphics[width=0.4\textwidth]{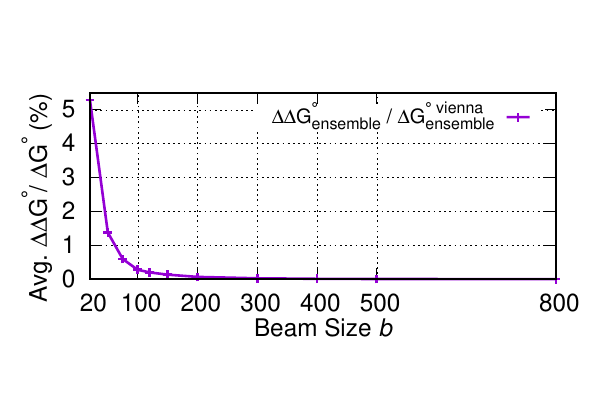} &
\raisebox{3.7cm}{\panel{B}} &
\hspace{-0.5cm}{\includegraphics[width=0.4\textwidth]{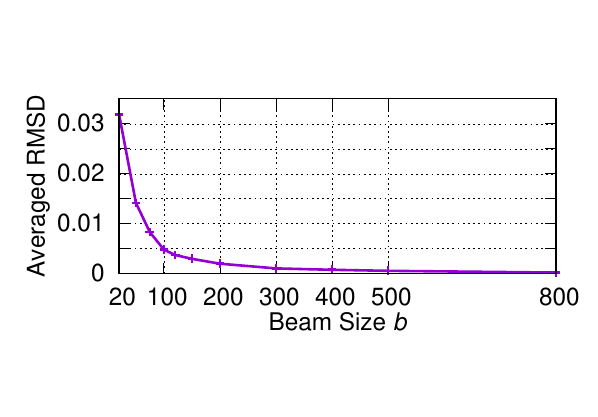}}
\end{tabular}
\\[-1.2cm]
\begin{tabular}{c@{}c@{}c@{}c@{}c@{}c}
% \hspace{-5.cm}\panel{C} & \hspace{-5.2cm}\panel{D} & \hspace{-5.8cm}\panel{E} \\[-0.2cm]
\hspace{0.6cm}\raisebox{3.8cm}{\panel{C}} & 
\hspace{-1cm}\includegraphics[width=.35\textwidth]{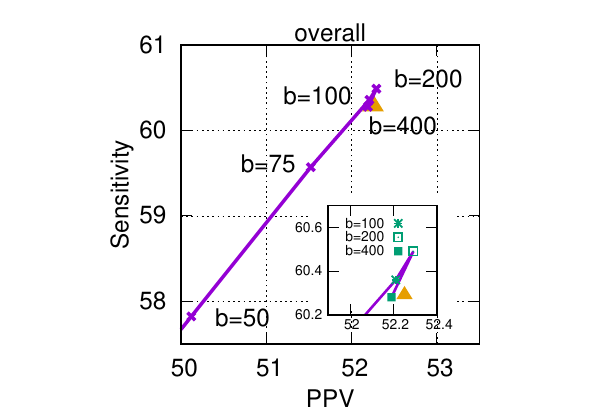} 
&
\hspace{-.5cm}\raisebox{3.8cm}{\panel{D}} & 
\hspace{-1.2cm}\includegraphics[width=0.35\textwidth]{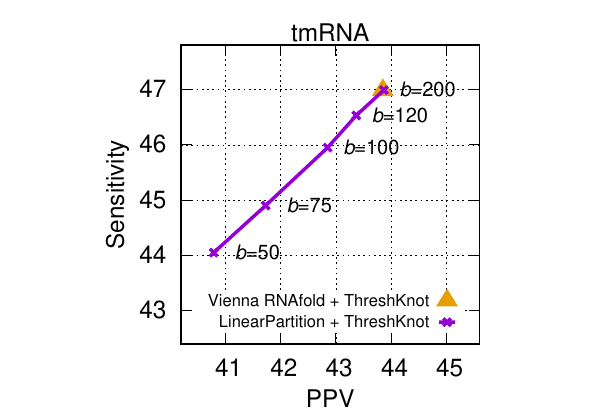}
&
\hspace{-0cm}\raisebox{3.8cm}{\panel{E}} & 
\hspace{-1.5cm}\includegraphics[width=0.35\textwidth]{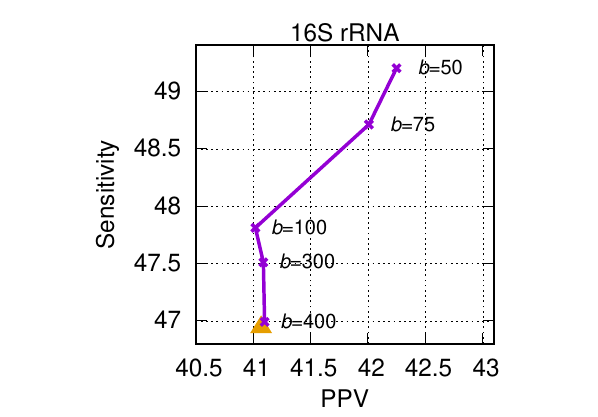}
\vspace{-0.4cm}
\end{tabular}
\caption{
Impact of beam size. 
{\bf A}: Relative difference of 
  ensemble folding free energy change, $\ddg/\Delta G^\circ_\text{ensemble}$, against  beam size. %, averaged over all families.
{\bf B}: RMSD against beam size. %, averaged over all families.
{\bf C}: Overall PPV and Sensitivity with beam size.
{\bf D--E}: tmRNA and 16S rRNA PPV and Sensitivity against beam size, respectively. 
%{\bf E}: 16S rRNA PPV and Sensitivity change with beam size. 
Note that the  results of \threshknot using \rnafold (yellow triangles in {\bf C--E}) are identical to \threshknot using
the {\em exact} version of \linearpartition ($b\!=\!\infty$).
%since \linearpartition with infinite beam size (i.e., no beam pruning) does $O(n^3)$ exact partition function calculation as \viennarnafold.
\label{fig:beamsize}
\vspace{-0.7cm}
}
\end{figure*}

Figs.~\ref{fig:rmsd}A and B confirm that our \linearpartition algorithm (with default beam size 100) can indeed approximate the base pairing probability matrix reasonably well. 
Fig.~\ref{fig:rmsd}A shows the heatmap of probability matrices for {\it E.~coli} tRNA$^\textit{Gly}$. 
\rnafold (lower triangle) and \linearpartition (upper triangle)
yield identical matrices (i.e., $\RMSD=0$). % (i.e., RMSD$\approx$0). 
Fig.~\ref{fig:rmsd}B shows that the \RMSD of each sequence in ArchiveII and RNAcentral datasets, 
and randomly generated artificial RNA sequences, %with length 100–16,000~\nts.
is relatively small. % across all sequences. 
% To verify approximation quality of 
The highest deviation is 0.065 for {\it A.~truei} RNase P RNA, 
which means on average 
each base pair's probability deviation in that worst-case sequence is about 0.065 between the cubic algorithm (\rnafold) and our linear-time one (\linearpartition). 
On the longest 23S rRNA family, the \RMSD is about 0.015. 
We notice that tmRNA is the family with biggest average \RMSD. %, and its accuracy is also the worst.
The random RNA sequences behave similarly to natural sequences in terms of \RMSD, 
i.e., \RMSD is  close to 0 ($\!<\!10^{-5}$) for short ones, then becomes bigger around length 500 and decreases after that, 
but for most cases their \RMSD's are slightly larger than  
the natural sequences. %in similar length range.
This indicates that the approximation quality is relatively better for natural sequences.
For RNAcentral-sampled sequences, \RMSD's are all small and around 0.01.

% With sequence length increasing, RMSD gradually decreases, 
% since the number of possible pairs grows in $O(n^2)$ but the number of highly probable pairs grows in $O(n)$. 
% To avoid RMSD is dominented by most base pairs with small probabilities,
% we consider a more strict circumstance. 
% Instead of divided by the number of all possible base pairs,
% We only consider the ones which have probability $p>0.01$.
% The RMSD resutls with such constrain are presented in Fig.~\ref{fig:rmsd_threshold}.
% It shows that for sequence shorter than 300$nt$, RMSD($p>0.01$) is still 0. 
% This also confirms that our \linearpartition gives exactly the same probability matrix as cubic algorithm.
% RMSD($p>0.01$) fluctuates from 0 to about 0.43 for sequences whose lengthes are in the range [300,1000]. Beyond 1,000$nt$, RMSD($p>0.01$) becomes stable between 0.2 and 0.4.

% Fig.~\ref{fig:rmsd}C and D show
We hypothesize that \linearpartition reduces the uncertainty of the output  distributions %(both base pairing probabilities distribution
%is shifted to higher probability 
because it filters out states with lower partition function.
We measure this using average positional structural entropy $H(p)$, %$~\cite{Garcia-Martin+Clote:2015}, 
% which is defined as:
which is the average of positional structural entropy $H_2(i)$ for each nucleotide $i$~\cite{Huynen+:1997,Garcia-Martin+Clote:2015}:
\vspace{-0.2cm}
\begin{equation}
  \vspace{-0.3cm}
\begin{split}
H(p) &= \frac{1}{n}\sum_{i=1}^{n}{H_2(i)} =\frac{1}{n}\sum_{i=1}^{n}({-\sum_{j=0}^{n}p_{i,j}\mathrm{log}_{2}p_{i,j}}) \\
		&=-\frac{1}{n}\sum_{i=1}^{n}{\sum_{j=0}^{n}p_{i,j}\mathrm{log}_{2}p_{i,j}} \notag
\end{split}
\end{equation}
where $p$ is the base pairing probability matrix,
% in which 
%$p_{i,j}$ is the probability of nucleotide $i$ paired with $j$ when $j\neq 0$, 
and $p_{i,0}$ is the probability of nucleotide $i$ being unpaired ($q_i$ in Eq.~\ref{eq:phi}).
%$n$ is the sequence length.
The lower entropy indicates that the distribution is dominated by fewer base pairing probabilities.
Fig.~\ref{fig:rmsd}C confirms \linearpartition distribution shifted to higher probabilities (lower average entropy) than \rnafold for most sequences.

Fig.~\ref{fig:rmsd}D uses \ecoli 23S rRNA to exemplify the difference in base pairing probabilities.
We sort all these probabilities from high to low and take the top 3,000.
The \linearpartition curve starts higher and finishes lower, confirming a lower entropy. %the distribution shifts to higher probabilities. % (rich get richer).

% E F
Figs.~\ref{fig:rmsd}E and F follow a previous analysis method~\cite{Zuber+:2017} to estimate the approximation quality with a different perspective. 
We divide the base pairing probabilities range [0,1] into 100 bins, i.e., the first bin is for base pairing probabilities [0,0.01), and the second is for [0.01, 0.02), so on so forth. 
In Fig.~\ref{fig:rmsd}E we visualize the averaged change of base pairing probabilities between \rnafold and \linearpartition for each bin.
We can see that larger probability changes are in the middle (bins with probability around 0.5),
and smaller changes on the two sides (with probability close to either 0 or 1).
%while both on the left (bins with probability near 0) and on the right (bins with probability near 1) the changes are smaller.
In Fig.~\ref{fig:rmsd}F we illustrate the counts in each bin based on \rnafold base pairing probabilities.
We can see that most base pairs have low probabilities (near 0) or very high probabilities (near 1).
Combine Figs.~\ref{fig:rmsd}E and F together, we can see that probabilities of most base pairs are near 0 or 1, where the differences between \rnafold and \linearpartition are relatively small. 
Fig.~\ref{fig:bin_counts} provides the comparison of counts in each bin between \rnafold and \linearpartitionv. The count of \linearpartitionv in bin [99,100) is slightly higher than \rnafold, 
while the counts in bins near 0 (being capped at 50,000) are much less than \rnafold.
This  also confirms that \linearpartition prunes base pairs with tiny probabilities.

\vspace{-0.4cm}
\subsection{Adjustable Beam Size}

Beam size in \linearpartition is a user-adjustable hyperparameter controlling beam prune, 
and it balances the approximation quality and runtime.
A smaller beam size shortens runtime, but sacrifices approximation quality.
With increasing beam size, \linearpartition gradually approaches the classical $O(n^3)$-time algorithm and 
the output is finally identical to the latter when the beam size is $\infty$ (no pruning).
Fig.~\ref{fig:beamsize}A shows the changes in approximation quality of the ensemble free energy change,
$\Delta G^\circ_\text{ensemble}(\vecx)$, with $b=20\rightarrow 800$.
Even with a small beam size ($b=20$) the difference is only about 5\%, which quickly shrinks to 0 as $b$ increases.
Fig.~\ref{fig:beamsize}B shows the changes in \RMSD with changing $b$. %=20\rightarrow 800$.
% We use averaged length and averaged RMSD for each family with a certain beam size.
%We observe that \RMSD decreases when beam size increases. 
With a small beam size $b=20$ the average \RMSD is lower than 0.035
over all ArchiveII sequences,
which shrinks to less than 0.005 at the default beam size ($b=100$),
and almost 0 with $b=500$.% the average \RMSD is
% and tmRNA remains relatively high averaged RMSD compared with other short families.
% This is consistant with Fig.~\ref{fig:rmsd}B.
%% With a larger beam size, $b=500$, 
%% % which means no beam prune for partition function calculation, 
%% the average \RMSD decreases to almost 0.
% It is also clear that shorter families' averaged RMSD decreases faster. 

Beam size also has impact on PPV and Sensitivity.
Fig.~\ref{fig:beamsize}C gives the overall PPV and Sensitivity changes with beam size.
We can see both PPV and Sensitivity improve from $b=50$ to $b=100$,
and then become stable beyond that.
%Therefore, we choose beam size 100 as the default beam size.
Figs.~\ref{fig:beamsize}D and E present this impact for two selected families.
% tmRNA is the worst family in the sense of accuracy for \linearpartition.
Fig.~\ref{fig:beamsize}D shows tmRNA's PPV and Sensitivity both increase when enlarging beam size.
Using beam size 200, \linearpartition achieves similar PPV and Sensitivity as \rnafold.
% This indicates that it is better to use a larger beam size, for example, $b=200$ when running 
% \linearpartition on tmRNA sequence.
However, increasing beam size is not benefical for all families.
% especially for long families.
Fig.~\ref{fig:beamsize}E gives the counterexample of 16S rRNA. 
We can see both PPV and Sensitivity decrease with $b$ from 50 to 100.
After that, Sensitivity drops with no PPV improvement.
% PPV increases while Sensitivity decreases when changing beam size from 100 to 120. 
% Continue increasing beam size, Sensitivity drops with no PPV improvement.

\smallskip
% discussion about k-best parsing
\linearfold uses $k$-best parsing~\cite{huang+:2005} to reduce runtime from $O(nb^2)$ to $O(nb{\mathrm {log}}b)$ without losing accuracy.
% Basically, $k$-best parsing is to find the top-$k$ states 
% The reason of the difference is that 
Basically, 
$k$-best parsing is to find the exact top-$k$ (here $k\!=\!b$) states out of $b^2$ candidates
% with no need of approximation
in $O(b{\mathrm {log}}b)$ runtime.
%by using a heap.
% instead of .
% while the naive algorithm needs to 
If we applied $k$-best parsing here,
\linearpartition would sum the partition function of only these top-$b$ states
instead of the partition function of $b^2$ states.
This change would introduce a larger approximation error,
especially when the differences of partition function between the top-$b$ states 
and the following states near the pruning boundary are small.
Therefore, in \linearpartition we do not use $k$-best parsing as in \linearfold,
and the runtime is $O(nb^2)$ instead of $O(n b\log b)$.
% For \linearfold, since it only needs to predict one MFE structure,
% the rest candidates are much less promising to be part of the MFE structure.
% However, \linearpartition needs to consider the ensumble at equilibrium, 
% so the partition function of the candidate states may also make unneglactable contributions to total partition function.

% \begin{figure}[b]
% \center
% \includegraphics[width=0.48\textwidth]{figs/RMSD_beamsize}
% \caption{
% Impact of beam size. Add beamsize-accuracy figures???
% 	\label{fig:beamsize}.
% }
% \end{figure}

%\subsection{Example}

Finally, we note that the default beam size $b\!=\!100$ follows \linearfold and %is a simple round number;
we do not tune it.

\label{sec:discussion}
% !TEX root = \main.tex
\vspace{-.4cm}
\section{Discussion}
\vspace{-0.1cm}

\subsection{Summary}
% In this paper, we present LinearPartition, which inherits LinearFold main idea and applies it to partition function and base pairing probability calculation, and leads to a small accuracy improvement in both linear runtime and linear memory space. 
% LinearPartition reduces classical cubic runtime by pruning states with lower energy. 
% Although it filters some substructure, but only the ones with worse free energy are given up, and results in a similar partition function as exact search. 
% LinearPartition is 10× faster than Vienna RNAfold for the longest sequence (about 3000 nucleotides) in the dataset. 
% Not only being fast, LinearPartition is as accurate as Vienna RNAfold when comparing MEA and ProbKnot output structure. 
% Surprisingly, even though LinearPar- tition uses an inexact search, it achieves better accuracy on longer families (16S and 23S rRNA).

The classical McCaskill (1990) algorithm for partition function and base pairing probabilities calculations 
are widely used in many studies of RNA sequences, 
but its application has been impossible for long sequences 
(such as full length mRNA)
due to its cubic runtime.
%because of the limitation imposed by their cubic scaling.
To address this issue, we present LinearPartition, a linear-time algorithm that dramatically reduces the runtime without sacrificing output quality. % of partition function and base pairing probabilities calculations.
We confirm that:
\begin{enumerate}
  \vspace{-0.2cm}
  \setlength{\itemsep}{0pt}
 \item 
 \linearpartition takes only linear runtime and memory usage, and is orders of magnitude faster on longer sequences (Fig.~\ref{fig:runtime}); %, for example, 
% about 11$\times$ faster than \viennarnafold on {\it H.~pylori} 23S rRNA (2,968~{\it nt}). 
%For example, it enjoys a 2,771$\times$ speedup (2.5 days vs.~1.3~min.)
%over \contrafold on the longest sequence (32,753~{\it nt}) that \contrafold can run in the RNAcentral dataset (Fig.~\ref{fig:runtime}).
 \item
 The base pairing probabilities produced by \linearpartition are better correlated with the ground truth structures on average (Figs.~\ref{fig:ensemble}--\ref{fig:example}); 
% For instance, 
% \linearpartition's ensemble distribution leads to 211.4 (7.3\%) more correctly predicted nucleotides %decrease in ensemble defect
% on \ecoli 23S rRNA compared with \viennarnafold, 
% and 47.1 (9.3\%) more on {\it C.~ellipsoidea} Group I Intron 
% (Figs.~\ref{fig:ensemble} and~\ref{fig:example}).
 \item
When used with downstream structure prediction methods such as MEA and ThreshKnot,
  LinearPartition's base pair probabilities have similar overall accuracy (or even a small improvement on MEA structures) compared with \rnafold,
  as well as better accuracies on longer families and long-distance base pairs (Fig.~\ref{mea});
%   and slightly better accuracies on longer families. 
% \linearpartition is also substantially better on long-distance base pairs (500+~\nts apart) in both MEA and ThreshKnot predictions (Fig.~\ref{mea}).
\item
 \linearpartition has a reasonable approximation quality (Figs.~\ref{fig:partition}--\ref{fig:rmsd}) in terms of \RMSD. 
 
%and can serve multiple downstream tasks.
% Although filtering out some structures, the ensemble free energy change of \linearpartition is either the same or only slightly smaller than \viennarnafold
% e.g., the largest fraction of total free energy change is 2.5\% in the ArchiveII dataset 
% (Fig.~\ref{fig:partition}).  
% Additionally, RMSD of base pairing probabilities between \linearpartition and \viennarnafold is small 
% e.g., the largest RMSD in the ArchiveII dataset is 0.065 
% (Fig.~\ref{fig:rmsd}).
% \item
 % (5) With increasing beam size, \linearpartition gradually approaches the classical McCaskill algorithm;
 %  both the difference in ensemble free energy change and RMSD quickly shrink to 0 (Fig.~\ref{fig:beamsize}).
%  and eventually produces identical outputs.
%%   averaged RMSD decreases. The change is more pronunced from beam size 20 to 100. 
%% Above 100, averaged RMSD is smaller than 0.05, and overall PPV and Sensitivity are stable. 
%% For tmRNA, PPV and Sensitivity increase with beam size and are very close to \viennarnafold at beam size 200. 
%% % Other families, especially for long families such as 16S rRNA, accuracy may drop when increase beam size.
%% But for 16S rRNA, accuracy drops with an increase beam size (Fig.~\ref{beamsize}).
 \end{enumerate}

  % \subsection{Analysis}
%  \smallskip
\vspace{-0.2cm}
There are two possible reasons why our approximation 
results in better base pairing probabilities: 
%in the sense of correlating with ground truth structure:
\begin{enumerate}
  \vspace{-0.2cm}
  \setlength{\itemsep}{0pt}%
 
\item
  This is consistent with the findings in \linearfold, where
  approximate folding via beam search yields more accurate structures.

%% It has been studied that 
%% a structure containing 86.1\% of the ground truth base pairs can be found
%% in a free energy gap of 4.8\% from the optimal structure~\cite{zuker+:1991,mathews+:1999}.
%% With well-designed heuristic,
%% \linearpartition captures the bulk of the free energy of the ensemble, 
%% especially sums up the suboptimal structures within a small gap from the MFE structure.

\item
%% Classical partition function sums up all possible secondary structures,
%% but in reality not all these structures exists at equilibrium.
\linearpartition's pruning of low-probability (sub)structures
has a ``regularization'' effect.
It eliminates some noise in the current energy model 
which is highly inaccurate, especially for long-distance interactions.
\end{enumerate}

\vspace{-0.2cm}
The success of \linearpartition is arguably more striking than \linearfold,
since the former needs to sum up exponentially many structures
that capture the bulk part of the ensemble free energy,
while the latter only needs to find one single optimal structure.

\vspace{-0.2cm}
\subsection{Extensions}

Our work has potential extensions.
\begin{enumerate}
  \vspace{-0.2cm}
  \setlength{\itemsep}{0pt}%
\item 
% Many ncRNAs function by interacting with other RNA sequences by base pairing.
  Existing methods and tools for % these problems
  bimolecular and multistrand 
  base pairing probabilities as well as accessibility computation % could be accelerated and improved.
%calculating two-strand (bimolecular) or multi-strand folding partition functions and
%base pairing probability matrices
\cite{chitsaz+:2009, bernhart+:2006b, Dirks+:2007, DiChiacchio+:2016} are rather slow, %suffer from slowness, 
which limits their applications on long sequences. 
\linearpartition will likely provide a much faster solution for these problems. 
% and will have immediate impact on the ability to predict bimolecular or multi-strand structures 
% and also providing additional structure information (both intra- and inter-molecular pairs) to users.

\item 
We will linearize the partition function-based heuristic methods for pseudoknot prediction such as 
% ProbKnot, 
IPknot and Dotknot. 
These heuristic methods use rather simple criteria to choose pairs from the base pairing probability matrix,
and their runtime bottleneck is $O(n^3)$-time calculation of the base pairing probabilities.
% For example, 
% IPknot %first computes base pairing probabilities 
% selects base pairs by solving an Integer Linear Program (ILP)
% to maximize the total probabilities of chosen pairs
% with well-designed constrains.
% Actually the first step of computing base pairing probabilities
% takes the vast majority of total computation time; e.g., on 23S rRNAs it accounts for 99.3\% of total time. % takes much more time.
% This might appear as $O(n^2)$ in the worst case, 
% but since the linear-time beam search used in \linearpartition only returns $O(nb)$ pairs where $b$ is the constant beam size, 
% this second step is still $O(n)$, 
% giving an overall linear-time method, LinearProbKnot. 
With \linearpartition we can overcome the costly bottleneck %$O(n^3)$-time computation
and get an overall much faster tool. %, such as IPknot or FastDotKnot.
% We can similary get FastDotKnot, etc.
% With the promising results of \linearpartition, 
% we believe \linearpartition-powered IPknot (or DotKnot) might be as accurate as, 
% if not more accurate than, 
% their original $O(n^3)$ versions.

\item 
We can also speed up stochastic sampling of RNA secondary structures from Boltzmann distribution.
The standard stochastic sampling algorithm runs in worst-case $O(n^2)$ time~\cite{Ding+Lawrence:2003},
but relies on the classical $O(n^3)$ partition function calculation.
% The slowness of 
% which becomes the bottleneck for sampling on long sequences.
With \linearpartition, 
we can apply stochastic sampling to full length sequences such as mRNAs, 
and compute their accessbility based on sampled structures.

\end{enumerate}

%\vspace{-.2cm}
% \label{sec:methods}
% \input{methods}

% \section*{Code availability}\label{code}

% Our \linearpartition source code can be downloaded  from\\ 
% {\tt\url{https://github.com/LinearFold/LinearPartition}}.

% \section*{Data availability}

%\small

% The data that support the findings of this study are available from the corresponding author upon request.

%% \label{sec:data}
%% \input{data}

% \showmatmethods{
%   % \ssmall
%   \section*{Methods}
%   \label{sec:method}  
% Detailed description of our algorithms, datasets, and evaluation metrics %statements of data availability and any associated accession codes and references,
% are available in the online version of the paper. 
% % \input{method}
% } % Display the Materials and Methods section

% \vspace{-0.5cm}
% \acknow{\small 
% This work was partially supported by 
% NSF grant IIS-1817231 (L.H.) and NIH grant R01 GM076485 (D.H.M.).
% We thank Rhiju Das for the early adoption of
% our software in the EteRNA game.
% \vspace{-0.5cm}
% }

% \showacknow{} % Display the acknowledgments section
% \smallskip

% \pnasbreak splits and balances the columns before the references.
% Uncomment \pnasbreak to view the references in the PNAS-style
% If you see unexpected formatting errors, try commenting out \pnasbreak
% as it can run into problems with floats and footnotes on the final page.
%\pnasbreak

% Bibliography
\vspace{-0.6cm}
\section*{References}
\balance
\bibliography{main}

\appendix

\newpage

% \input{method}

% !TEX root = main.tex
\onecolumn
\newpage

%\begin{titlepage}
  \begin{centering}
    \vspace*{1cm}
    
    \textbf{\Large Supporting Information}\\
    \vspace{0.5cm}
    \textbf{\Large \linearpartition: Linear-Time Approximation of RNA Folding Partition Function and Base Pairing Probabilities}\\
    \vspace{0.5cm}
    \textbf{\large
    He Zhang, Liang Zhang, David H.~Mathews and Liang Huang}
    \vspace{1.5cm}
    
  \end{centering}
%\end{titlepage}

\setcounter{figure}{0}
\renewcommand{\thefigure}{SI\,\arabic{figure}} % spacing
\setcounter{table}{0}
\renewcommand{\thetable}{SI\,\arabic{table}}
\setcounter{page}{1}

\setcounter{section}{0}
\renewcommand\thesection{\Alph{section}}
\setcounter{subsection}{0}
\renewcommand\thesubsection{\Alph{section}.\arabic{subsection}}

\section{Details of the Efficient Implementation}
\label{sec:si:algdetails}

\titleformat{\subsection}%[runin]
{\normalfont\bfseries}% formatting commands to apply to the whole heading
        {\thesubsection}% the label and number
        {0.5em}% space between label/number and subsection title
        {#1}% formatting commands applied just to subsection title
        []
        
\subsection{Data Structures}

%\begin{table}
In the main text, for simplicity of presentation, $Q$ is described as a hash from span $[i,j]$
to $\Qf{i}{j}$, but in our actual implementation,
to make sure the overall runtime is $O(n b^2)$,  we implement $Q$ as
an array of $n$ hashes, where each $Q[j]$ is a hash
mapping $i$ to $Q[j][i]$ which is conveniently notated as \Qf{i}{j} in the main text. It is important to note that the first dimension $j$ is the right boundary
and the second dimension $i$ is the left boundary of the span $[i,j]$. See the following table for a summary of notations and the corresponding actual implementations.
Here we use Python notation for simplicity, but in actual system we implement with C++.
\begin{center}
%\resizebox{0.5\textwidth}{!}{
\begin{tabular}{l|l}
notations in this paper & Python implementation\\
\hline
%$Q\gets$ hash() & \verb|Q = defaultdict(lambda : defaultdict(float))|\\
$Q\gets$ hash() & {\tt Q = [defaultdict(float) \textbf{for} \_ \textbf{in} range(n)]}\\
\Qf{i}{j} & \verb|Q[j][i]| \\
$[i,j]$ in $Q$ & {\tt i \textbf{in} Q[j]}\\
{\bf for each} $i$ such that $[i,j]$ in $Q$ & {\tt \textbf{for} i \textbf{in} Q[j]}\\
{\bf delete} $[i,j]$ from $Q$ & {\tt \textbf{del} Q[j][i]}
\end{tabular}
%}
\end{center}
%\end{table}

\subsection{Complexity Analysis}

In the partition function calculation (inside phase) in Fig.~\ref{fig:algorithm},
the number of states is $O(nb)$ because each $Q[j]$ contains at most $b$ states (\Qf{i}{j}'s) after pruning. Therefore the space complexity is $O(nb)$.
For time complexity, there are three nested loops, the first one ($j$) with $n$ iterations,
the second ($i$) and the third ($k$) loops both have $O(b)$ iterations thanks to pruning, so the overall runtime is $O(nb^2)$.

\subsection{Outside Partition Function and Base Pairing Probability Calculation}
\label{sec:outside}

After we compute the partition functions \Qf{i}{j} on each span $[i,j]$ (known as the ``inside partition function''),
we also need to compute the complementary function \Qhatf{i}{j} for each span known as
the ``outside partition function'' in order to derive the base-pairing probabilities. 
Unlike the inside phase, this outside partition function is calculated from top down,
with $\Qhatf{1}{n} = 1$ as the base case.
\begin{equation*}
\begin{split}
\Qhatf{i}{j} & = \Qhatf{i}{j+1} \cdot e^{-\frac{\delta(\vecx, j+1)}{RT}} \\
                  & + \sum_{k < i} \Qhatf{k}{j+1} \cdot \Qf{k}{i-2} \cdot e^{-\frac{\xi(\vecx,i-1,j+1)}{RT}} \\
                  & + \sum_{k > j+1} \Qhatf{i}{k} \cdot \Qf{j+2}{k-1} \cdot e^{-\frac{\xi(\vecx,j+1,k)}{RT}}
\end{split}
\end{equation*}
Note that the second line is only possible when $x_{i-1} x_{j+1}$ can form a base pair
(otherwise $e^{-\frac{\xi(\vecx,i-1,j+1)}{RT}} = 0$)
and the third line has a constraint that $x_{j+1}x_k$ can form a base pair 
(otherwise $e^{-\frac{\xi(\vecx,j+1,k)}{RT}} = 0$).

For each $(i,j)$  where $x_i x_j$ can form a base pair, we compute its pairing probability:
\[
p_{i,j}  = \sum_{k \leq i} \Qhatf{k}{j} \cdot \Qf{k}{i-1} \cdot e^{-\frac{\xi(\vecx,i,j)}{RT}} \cdot \Qf{i+1}{j-1}
\]

The whole ``outside'' computation takes $O(n^3)$ without pruning,
but also $O(nb^2)$ with beam pruning.
See Fig.~\ref{fig:outside} for the pseudocode to compute the outside partition function and base pairing probabilities.

%\subsection{Supporting Pseudocode}

\section{Details of datasets, baselines and methods}

% \newpage
\subsection{Datasets}
\label{sec:datasets}

We use sequences from two datasets, ArchiveII and RNAcentral.
The archiveII dataset 
(available in \url{http://rna.urmc.rochester.edu/pub/archiveII.tar.gz}) is % \cite{sloma+mathews:2016}, 
a diverse set with 3,857 RNA sequences and their secondary structures.
It is first curated in the 1990s to contain sequences with structures that were well-determined by comparative sequence analysis~\cite{mathews+:1999}% [72] 
and updated later with additional structures~\cite{sloma+mathews:2016}. % [96].  
% and is available in \url{http://rna.urmc.rochester.edu/pub/archiveII.tar.gz}.
We remove 957 sequences that appear both in the ArchiveII and the S-Processed datasets~\cite{Andronescu+:2007}, 
because CONTRAfold uses S-Processed for training. 
We also remove all 11 Group II Intron sequences 
because there are so few instances of these that are available electronically.
Additionally, we removed 30 sequences in the tmRNA family because the annotated structure for each of these sequences contains fewer than 4 pseudoknots, 
which suggests the structures are incomplete. 
These preprocessing steps lead to a subset of ArchiveII with 2,859 reliable secondary structure 
% RNA sequence and structure pairs 
examples 
distributed in 9 families. 
See~\ref{tab:archiveII} for the statistics of the sequences we use in the ArchiveII dataset.
% But since CONTRAfold (v2.02) machine-learned model is trained on the S-Processed dataset \cite{},
% we removed those overlap sequences. 
% As in \linearfold paper, we also remove those sequences CONTRAfold (v2.02) used for training.
% The remaining dataset contains 2889 RNA sequences from 9 families, 
% with average length 222 $nt$ and max length 2968 $nt$.
Moreover, we randomly sampled 22 longer RNA sequences (without known structures) from RNAcentral~\cite{rnacentral:2017} 
% (The RNAcentral Consortium, 2017) 
(\url{https://rnacentral.org/}),
with sequence lengths ranging from 3,048~{\it nt} to 244,296~{\it nt}.
For the sampling, we evenly split the range from $3,000$ to $244,296$ (the longest) into 24 bins by log-scale, and for each bin we
randomly select a sequence (there are bins with
no sequences).
% (Homo Sapiens Transcript NONHSAT168677.1, from the NONCODE database (Zhao et al., 2016)).
% We run all experiments 
% % (compiled by GCC 4.9.0) 
% on a Linux machine, 
% with 2.90GHz Intel Core i9-7920X CPU and 64G memory.

To show the approximation quality on random RNA sequences, 
we generated 30 sequences with uniform distribution over \{A, C, G, U\}.
% The lengths of these sequences are 100, 200, ..., 3000.
The lengths of these sequences are spaced in 100 nucleotide intervals from 100 to 3,000.
% with lengths range from 100~{\it nt} to 3,000~{\it nt}. 

\begin{table}[!h] % hzhang: redo with new data
  \centering
  \large
  \setlength{\tabcolsep}{12pt}
  \begin{tabular}{r|rr|rrr}
    & \multicolumn{2}{c|}{\# of seqs} & \multicolumn{3}{c}{length} \\
    Family & total & used & avg & max & min \\
    \hline
    tRNA  & 557 & 74  & 77.3  & 88 &58  \\
    5S rRNA & 1,283 & 1,125 & 118.8 & 135 &102  \\
    SRP RNA & 928 & 886 & 186.1 & 533 & 28\\
    RNase P RNA & 454 & 182 & 344.1 & 486 & 120 \\
    tmRNA & 462 & 432 & 369.1 & 433 & 307 \\
    Group I Intron  & 98  & 96  & 424.9 & 736 &210  \\
    Group II Intron  & 11  & 0  & - & -&-  \\
    telomerase RNA  & 37  & 37  & 444.6 & 559 &382\\
    16S rRNA  & 22  & 22  & 1,547.9  & 1995 & 950\\
    23S rRNA  & 5 & 5 & 2,927.4  & 2968& 2904\\
    \hline
    {\em Overall} & 3,846 & 2,859 & 221.1 &2968 &28 \\
  \end{tabular}
  \\[0.3cm]
  % \smallskip
  \caption{Statistics of the sequences in the ArchiveII dataset used in this work.
    \label{tab:archiveII}}
\end{table}

\subsection{Baseline Software}

We use two baseline software packages: 
(1) \viennarnafold %~\cite{lorenz+:2011}
(Version 2.4.11) 
from 
\url{https://www.tbi.univie.ac.at/RNA/download/sourcecode/2_ 4_x/ViennaRNA-2.4.11.tar.gz} 
and 
(2) \contrafold %~\cite{do+:2006}
(Version 2.0.2) 
from
\url{http://contra.stanford.edu/}.
\viennarnafold is a widely-used RNA structure prediction package,
while \contrafold is a successful machine learning-based RNA structure prediction system.
Both provide partition function and base pairing probability calculations based on 
the classical cubic runtime algorithm.
Our comparisons mainly focus on the systems with the same model, 
i.e., \linearpartitionv vs. \viennarnafold and \linearpartitionc vs. \contrafold.
In this way the differences are based on algorithms themselves rather than models.
% bugs in contrafold
We found a bug in \contrafold by comparing our results to CONTRAfold, 
which led to overcounting multiloops in the partition function calculation.
We corrected the bug, and all experiments are based on this bug-fixed version of \contrafold.

\subsection{Evaluation Metrics and Significance Test}

Due to the uncertainty of base-pair matches existing in comparative analysis
and the fact that there is fluctuation in base pairing at equilibrium,
we consider a base pair to be correctly predicted if it is also displaced by one
nucleotide on a strand~\cite{mathews+:1999}.
Generally, if a pair $(i,j)$ is in the predicted structure, we consider it a
correct prediction if one of $(i,j)$, $(i-1,j)$, $(i+1,j)$, $(i,j-1)$, $(i,j+1)$ is in the
ground truth structure.
% We also report the accuracy using exact base pair matching instead of this
% method, in Figure~\ref{tab:accuracy_nos}. 
%To evaluate the accuracy, 
% Both sensitivity and PPV are reported.
%Positive Predictive Value
%(PPV), 

We use Positive Predictive Value (PPV)
and sensitivity 
as accuracy measurements. 
Formally, denote $\vecy$ as the predicted structure and $\vecy^{*}$ as the ground
truth, we have:
% $\ppv = \frac{\text{true positives}}{\text{true positives} +
%   \text{false positives}} = \frac{|\vecy \cap \vecy^*|}{|\vecy|} $
% $ \sens = \frac{\text{true positives}}{\text{true positives} +
%   \text{false negatives}} =
% \frac{|\vecy \cap \vecy^*|}{|\vecy^*|}$

$$\ppv = \frac{\#_{\text {TP}}}{\#_{\text {TP}} + \#_{\text {FP}}}  = 
\frac{|\pairs(\vecy) \cap \pairs(\vecy^*)|}{|\pairs(\vecy)|} $$

$$ \sens = \frac{\#_{\text {TP}}}{\#_{\text {TP}} + \#_{\text {FN}}}  =
\frac{|\pairs(\vecy) \cap \pairs(\vecy^*)|}{|\pairs(\vecy^*)|}$$
where $\#_{\text {TP}}$ is the number of true positives (correctly predicted pairs),
$\#_{\text {FP}}$ is the number of false positives (wrong predicted pairs)
and $\#_{\text {FN}}$ is the number of false negatives (missing ground truth pairs).

We test statistical significance using a paired, two-sided permutation test~\cite{Aghaeepour+Hoos:2013}.
We follow the common practice, choosing $10,000$ as the repetition number
and $\alpha=0.05$ as the significance threshold.
% following previous work~\cite{Aghaeepour+Hoos:2013}.

\subsection{Curve Fitting}
We determine the best exponent $a$ for the scaling curve $O(n^a)$ for each data series in Figures~\ref{fig:linearpairs} and \ref{fig:runtime}.
Specifically, we use $f(x) = a x + b$ to fit the log-log plot of those series in Gnuplot;
e.g., fitting $\log t_n = a \log n + b$, where $t_n$ is the running time on a sequence of length $n$,
so that $t_n = e^b n^a$.
Gnuplot uses the nonlinear least-squares Marquardt-Levenberg algorithm.

\newpage
\section{Supporting Figures}

\begin{figure}[h]%[b]
\center
\small
% \hspace{-0.23cm}\includegraphics[scale=.16]{figs/index} \\[-3.cm]
%\hspace{-0.23cm}\includegraphics[scale=1.2]{figs/beam_prune_alg} \\[0.5cm]
  \algrenewcommand\algorithmicindent{1.5em}%
\begin{minipage}{0.85\textwidth}
\begin{algorithmic}[1]
  \newcommand{\INDSTATE}[1][1]{\State\hspace{#1\algorithmicindent}}
  \setstretch{1.2} % lhuang: usepackage setspace
\Function{beamprune}{$Q, j, b$}
    \State $\candidates \gets$ hash() \Comment{hash table: from candidates $i$ to score}
    \ForEach{$i$ such that $[i,j]$ in $Q$}
        \State $\candidates[i] \gets \Qf{1}{i-1} \cdot \Qf{i}{j}$ \Comment{like \linearfold, use $\Qf{1}{i-1} $ as prefix score}
    \EndFor
    \State $\candidates \gets \textsc {SelectTopB}(candidates, b)$ \Comment{select top-$b$ states by score}
    \ForEach{$i$ such that $[i,j]$ in $Q$}
        \If{key $i$ not in $candidates$}
            \State {\bf delete} $[i,j]$ from $Q$ \Comment{prune low-scoring states}
        \EndIf
    \EndFor
\EndFunction
\end{algorithmic}
% \end{algorithm}
\end{minipage}
\caption{
The {\sc BeamPrune} function from the Pseudocode of our main algorithm (Fig.~\ref{fig:algorithm}).
\label{fig:beam_prune_alg}}
% \vspace{-0.3cm}
% \end{figure*}
\end{figure}x

\begin{figure}[h]%[b]
\center
\small
\algrenewcommand\algorithmicindent{1.5em}%
\algdef{S}[FOR]{ForEach}[1]{\algorithmicforeach\ #1\ \algorithmicdo}
% \hspace{-0.23cm}\includegraphics[scale=.16]{figs/index} \\[-3.cm]
%\hspace{-0.23cm}\includegraphics[scale=.83]{figs/algorithm} \\[0.2cm]
\begin{minipage}{.85\textwidth}
\begin{algorithmic}[1]
  \newcommand{\INDSTATE}[1][1]{\State\hspace{#1\algorithmicindent}}
  \setstretch{1.2} % lhuang: usepackage setspace
\Function{BasePairingProbs}{$\vecx, Q$}  \Comment{outside calculation}
% \bindent
    \State $n \gets$ length of $\mathbf x$
    \State $\Qhat \gets$ hash() \Comment{hash table: from span $[i,j]$ to $\Qhatf{i}{j}$: outside partition function}
    \State $p \gets$ hash() \Comment{hash table: from span $[i,j]$ to $p_{i,j}$: base-pairing probs}
    % \State $Q[j,\,j-1] \gets 1$ for all $j$ in $1...n$ \Comment{base cases} \label{line:base}
    \State $\Qhatf{1}{n} \gets 1$ \Comment{base case}
    \For{$j=n$ {\bf downto } $1$}
        \ForEach {$i$ such that $[i,\,j-1]$ in $Q$}\smallskip
            \State $\Qhatf{i}{j-1} \pluseq \Qhatf{i}{j} \cdot e^{-\frac{\delta(\vecx,j)}{RT}} $ \Comment{\nskip} \label{line:skip}
            \If{$x_{i-1}x_j$ in \{AU, UA, CG, GC, GU, UG\}}  \label{line:pair}
                % \State $Q_{i,\,j+1} \gets  C(i,\,j) \cdot e^{-\frac{\xi(\vecx,i,\,j)}{RT}} $
                \ForEach{$k$ such that $[k,\,i-2]$ in $Q$}\smallskip
                    \State $\Qhatf{k}{i-2} \pluseq {\Qhatf{k}{j} \cdot \Qf{i}{j-1} \cdot e^{-\frac{\xi(\vecx,i-1,j)}{RT}}}$ \Comment{\pop: left} %\label{line:pop}
                    % \State $C(0,j+1) \pluseq {C(0,k) \cdot C(k,j+1) \cdot e^{-\frac{\xi(\vecx,i,\,j)}{RT}}} $
                    \State $\Qhatf{i}{j-1} \pluseq {\Qhatf{k}{j} \cdot \Qf{k}{i-2} \cdot e^{-\frac{\xi(\vecx,i-1,j)}{RT}}}$ \Comment{\pop: right} %\label{line:pop}
                    \State $p_{i-1,\,j} \pluseq \displaystyle\frac{\Qhatf{k}{j} \cdot \Qf{k}{i-2}  \cdot  e^{-\frac{\xi(\vecx,i-1,j)}{RT}} \cdot \Qf{i}{j-1}}{\Qf{1}{n}}$ \Comment{accumulate base pairing probs}
                \EndFor
                % \State $C(0,j+1) \pluseq C(0,i) \cdot Q_{i,j+1}$ \Comment{COMBINE}
            \EndIf
        \EndFor
        %% \ForEach {$i$ such that $[i,\,j]$ in $P$}\smallskip
        %%          \State $p_{i,j} = \frac{\hatp[i,\,j] \cdot Q[i+1,\, j-1]}{Q[1,n]}$
        %% \EndFor
        % \State $\textsc {BeamPrune}(Q,j+1, b)$ \Comment{see Fig.~\ref{fig:beam_prune_alg}}
    \EndFor
    \State \Return $p$ \Comment{return the (sparse) base-pairing probability matrix}
% \eindent
\EndFunction
%% \Function{BasePairingProbabilities}{$\vecx, Q, \hatp$} % \Comment{$Q$ is the beam size}
%% % \bindent
%%     \State $p \gets$ hash() \Comment{hash table: from span $[i,j]$ to base pairing probability $p_{i,j}$}
%%     \For{$j=2 ... |\vecx|$}
%%       \ForEach {$i$ such that $[i,j]$ in $\hatp$}
%%           \State $p_{i,j} = \displaystyle\frac{\hatp[i,\,j] \cdot Q[i+ 1,\, j-1] \cdot  e^{-\frac{\xi(\vecx,i,j)}{RT}}}{Q[1,n]}$
%%       \EndFor
%%     \EndFor
%%     \State \Return the (sparse) base-pairing probability matrix $p$
%% \EndFunction
\end{algorithmic}
\end{minipage}
\caption{
  Outside partition function and base pairing probabilities calculation for a simplified version of the \linearpartition.
  $Q$ is the (inside) partition function calculated by the pseudocode in Fig.~\ref{fig:algorithm}, and $\Qhat$ is the outside partition function. 
% as well as a beam prune algorithm. 
% Here we model hash tables following Python dictionaries, where $(i, j) \in C$ checks whether the key $(i, j)$ is in the hash $C$; 
% this is needed to ensure linear runtime. 
% Quick select algotithm is used in beam prune, 
% and we skip the details for quick select here since it is well known.
% Real \linearpartition system is much more involved, but the pseudocode illustrates the left-to-right partition function calculation idea using a Nussinov-like fashion.
The actual algorithm using the Turner model is in our \href{https://github.com/LinearFold/LinearPartition}{GitHub codebase}.
%See Fig.~\ref{fig:beam_prune_alg} for {\sc BEAMPRUNE} function.
\label{fig:outside}}
\vspace{-0.3cm}
% \end{figure*}
\end{figure}

\iftrue
\begin{figure}[h]
  \centering
  \captionsetup{singlelinecheck=off}
%\hspace{-0.5cm}
\begin{tabular}{ll}
\hspace{-.5cm}{\panel A} & {\panel B}\\[-1cm]
    \includegraphics[width=.3\textwidth]{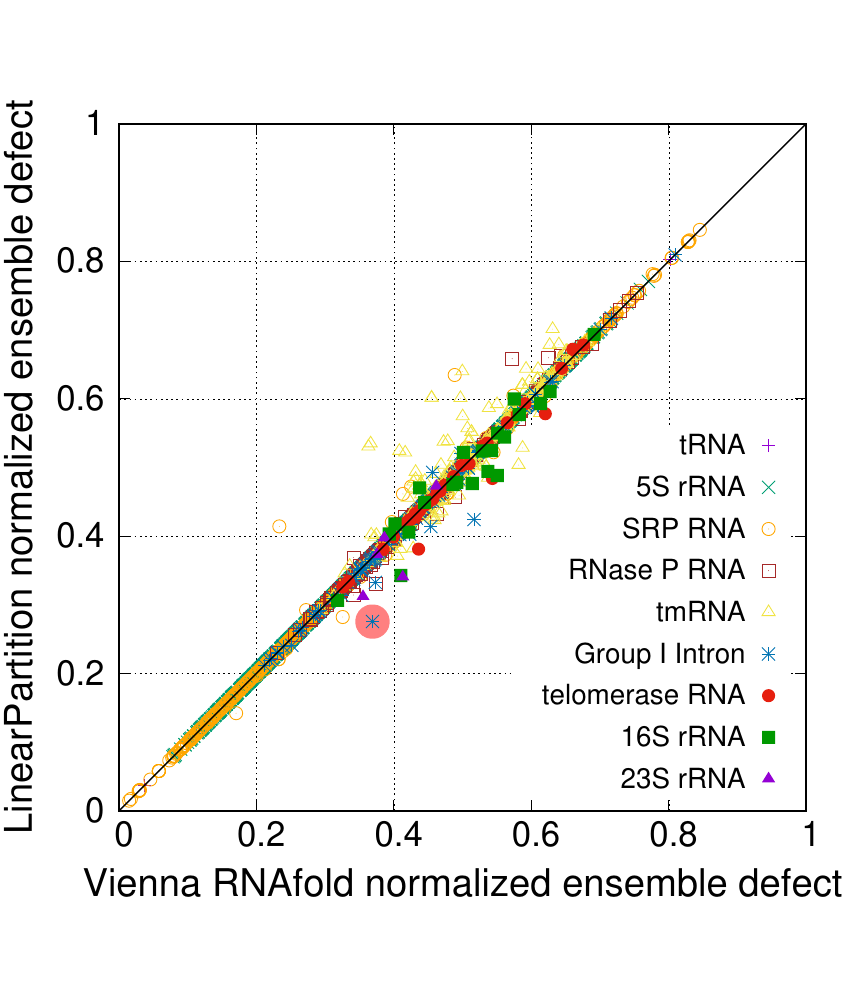}
    &
    \hspace{-0.1cm}
    \raisebox{.5cm}{\includegraphics[width=.5\textwidth]{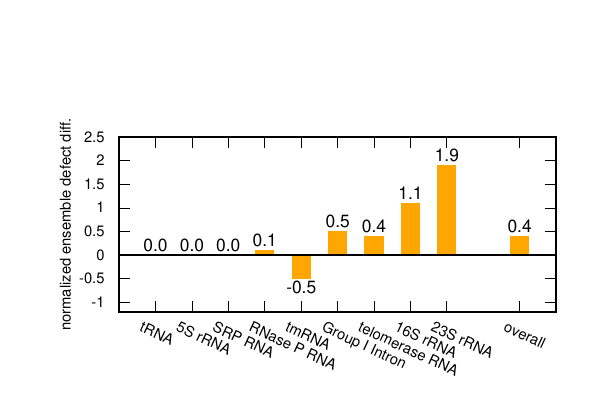}}
  \end{tabular} \\[-.3cm]
  \caption[.]{
  The 
  comparison of normalized ensemble defects (normalized by sequence length)
  % between \viennarnafold and \linearpartitionv 
  on the ArchiveII dataset.
  % Ensemble defects are normalized by sequence length. 
  {\bf A}: 
  Normalized ensemble defect between \viennarnafold and \linearpartitionv for each sequence; the trend is similar as Fig.~\ref{fig:ensemble}A, but the deviations for tmRNAs are more apparent; 
  the point with red shaded are the example in Fig.~\ref{fig:example}.
  {\bf B}: Normalized ensemble defect difference for each family; for longer families, 
  e.g., Group I Intron, telomerase RNA, 16S and 23S rRNA, 
  \linearpartition has lower normalized ensemble defect differences;
  note that \linearpartition's normalized ensemble defects are significantly better than \viennarnafold on Group I Intron ($p < 0.01$), 
  but significantly worse on tmRNA ($p < 0.01$).
  \label{fig:ensemble_defect}}
%\vspace{1cm}
\end{figure}
\fi

\iffalse
\begin{figure}[h]
  \centering
  \captionsetup{singlelinecheck=off}
\begin{tabular}{cccc}
\hspace{-4.4cm} \panel{A} & \hspace{-4.6cm}\panel{B} & \hspace{-4.6cm}\panel{C} & \hspace{-4.6cm}\panel{D}\\[-0.2cm]
\hspace{-0.2cm}\includegraphics[width=0.25\textwidth]{figs/grp1_gold} &
\hspace{-0.35cm}\includegraphics[width=0.25\textwidth]{figs/grp1_vienna_plfold_example.pdf} &
\hspace{-0.35cm}\includegraphics[width=0.25\textwidth]{figs/grp1_vienna_example} &
\hspace{-0.35cm}\includegraphics[width=0.25\textwidth]{figs/grp1_lpv_example.pdf}
\end{tabular}
  \caption[.]{Circular plots of {\it C.~ellipsoidea} Group I Intron.  
  Blue denotes pairs in the known structure and Red denotes predicted pairs not in the known structure.  
  The darkness of the line indicates pairing probability, 
  which the darkest lines close to a portability of 1. 
  {\bf A}: 
  [Please explain what is shown in each panel.] [I also think a key with blue and red in the figure would be helpful here as well.]
  \label{fig:circular_grp1}}
%\vspace{1cm}
\end{figure}
\fi

%\newpage
% mea
\iftrue
\begin{figure}[h]
  \centering
%\hspace{-0.5cm}
\begin{tabular}{ll}
{\large\sf A} & {\large\sf B}\\[-1cm]
    \includegraphics[width=.45\textwidth]{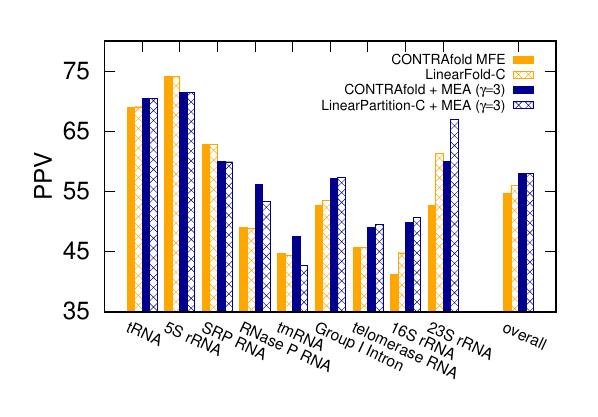}
    &
    \hspace{-0.1cm}
    \includegraphics[width=.45\textwidth]{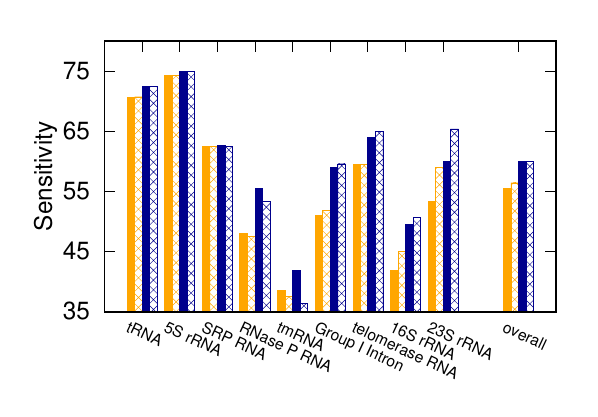}
  \end{tabular} \\[-0.5cm]
  \caption{Accuracy comparison of MEA structures ($\gamma=3$) between \contrafold and \linearpartitionc on the ArchiveII dataset. 
  $\gamma$ is the hyperparameter balances PPV and Sensitivity. Note that \linearpartitionc + MEA is significantly worse than \contrafold + MEA on two families in both PPV and Sensitivity, tmRNA and RNase P RNA ($p < 0.01$).
  \label{fig:mea_lpc}}
%\vspace{1cm}
\end{figure}
\fi

% threshknot
\iftrue
\begin{figure}[h]
  \centering
%\hspace{-0.5cm}
\begin{tabular}{ll}
{\large\sf A} & {\large\sf B}\\[-1cm]
    \includegraphics[width=.45\textwidth]{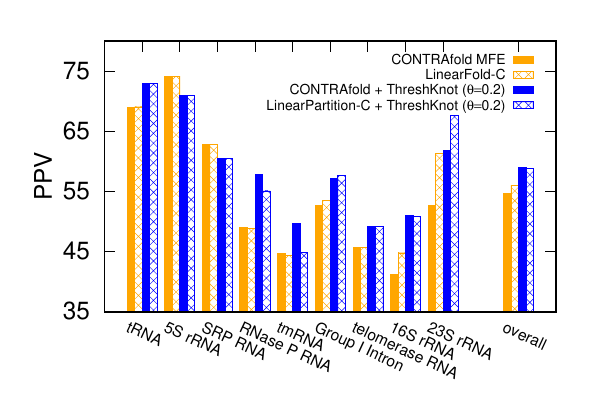}
    &
    \hspace{-0.1cm}
    \includegraphics[width=.45\textwidth]{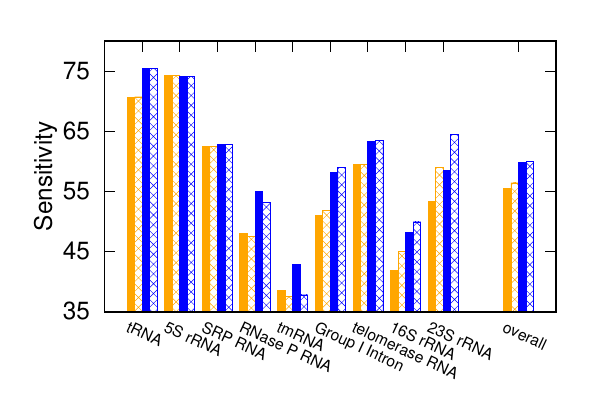}
  \end{tabular} \\[-0.5cm]
  \caption{Accuracy comparison of ThreshKnot structure ($\theta=0.2$) between \contrafold and \linearpartitionc on ArchiveII dataset. $\theta$ is the hyperparameter that balances PPV and Sensitivity. Note that \linearpartitionc + ThreshKnot is significantly worse than \contrafold + ThreshKnot on two families in both PPV and Sensitivity, tmRNA and RNase P RNA ($p < 0.01$), and significantly better on three longer families in Sensitivity, Group I Intron ($p < 0.01$), telomerase RNA and 16S rRNA ($0.01 \leq p < 0.05$).
  \label{fig:threshknot_lpc}}
%\vspace{1cm}
\end{figure}
\fi

\begin{figure}[h]
  \centering
  \captionsetup{singlelinecheck=off}
\begin{tabular}{cc}
\hspace{-6.cm} \panel{A} & \hspace{-6cm}\panel{B} \\[-0.5cm]%& \hspace{-4.6cm}\panel{C} & \hspace{-4.6cm}\panel{D}\\[-0.2cm]
\hspace{-0.2cm}\includegraphics[width=0.35\textwidth]{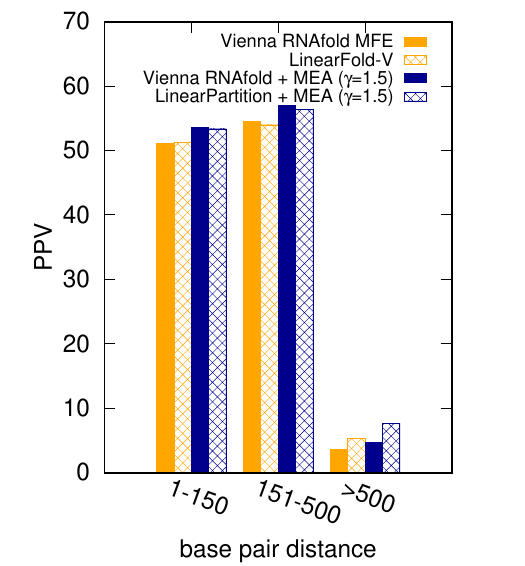} &
\hspace{-0.35cm}\includegraphics[width=0.35\textwidth]{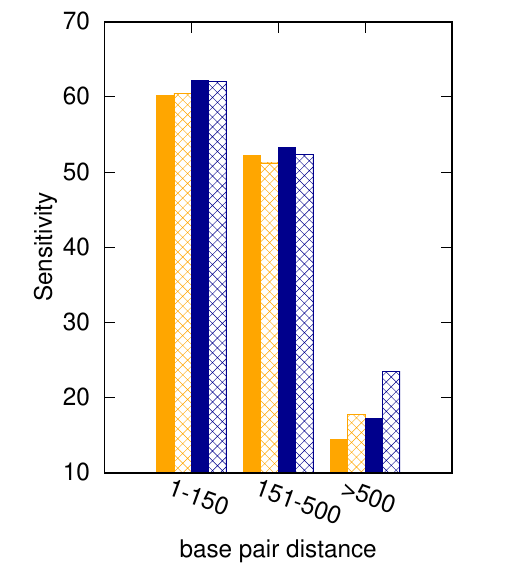} \\
\hspace{-6.cm} \panel{C} & \hspace{-6cm}\panel{D} \\[-0.5cm]
\hspace{-0.35cm}\includegraphics[width=0.35\textwidth]{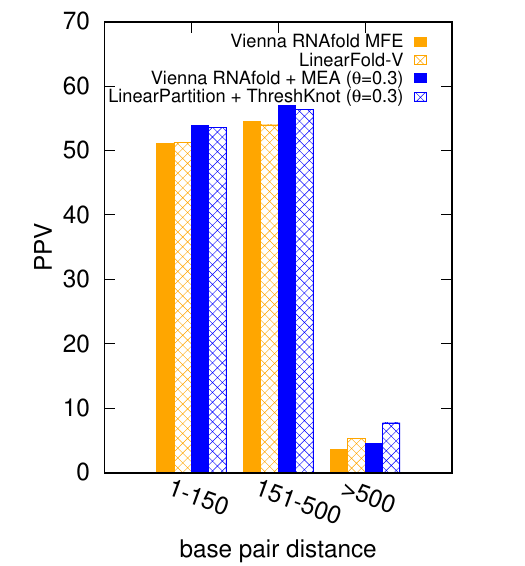} &
\hspace{-0.35cm}\includegraphics[width=0.35\textwidth]{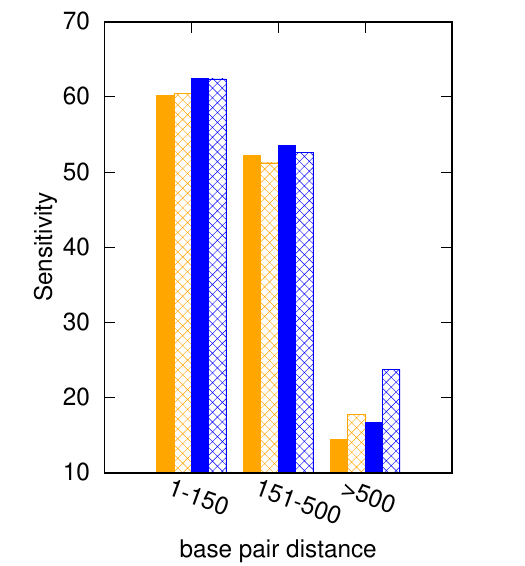}
\end{tabular}
  \caption[.]{Accuracy comparison of base pair prediction with different base pair distances. 
  Each bar represents the overall PPV/sensitivity of all predicted 
base pairs in a certain length range across all sequences. 
\linearpartition performs best on long base pairs over four systems. 
{\bf A} and {\bf B}: Comparison using MEA structures. 
{\bf C} and {\bf D}: Comparison using \threshknot structures.
In all cases, \linearpartition's base pair probabilities lead to substantially better accuracies on long-distance pairs (500+ \nts apart).
  \label{fig:distance}}
%\vspace{1cm}
\end{figure}

\iftrue
\begin{figure}[h]
  \centering
%\hspace{-0.5cm}
\begin{tabular}{ll}
{\large\sf A} & {\large\sf B}\\
    \includegraphics[width=.45\textwidth]{figs/overall_vienna_prob_bin_count}
    &
    \hspace{-0.1cm}
    \includegraphics[width=.45\textwidth]{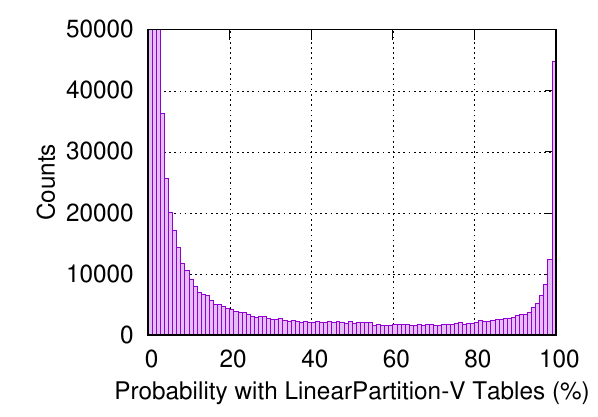}
  \end{tabular} 
  \caption{Pair probability distributions of \viennarnafold and \linearpartitionv are similar.
  {\bf A}: 
  Pair probability distribution of \viennarnafold;
  {\bf B}: 
  Pair probability distribution of \linearpartitionv.
  The count of \linearpartitionv in bin [99,100) is slightly bigger than \viennarnafold, 
  while the count in bin [0,1) (cut here at 50,000) is much less than \viennarnafold 
  (2,068,758 for \linearpartitionv and 48,382,357 for \viennarnafold).
  % This comparison also confirm that \linearpartition prune out lots of base pairs with probabilities close to 0, and the base pairing probability distribution of \linearpartition is peakier.
  \label{fig:bin_counts}}
%\vspace{1cm}
\end{figure}
\fi

% \vspace{-0.6cm}
% \section*{References}
% \balance
% % \bibliographystyle{elsarticle-harv} % can't use any; already used unsrt in pnas-new.cls
% \bibliography{si}

%\includepdf[pages=-]{si.pdf}

\end{document}